\documentclass[a4paper,11pt]{article}

\usepackage{amsmath,amsfonts}
\usepackage{bm}
\usepackage{txfonts}
\usepackage{eurosym}
\usepackage{float}
\usepackage{color}
\usepackage{typearea}
\typearea{12}

\usepackage{graphicx}

\begin{document}

\title{Constructing Copulas Using Corrected Hermite Polynomial Expansion for Estimating Cross Foreign Exchange Volatility}
\author{Kenichiro Shiraya\thanks{Graduate School of Economics, The University of Tokyo. Kenichiro Shiraya is supported by Center for Advanced Research in Finance (CARF).} \and Tomohisa Yamakami\thanks{Graduate School of Economics, The University of Tokyo. Mizuho-DL Financial Technology Co., Ltd. The opinions expressed herein are only those of the authors. They do not represent the official views of the Mizuho-DL Financial Technology Co., Ltd.}}

\date{January 24, 2023}
\maketitle

\begin{abstract}
  Copulas are used to construct joint distributions in many areas.
  In some problems, it is necessary to deal with correlation structures
that are more complicated than the commonly known copulas.
  A finite order multivariate Hermite polynomial expansion, as an approximation of a joint density function,
  can handle complex correlation structures.
  However, it does not construct copulas because the density function can take negative values.
  In this study, we propose a method to construct a copula
  based on the finite sum of multivariate Hermite polynomial expansions by applying corrections to the joint density function.
  
  Furthermore, we apply this copula to estimate the volatility smile of cross currency pairs in the foreign exchange option market.
  This method can easily reproduce the volatility smile of cross currency pairs by appropriately
adjusting the parameters
  and following the daily volatility fluctuations even if the higher-order parameters are fixed. 
  In the numerical experiments, we compare the estimation results of the volatility smile of EUR-JPY
  with those of USD-JPY and EUR-USD for the proposed and other copulas,
  and show the validity of the proposed copula.  

\end{abstract}

\section{Introduction}
A copula is a function that expresses the relationship between joint
and marginal distributions in multivariate random variables. It allows us to treat the complex correlation structure among random variables
separately from the marginal distributions.
Given this property, copulas are used to compose a joint distribution from marginal distributions.
In the field of derivatives pricing, 
a variety of applications have been proposed by composing a joint distribution with a copula and distributions of underlying assets prices,
such as basket options, worst of options, and options of cross currency pairs
(see e.g., Cherubini et al. \cite{cherubini2004copula}, Taylor and Wang \cite{taylor2010option}).
For the valuation of Collateralized Debt Obligations (CDOs), Li \cite{li2000default} and Laurent and Gregory \cite{laurent2005basket} have proposed a method to obtain the simultaneous survival probabilities using copulas of the survival probabilities of each entity.
In the field of portfolio risk management,
copulas are used to calculate conditional value-at-risk (CVaR) (see e.g., Stoyanov et al. \cite{stoyanov2010stochastic}).
An extension of the CVaR using a copula has been studied by Krzemienowski and Szymczyk \cite{krzemienowski2016portfolio}.
In X-Value Adjustment (XVA),  such as Credit Valuation Adjustment,
a copula is used to construct the simultaneous survival probability of self and a counterparty (see e.g., Brigo et al. \cite{brigo2013counterparty}).
Cherubini \cite{cherubini2013credit} have introduced a copula to explain the correlation between a swap rate and survival probability to evaluate wrong way risk. 

Most of these examples have used Archimedes, Gaussian, and $t$ copulas which are classical and commonly known.
However, in some cases, these copulas cannot appropriately represent complex correlation structures.
Therefore, many attempts have been made to apply copulas which can represent more complex correlation structures depending on problems.
For example, Tavin \cite{tavin2013application} have proposed to use Bernstein polynomial copulas for multi-asset option pricing.
Tavin \cite{tavin2018measuring} have applied copulas to measure dependence risk of a portfolio.

However, an extension of the Gaussian copula is more suitable for representing a joint distribution,
which is close to a multidimensional normal distribution.
Tsionas and Andrikopoulos \cite{tsionas2020high} have used a copula derived from a multivariate mixture of normal distributions in their study.
This copula can express a complex correlation structure using many parameters.
As an approximation of a joint distribution by extension of a multidimensional normal distribution,
a method using multivariate Hermite polynomial expansion,
 has been studied by Marumo and Wolff \cite{Marumo2013non} and Amengual et al. \cite{amengual2020normal}.
To construct copulas, multivariate Hermite polynomial expansions must be non-negative; however,
non-negativity may not be satisfied in a general case truncated by finite terms.
This point has not been resolved in these studies.
Quatto et al. \cite{QUATTO2021101529} have proposed a constraint on the coefficients of the fourth-order multivariate Hermite polynomial expansion. While it satisfies the copula's condition,
the constraint is very strong and difficult to apply to a variety of issues.

Takahashi and Tsuzuki \cite{takahashi2016new} have proposed a method to correct a one-dimensional function
to satisfy the properties of the probability density, such as the non-negativity and normalized conditions, which means that the total probability is 1.
The functions satisfying these properties are represented as closed convex sets on Hilbert spaces.
The correction is applied by calculating the projection onto this set.
Dykstra's algorithm calculates this projection.
In our study, we formulate a multidimensional Hermitian polynomial expansion with a correlation matrix in a weight function
and propose a method to construct copulas by extending Takahashi and Tsuzuki's method to $N$ dimensions.
Furthermore, we apply this copula to the problem of option-pricing of cross currency pairs and show its validity.

Foreign exchange and its derivatives are the second most heavily traded category of financial instruments after interest rates. Therefore,
market participants need to quickly reflect market information and their views on their prices.
A triangular relationship in the foreign exchange rate among the three currencies is a typical feature of the foreign exchange market.
For example, let the foreign exchange rates of EUR-USD, USD-JPY, and EUR-JPY be $S_{\mbox{\tiny \euro \$}}$, $S_{\mbox{\tiny \$Y\llap{=}}}$, and $S_{\mbox{\tiny \euro Y\llap{=}}}$.
A currency pair with the key currency USD is called a straight currency pair, while a currency pair between two non-key currencies is called a cross currency pair. 
Assuming no arbitrage, $S_{\mbox{\tiny \euro Y\llap{=}}}=S_{\tiny \mbox{\euro \$}} \times S_{\tiny \mbox{\$Y\llap{=}}}$ must be satisfied.
Therefore, if $S_{\mbox{\tiny \euro \$}}$ is moved, traders also have to change the rates of $S_{\mbox{\tiny \euro Y\llap{=}}}$ and $S_{\tiny \mbox{\$Y\llap{=}}}$ to satisfy this equation.
This relationship among foreign exchange rates constrains the relationship among the implied volatilities of each currency pair.
When two foreign exchange rates follow a multi-asset Black-Scholes model, in which a correlation between two Brownian motions is constant,
the joint distribution of log foreign exchange rates at option maturity follows a bivariate normal distribution.
In this situation, the correlation structure is expressed using a Gaussian copula,
and a simple relationship among the three volatilities is derived.
In the real market, implied volatility depends on the strike price, called the volatility smile.
A relationship among the implied volatilities of currency pairs, under the assumption of volatility smiles, is complex and has been studied using several approaches.

One approach involves obtaining the option prices of cross currency pairs by focusing only on the risk-neutral probability distributions of the underlying asset prices at maturity.
According to Breeden and Litzenberger, risk-neutral probability distributions can be obtained using implied volatilities.
By considering the risk-neutral probability distributions of straight currency pairs as marginal distributions
and introducing an appropriate correlation structure using a copula,
the joint distribution is constructed and the option price of the cross-currency pair is obtained as a two-asset option.
Taylor and Wang \cite{taylor2010option} have studied this approach.

Given that the volatility of cross currency pairs is less liquid than that of straight currency pairs in the market,
there is a need to estimate volatility of cross currency pairs using volatilities of straight currency pairs.
Combinations of European options such as At-The-Money straddle (ATM), Risk Reversal (RR), and Butterfly (BF) are standard in the currency options market.
The liquidity of RR and BF is lower than that of ATM.
Therefore, 
when only the ATM of the cross currency pairs and volatility smiles of the straight currency pairs are available, it is required that the volatility smile of the cross-currency pairs be more precisely estimated.
Most of the existing studies that have considered the copula approach have used classical copulas with a small number of parameters,
which are difficult to fit the entire volatility smile of cross currency pairs.
In this study, we derive a copula from a joint density using a multivariate Hermite polynomial expansion and apply it to estimate the volatility smiles of cross currency pairs.
We present a procedure for calculating option prices suitable for such a copula and
show that it has enough parameters that can be fit to the volatility smile of cross currency pairs.
The numerical examples generated using volatility data for currency pairs among EUR, USD, and JPY show that the proposed copula is a better fit to volatility smiles of cross currency pairs than other classical copulas.
We also show that estimating the volatility smile of cross currency pairs using parameters calibrated on month-ends
and volatility smiles of straight currency pairs calibrated during the month, yield better estimation results than other classical copulas. 

The remainder of this study is organized as follows. 
In Section 2, we describe the properties and examples of classical copulas and introduce approximation method of the probability density using orthogonal polynomial expansion to prepare the proposed method. 
In Section 3, we propose the construction of copulas using a multivariate Hermite polynomial expansion and its parameter estimation,
and a correction method using a projection onto a convex set in a case where the definition of probability density is not satisfied.
In Section 4, we discuss a method for estimating the volatility smile of cross currency pairs using the proposed copula.
In Section 5, we present numerical examples of the comparison between the proposed and other classical copulas. We do this 
by approximating the joint distribution and estimating volatility smiles of cross currency pair using the volatility data on EUR, USD, and JPY.
We conclude the study in Section 6. 

\section{Preliminary}
In this section, we introduce the properties of copulas and approximate probability density using the orthogonal polynomial expansion to prepare for the proposed method.
We assume a probability space $(\Omega,\mathcal{F},\mathbb{P})$ satisfying appropriate conditions and use the following symbols throughout this study.

$\mathbb{N}_{+}$: the set of natural numbers.
$\mathbb{N}_{0}$: the set of natural numbers containing 0.
$\bm{x}:=(x_{1},\cdots,x_{N})'$: $N$-dimensional vectors,
and we use the notation that a function with a vector argument is identical to an $N$-variable function as $f(\bm{x})=f(x_{1},\cdots,x_{N})$.
$\partial_{\bm{x}}:=\frac{\partial^{N}}{\partial x_{1}\cdots\partial x_{N}}$: differentiation operator that differentiates partials once, using an element of $\bm{x}$. 
Particularly, if the subscript is scalar, the differential operation $\partial_{x}:=\frac{\partial}{\partial x}$ is used.
$\Phi_{\Sigma}(\bm{x})$: $N$-variate standard normal distribution with correlation matrix $\Sigma$.
$\Phi(x)=\Phi_{1}(x)$: standard normal distribution, the latter is used when we want to especially emphasize that it is one dimension.
$\int_{A} f(\bm{x}) d\bm{x}:=\int_{A} f(\bm{x}) dx_{1}\cdots dx_{N}$: multiple integrals in the domain $A$.
In the context of functional, function arguments are omitted and real numbers (such as $0$, $1$) are used as constant functions that take their values.

\subsection{Properties of Copula}
A copula is a function that expresses the relationship between a multivariate joint distribution and marginal distributions. It
is often used as a method to construct a joint distribution based on marginal distributions.

We assume that $N$-dimensional random variables $\bm{X}:=(X_{1},\cdots,X_{N})'$ have the joint distribution $F(x_{1},\cdots,x_{N}):=\mathbb{P}\left(X_{1}\leq x_{1},\cdots,X_{N}\leq x_{N}\right)$
and also have the continuous marginal distributions $G_{1}(x_{1}):=\mathbb{P}\left(X_{1}\leq x_{1}\right),\cdots,G_{N}(x_{N}):=\mathbb{P}\left(X_{N}\leq x_{N}\right)$.
According to Sklar's theorem (see e.g., Cherubini et al. \cite{cherubini2004copula}), there exists a unique copula $\mathcal{C}_{F}(u_{1},\cdots,u_{N})$ satisfying
\begin{eqnarray}
    F(x_{1},\cdots,x_{N})=\mathcal{C}_{F}(G_{1}(x_{1}),\cdots,G_{N}(x_{N})).\label{eq:sklar}
\end{eqnarray}
Furthermore, if $G_{1}(x_{1}),\cdots,G_{N}(x_{N})$ have inverse functions, we can write
\begin{eqnarray}
  \mathcal{C}_{F}(u_{1},\cdots,u_{N}):=F\left(G_{1}^{-1}(u_{1}),\cdots,G_{N}^{-1}(u_{N})\right),
\end{eqnarray}
which is called the copula derived from the joint distribution $F(\bm{x})$.
Based on this formula, copulas can be considered as joint distributions whose marginal distributions follow the uniform distribution of $[0,1]$.

For simplicity of notation, we use lower-case in bold to denote $N$-dimensional vectors as
$\bm{x}:=(x_{1},\cdots,x_{N})'$, $\bm{u}:=(u_{1},\cdots,u_{N})'$ and identify a $N$-dimensional vector argument 
with $N$ variable arguments (e.g., $F(\bm{x}):=F(x_{1},\cdots,x_{N})$).
We use bold capital letters to map $N$-dimensional vectors onto $N$-dimensional vectors
(e.g., $\bm{G}(\bm{x}):=(G_{1}(x_{1}),\cdots,G_{N}(x_{N}))'$). 

If $F(\bm{x})$ in \eqref{eq:sklar} has a joint density, 
it can be differentiated as
\begin{eqnarray}
  \partial_{\bm{x}}F(\bm{x})=\partial_{\bm{u}}C_{F}(\bm{G}(\bm{x}))\prod_{i=1}^{N}\partial_{x_{i}}G_{i}(x_{i}),
\end{eqnarray}
where we use the differential operator $\partial_{\bm{x}}:=\frac{\partial^{N}}{\partial x_{1}\cdots\partial x_{N}}$ ($\partial_{x}:=\frac{\partial}{\partial x}$ for scalar),
and $\partial_{\bm{u}}\mathcal{C}_{F}(\bm{u})$ is called the copula density.
Given the non-negativity of the joint density, the copula density must also be non-negative. 

Next, we define a monotonically increasing function.$\alpha_{1}(x),\cdots,\alpha_{N}(x)$ which has inverse functions in the whole regions
and considers a random variable $\bar{X}_{1}:=\alpha_{1}(X_{1}),\cdots,\bar{X}_{N}:=\alpha_{N}(X_{N})$.
Given the joint distribution $\bar{F}(\bar{x}_{1},\cdots,\bar{x}_{N})$ of these random variables are expressed as
\begin{eqnarray}
  \bar{F}(\bar{x}_{1},\cdots,\bar{x}_{N})&:=&\mathbb{P}\left(\bar{X}_{1}\leq\bar{x}_{1},\cdots,\bar{X}_{N}\leq\bar{x}_{N}\right)\nonumber\\
  &=&\mathbb{P}\left(X_{1}\leq\alpha_{1}^{-1}(\bar{x}_{1}),\cdots,X_{N}\leq\alpha_{N}^{-1}(\bar{x}_{N})\right)\nonumber\\
  &=&F\left(\alpha_{1}^{-1}(\bar{x}_{1}),\cdots,\alpha_{N}^{-1}(\bar{x}_{N})\right),
\end{eqnarray}
and the marginal distributions are
\begin{eqnarray}
  \bar{G}_{1}(\bar{x}_{1})&:=&\mathbb{P}\left(\bar{X}_{1}\leq\bar{x}_{1}\right)=G_{1}\left(\alpha_{1}^{-1}(\bar{x}_{1})\right),\\
  &\vdots&\nonumber\\
  \bar{G}_{N}(\bar{x}_{N})&:=&\mathbb{P}\left(\bar{X}_{N}\leq\bar{x}_{N}\right)=G_{N}\left(\alpha_{N}^{-1}(\bar{x}_{N})\right).
\end{eqnarray}
The copula derived from $\bar{F}$ is
\begin{eqnarray}
  \mathcal{C}_{\bar{F}}(\bm{u})&:=&\bar{F}\left(\bar{G}_{1}^{-1}(u_{1}),\cdots,\bar{G}_{N}^{-1}(u_{N})\right)\nonumber\\
  &=&F\left(G_{1}^{-1}(u_{1}),\cdots,G_{N}^{-1}(u_{N})\right),
\end{eqnarray}
which is the same as $\mathcal{C}_{F}(\bm{u})$. 
Therefore, when considering copulas derived from the probability distribution $F(\bm{x})$,
no generality is lost and it is assumed that the probability distribution is standardized by linear transformation scaled to mean 0 and variance 1. 
Furthermore, when we consider the joint distribution of exponential or logarithmic random variables on an appropriate domain, the same copula is obtained.
Moreover, the same copula is derived from a joint distribution obtained by transforming the random variables 
such that the marginal distributions are standard normal distributions.

The following are well-known classical copulas (see e.g., Cherubini et al. \cite{cherubini2004copula}, Taylor and Wang \cite{taylor2010option}).
\begin{itemize}
  \item Archimedean copula\\
  Using a function $\phi$ that satisfies appropriate conditions, which is called a generator function,
  a copula that can be expressed as
  \begin{eqnarray}
    \mathcal{C}(u_{1},\cdots,u_{N})=\phi^{-1}(\phi(u_{1})+\cdots+\phi(u_{N})),
  \end{eqnarray}
 is called an Archimedean copula. Table \ref{table:ArcCopula} shows typical variations.
  
  \begin{table}[hbtp]
    \caption{Typical Archimedean copulas' generator functions and two-dimensional representations}
    \label{table:ArcCopula}
    \centering
    \begin{tabular}{c|c|c}
      Name & $\phi(u)$ & $C(u_{1},u_{2})$\\\hline
      Clayton copula & $\frac{1}{\theta}(t^{-\theta}-1)$ & $\left\{\max\left(u_{1}^{-\theta}+u_{2}^{-\theta}-1,0\right)\right\}^{\frac{1}{\theta}}$\\
      Frank copula & $\log\left(\frac{e^{\theta u}-1}{e^{\theta}-1}\right)$ & $-\frac{1}{\theta}\log\left(1+\frac{(e^{-\theta u_{1}}-1)(e^{-\theta u_{2}}-1)}{e^{-\theta}-1}\right)$\\
      Gumbel copula & $(-\log(u))^{\theta}$ & $e^{-\left[(-\log(u_{1}))^{\theta}+(-\log(u_{2})^{\theta})\right]^{1/\theta}}$\\
    \end{tabular}
  \end{table}

  \item Plackett copula\\
    A copula denoted by
    \begin{eqnarray}
      C_{Pl(\theta)}:=\frac{[1+(\theta-1)(u_{1}+u_{2})]-\sqrt{[1+(\theta-1)(u_{1}+u_{2})]^{2}-4u_{1}u_{2}\theta(\theta-1)}}{2(\theta-1)},
    \end{eqnarray}
    is called a two-dimensional Plackett copula.

    \item Gauss copula\\
    A copula derived from the distribution $\Phi_{\Sigma}$ of A multivariate standard normal distributions with correlation matrix $\Sigma$, is called a Gauss copula.
    \begin{eqnarray}
      C_{\Phi_{\Sigma}}(u_{1},\cdots,u_{N}):=\Phi_{\Sigma}\left(\Phi^{-1}(u_{1}),\cdots,\Phi^{-1}(u_{N})\right).
    \end{eqnarray}
  \end{itemize}

\subsection{Orthogonal Polynomial Expansion}
Let $\mathcal{M}_{N}$ be the set of all Lebesgue measurable subsets of $\mathbb{R}^{N}$ $(N\in \mathbb{N}_{+})$
and $\mu$ be the measure with a density $p_{\mu}(\bm{x})$. 
We consider the measure space $(\mathbb{R}^{N},\mathcal{M}_{N},\mu)$. 

The set $\mathcal{H}_{\mu}^{N}=L^{2}(\mathbb{R}^{N},\mathcal{M}_{N},\mu)$ of the square integrable real-valued functions on this measure space is a Hilbert space defined by the inner product
\begin{eqnarray}
  \left<f,g\right>_{\mu}&:=&\int_{\mathbb{R}^{N}} f(\bm{x})g(\bm{x})d\mu \ \ \ f,g\in \mathcal{H}_{\mu}^{N},\\
  &=&\int_{\mathbb{R}^{N}} f(\bm{x})g(\bm{x})p_{\mu}(\bm{x})d\bm{x}.
\end{eqnarray}
Hereafter, we consider only the measure $\mu$ whose $\mathcal{H}_{\mu}^{N}$ contains all $N$-variable polynomials of arbitrary degree.
We define that $\left\|f\right\|_{\mu}:=\sqrt{\left<f,f\right>_{\mu}}$ is the norm defined based on the inner product.

First, we consider a case where $N=1$.
A polynomial sequence $R_{n}$ of degree $n$ is called an orthogonal polynomial sequence if it satisfies $\left<R_{m},R_{n}\right>_{\mu}=0, (m\neq n\ m,n\in\mathbb{N}_{0}$).
The orthogonal polynomial sequence can be constructed, for example, by applying Gram-Schmidt orthogonalization to the monomial sequence $1,x,x^{2},\cdots$. 
Furthermore, the normalized polynomial sequence whose norm is 1 is defined as $\bar{R}_{n}:=\frac{R_{n}}{\left\|R_{n}\right\|_{\mu}}$ and called an orthonormal polynomial sequence.
A normalized orthonormal polynomial sequence exists uniquely if the coefficients of the highest order are assumed to be positive.

Using an orthonormal polynomial sequence $\bar{R}_{i}(x)$ and a real number sequence $m_{i},\ i\in\mathbb{N}_{0}$, an expansion
\begin{eqnarray}
  \tilde{p}_{\mu}(x):=\left\{\sum_{i=0}^{\infty}m_{i}\bar{R}_{i}(x)\right\}p_{\mu}(x),
\end{eqnarray}
is called the orthogonal polynomial expansion.
According to the Riesz-Fische's theorem, this series converges when $\sum_{i=0}^{\infty}m_{i}^{2}<\infty$ (see Krall \cite{krall2002hilbert}).

Next, we consider the case where $\mu$ is a probability measure and $m_{0}=1$.
Under this condition, 
\begin{eqnarray}
  \int_{\mathbb{R}}\tilde{p}_{\mu}(x)dx&=&\int_{\mathbb{R}}p_{\mu}(x)dx=1,
\end{eqnarray}
holds. Additionally, if the condition $\tilde{p}_{\mu}(x)\geq0$ is satisfied,
$\tilde{p}_{\mu}(x)$ is a probability density.

Based on the fact that $x^{j}$ is orthogonal to $\bar{R}_{i}$ $(i>j\geq0)$, 
\begin{eqnarray}
  \int_{\mathbb{R}}x^{j}\tilde{p}_{\mu}(x)dx&=&\int_{\mathbb{R}}x^{j}p_{\mu}(x)dx+\sum_{i=1}^{j}m_{i}\left<x^{j},\bar{R}_{i}(x)\right>_{\mu},
\end{eqnarray}
holds.
Therefore, $m_{i}$ $(i>j>0)$ are parameters that do not affect the moments of $j$ degrees.

When we consider the measure $\mu_{\Phi_{1}}$, whose density is the standard normal density function $\partial_{x}\Phi_{1}(x)$,
$R_{n}$ are identified as the Hermite polynomials $\mbox{He}_{n}$ calculated using
\begin{eqnarray}
  \mbox{He}_{0}(x)&=&1,\\
  \mbox{He}_{1}(x)&=&x,\\
  \mbox{He}_{n+1}(x)&=&x\mbox{He}_{n}(x)-n\mbox{He}_{n-1}(x).
\end{eqnarray}
The norm is
\begin{eqnarray}
  \left<\mbox{He}_{n},\mbox{He}_{n}\right>_{\mu_{\Phi_{1}}}=n!,
\end{eqnarray}
and $\bar{\mbox{He}}_{n}:=\frac{1}{\sqrt{n!}}\mbox{He}_{n}$ constitutes an orthonormal polynomial sequence.
Furthermore, when $m_{i}$ is expressed using the cumulant of $\tilde{p}_{\mu}(x)$,
it is called the Gram-Charlier expansion.

In $N(>1)$-dimensional case, an $n$-degree polynomial that is orthogonal to any $(n-1)$-degree polynomial is called an orthogonal polynomial.
The dimension of the linear space based on $n$-degree $N$-variable polynomials,
which is orthogonal to a $n-1$-degree $N$-variable polynomial, is expressed using the combination $M_{n}:=\begin{pmatrix}N+n-1\\N-1\end{pmatrix}$.
Therefore, there is no unique orthogonal polynomial and various bases are possible.
Among these variations, it is useful to consider the case where $n$-degree orthogonal polynomials are orthogonal to each other in application.

As a special case, let us consider an $N$-dimensional density of a measure which is expressed as a product of one-dimensional densities. 
Let the densities of the one-dimensional measures $\mu_{1},\cdots,\mu_{N}$ be $p_{\mu_{1}}(x),\cdots,p_{\mu_{N}}(x)$
and $\bar{R}_{\mu_{1},j},\cdots,\bar{R}_{\mu_{N},j}$ $(j\in\mathbb{N}_{0})$ be their $j$-th orthonormal polynomials, respectively.
Assuming the density of $N$-dimensional measure $\mu_{\perp}$ is defined by $p_{\mu_{\perp}}(\bm{x}):=\prod_{i=1}^{N}p_{\mu_{i}}(x_{i})$,
the set of polynomials.
\begin{eqnarray}
  E_{\mu_{\perp},n}&:=&\left\{\prod_{i=1}^{N}\bar{R}_{\mu_{i},j_{i}}\left(x_{i}\right) \middle| j_{i}\in \mathbb{N}_{0}, \sum_{i=1}^{N}j_{i}=n\right\}\ \ \ n \in \mathbb{N}_{0},\label{formula:nDimOrthogonalPolynomialsPart}
\end{eqnarray}
are $n$-degree orthonormal polynomials whose elements are orthogonal to each other with respect to the inner product.
$\left<\cdot,\cdot\right>_{\mu_{\perp}}$.
Note that the number of elements of \eqref{formula:nDimOrthogonalPolynomialsPart} is $M_{n}$,
indicating that the arbitrary $n$-degree $N$-variate orthogonal polynomial can be expressed using the linear combination of elements in \eqref{formula:nDimOrthogonalPolynomialsPart}.
Thus, 
\begin{eqnarray}
  E_{\mu_{\perp}}&:=&\bigcup_{n=0}^{\infty}E_{\mu_{\perp},n},\label{formula:nDimOrthogonalPolynomialsAll}
\end{eqnarray}
forms an $N$-variate orthonormal polynomial system, and any polynomial can be expressed by the linear combination of these elements.
See Dunkl and Xu \cite{dunkl2014orthogonal} for a detailed discussion of the aforementioned multivariate orthogonal polynomials.

Using the $N$-variate orthonormal polynomial system \eqref{formula:nDimOrthogonalPolynomialsAll},
we obtain an $N$-variate orthogonal polynomial expansion
\begin{eqnarray}
  \tilde{p}_{\mu_{\perp}}(\bm{x}):=\left\{\sum_{n=0}^{\infty}\sum_{j_{1}+\cdots+j_{N}=n}m_{j_{1},\cdots,j_{N}}\prod_{i=1}^{N}\bar{R}_{\mu_{i},j_{i}}(x_{i})\right\}\prod_{l=1}^{N}p_{\mu_{l}}(x_{l}),
\end{eqnarray}
with coefficients $m_{j_{1},\cdots,j_{N}}$.

Similar to the one-dimensional case, it converges if $\sum_{n=0}^{\infty}\sum_{j_{1}+\cdots+j_{N}=n}m_{j_{1},\cdots,j_{N}}^{2}<\infty$.
Furthermore, when $\mu_{\perp}$ is a probability measure and $m_{0,\cdots,0}=1$,
\begin{eqnarray}
  \tilde{p}_{\mu_{\perp}}(\bm{x}):=\left\{1+\sum_{n=1}^{\infty}\sum_{j_{1}+\cdots+j_{N}=n}m_{j_{1},\cdots,j_{N}}\prod_{i=1}^{N}\bar{R}_{\mu_{i},j_{i}}(x_{i})\right\}\prod_{l=1}^{N}p_{\mu_{l}}(x_{l}),
\end{eqnarray}
is a probability density if $\tilde{p}_{\mu_{\perp}}(\bm{x})\geq0$ is satisfied.
Particularly, when $\mu_{1}=\cdots=\mu_{N}=\mu_{\Phi_{1}}$,
we obtain the orthogonal polynomial expansions of the $N$-dimensional probability density using $N$-dimensional Hermite polynomials.

\subsection{Copulas derived from orthogonal polynomial expansion}
Several copulas derived from the probability density expressed using the orthogonal polynomial expansion have been proposed.

Bakam and Pommeret \cite{bakam2020nonparametric} have proposed a copula derived from a shifted Legendre polynomial expansion whose density of a probability measure is 1 on $[0,1]^{N}$.
Marumo and Wolff \cite{Marumo2013non} and Amengual et al. \cite{amengual2020normal} have discussed copulas in two dimensions using Hermite polynomials.
These probability densities using orthogonal polynomial expansions and derived copulas have a problem:
the probability density may take a negative value when truncated using finite terms in a general case,
and a copula cannot be derived from the density.
As a copula using non-orthogonal polynomials, Sancetta et al. \cite{sancetta2004bernstein} have proposed a copula using Bernstein polynomials
in which the conditions for coefficients to satisfy the copula are treated.
However, in a problem where the copula has a Gaussian-like joint distribution,
using a copula extended from a Gaussian copula may provide a better approximation with fewer parameters than
other complex copulas such as Bakam and Pommeret \cite{bakam2020nonparametric} and Sancetta et al. \cite{sancetta2004bernstein}.

In the next section, we discuss the formulation of multivariate Hermite polynomial expansions for probability density
and a method to solve the non-negativity of probability density when the expansion is truncated by finite terms.

\section{Construction of Copulas using Hermite Polynomial Expansion}
\label{section:CopulaWithHermiteExpansion}

Several studies have constructed a multidimensional probability density using multivariate Hermite polynomial expansion. 
However, there are variations for taking densities of measures and bases of the polynomial expansion.

In our formulation, we consider the relation between the inner product under the measures $\mu_{\Phi_{\Sigma}}$ and $\mu_{\Phi_{\mathbb{I}}}$
whose densities are $\partial_{\bm{x}}\Phi_{\Sigma}$ and $\partial_{\bm{x}}\Phi_{\mathbb{I}}$.
We derive a simple expression and calculation of the multivariate Hermite expansion on the measure $\mu_{\Phi_{\Sigma}}$
converting to that on the measure $\mu_{\Phi_{\mathbb{I}}}$.

When orthogonal polynomial expansions are truncated by finite terms,
the probability density may become negative. In this case, the definition of probability density and copulas are not satisfied.
To solve this problem,
we extend the correction method using a probability density proposed by Takahashi and Tsuzuki to $N$-dimensions. The method is
then applied to obtain the probability density and copulas.

\subsection{Hermite Polynomial Expansion for Joint Density}
Let $\Sigma$ be a regular $N$-dimensional correlation matrix and
$\Gamma$ be a decomposition that fills $\Sigma = \Gamma\Gamma^{T}$.
Cholesky decomposition is one example, but decomposition is not limited to it.

Let $\mu_{\Phi_{\Sigma}}$ be the measure with density $\partial_{\bm{x}}\Phi_{\Sigma}$
which is the $N$-dimensional standard normal distribution with the correlation matrix $\Sigma$.
By performing the variables transformation $\bm{v}:=\Gamma^{-1}\bm{x}$,
the inner product of any function$f,g\in\mathcal{H}_{\mu_{\Phi_{\Sigma}}}^{N}$ is
\begin{eqnarray}
  \left<f,g\right>_{\mu_{\Phi_{\Sigma}}}&=&\int_{\mathbb{R}^{N}}f(\bm{x})g(\bm{x})\partial_{\bm{x}}\Phi_{\Sigma}(\bm{x})d\bm{x}\nonumber\\
  &=&\int_{\mathbb{R}^{N}}f(\Gamma\bm{v})g(\Gamma\bm{v})\partial_{\bm{v}}\Phi_{\mathbb{I}}(\bm{v})d\bm{v}\nonumber\\
  &=&\left<f \circ \Gamma,g \circ \Gamma \right>_{\mu_{\Phi_{\mathbb{I}}}},\label{formula:iptransform}
\end{eqnarray}
where $\mu_{\Phi_{\mathbb{I}}}$ is a special case of $\Sigma=\mathbb{I}$
and the notation $(f \circ \Gamma)(\bm{v}):=f(\Gamma\bm{v})$ is used as a composition of maps. 
Therefore, the inner product is preserved by this variable transformation and change of measures. 

Let $\left\{\bm{x}\right\}_{i}$ be the $i$-th element of the vector $\bm{x}$.
By considering \eqref{formula:nDimOrthogonalPolynomialsPart} and \eqref{formula:iptransform},
\begin{eqnarray}
  E_{\mu_{\Phi_{\Sigma}},n}&:=&\left\{\prod_{i=1}^{N}\bar{\mbox{He}}_{j_{i}}\left(\left\{\Gamma^{-1}\bm{x}\right\}_{i}\right) \middle| j_{i}\in \mathbb{N}_{0}, \sum_{i=1}^{N}j_{i}=n\right\}\ \ \ n \in \mathbb{N}_{0},\\
  E_{\mu_{\Phi_{\Sigma}}}&:=&\bigcup_{n=0}^{\infty}E_{\mu_{\Phi_{\Sigma}},n},
\end{eqnarray}
consist of $n$-degree orthonormal polynomial sets and an orthonormal polynomial system under the inner product $\left<\cdot,\cdot\right>_{\mu_{\Phi_{\Sigma}}}$.

Next, let $\bm{e}_{n}:=(e_{n,0},e_{n,1},\cdots,e_{n,M_{n}-1})'$ be a vector of the elements of $E_{\mu_{\Phi_{\Sigma}},n}$.
Consider the vector $(\hat{e}_{n,0},\hat{e}_{n,1},\cdots,\hat{e}_{n,M_{n}-1})':=U_{n}\bm{e}_{n}$
is obtained by multiplying it with a suitable orthogonal matrix $U_{n}$ of $M_{n}$ dimensions.
Let $\hat{E}_{\mu_{\Phi_{\Sigma}},n}:=\left\{\hat{e}_{n,i}\middle|i=0\cdots,M_{n}-1\right\}$ be the set of multiplying vector elements.
$\hat{E}_{\mu_{\Phi_{\Sigma}}}:=\bigcup_{n=0}^{\infty}\hat{E}_{\mu_{\Phi_{\Sigma}},n}$ also forms an orthonormal system, and any polynomial can be expressed by the linear combination of the elements.
Next, we consider the expression of the multivariate Hermite polynomial expansion
\begin{eqnarray}
  a(\bm{x})&:=&1+\sum_{n=1}^{\infty}\sum_{i=1}^{M_{n}}\hat{m}_{n,i}\hat{e}_{n,i}(\bm{x}),\nonumber\\
  \tilde{p}_{\mu_{\Phi_{\Sigma}}}(\bm{x})&:=&a(\bm{x})\partial_{\bm{x}}\Phi_{\Sigma}(\bm{x}),\label{formula:HermiteExpansion}
\end{eqnarray}
with $\hat{m}_{n,i}$ as the corresponding coefficient of $\hat{e}_{n,i}(\bm{x})$.

Here, we assume $\tilde{p}_{\mu_{\Phi_{\Sigma}}}(\bm{x})$ satisfies the definition of probability density.
Let $\tilde{F}_{\mu_{\Phi_{\Sigma}}}(\bm{x})$ be the joint distribution of $\tilde{p}_{\mu_{\Phi_{\Sigma}}}(\bm{x})$
and $\tilde{\bm{G}}_{\mu_{\Phi_{\Sigma}}}(\bm{x}):=(\tilde{G}_{\mu_{\Phi_{\Sigma}},1}(x_{1}),\cdots,\tilde{G}_{\mu_{\Phi_{\Sigma}},N}(x_{N}))$
be the vector of marginal distributions of $\tilde{p}_{\mu_{\Phi_{\Sigma}}}(\bm{x})$.
We construct the copula
\begin{eqnarray}
  \mathcal{C}_{\tilde{F}_{\mu_{\Phi_{\Sigma}}}}(\bm{u}):=\tilde{F}_{\mu_{\Phi_{\Sigma}}}\left(\tilde{\bm{G}}_{\mu_{\Phi_{\Sigma}}}^{-1}(\bm{u})\right).
\end{eqnarray}
This copula is an extension of the Gaussian copula because it coincides with the Gaussian copula when all parameters $\hat{m}_{n,i}$ are 0.

Next, we consider the integral calculations for a distribution constructed using this copula.
Let $\bm{Y}:=(Y_{1},\cdots,Y_{N})$ be random variables and
$\bm{G}_{\bm{Y}}(\bm{y}):=(G_{Y,1}(y_{1}),\cdots,G_{Y,N}(y_{N}))$ be the marginal distributions of $\bm{Y}$.
The joint distribution of $Y$ composed using copula $\mathcal{C}_{\tilde{F}_{\mu_{\Phi_{\Sigma}}}}$ is
\begin{eqnarray}
  \tilde{F}_{\bm{Y}}(\bm{y})&:=&\tilde{F}_{\mu_{\Phi_{\Sigma}}}\left(\tilde{\bm{G}}_{\mu_{\Phi_{\Sigma}}}^{-1}\left(\bm{G}_{\bm{Y}}(\bm{y})\right)\right).
\end{eqnarray}
Considering the variables transformation of $\bm{x}:=\tilde{\bm{G}}_{\mu_{\Phi_{\Sigma}}}^{-1}\left(\bm{G}_{\bm{Y}}(\bm{y})\right)$,
the integral with the appropriate function $h(\bm{y})$ is defined by the domain $A\subseteq \mathbb{R}^{N}$ in which the $\bm{Y}$ values can be written by
\begin{eqnarray}
  \int_{A}h(\bm{y})d\tilde{F}_{\bm{Y}}(\bm{y})&=&\int_{A}h(\bm{y})\partial_{\bm{y}}\left[\tilde{F}_{\mu_{\Phi_{\Sigma}}} \left( \tilde{\bm{G}}_{\mu_{\Phi_{\Sigma}}}^{-1} \left( \bm{G}_{\bm{Y}}(\bm{y})\right)\right)\right]d\bm{y}\nonumber\\
  &=&\int_{\mathbb{R}^{N}}h\left(\bm{G}_{\bm{Y}}^{-1} \left( \bm{G}_{\mu_{\Phi_{\Sigma}}}(\bm{x})\right)\right)\tilde{p}_{\mu_{\Phi_{\Sigma}}}(\bm{x})d\bm{x}\nonumber\\
  &=&\int_{\mathbb{R}^{N}}h\left(\bm{G}_{\bm{Y}}^{-1} \left( \bm{G}_{\mu_{\Phi_{\Sigma}}}(\Gamma\bm{v})\right)\right)a(\Gamma\bm{v})\partial_{\bm{v}}\Phi_{\mathbb{I}}(\bm{v})d\bm{v},\label{formula:intgbysubstitution}
\end{eqnarray}
where we assume that $h\left(\bm{G}_{\bm{Y}}^{-1} \left( \bm{G}_{\mu_{\Phi_{\Sigma}}}(\bm{x})\right)\right)\in \mathcal{H}_{\mu_{\Phi_{\Sigma}}}^{N}$ is satisfied.
If $h\left(\bm{G}_{\bm{Y}}^{-1} \left( \bm{G}_{\mu_{\Phi_{\Sigma}}}(\Gamma\bm{v})\right)\right)a(\Gamma\bm{v})$
is sufficiently smooth and increases with the polynomial speed,
it can be calculated with high accuracy using the trapezoidal formula, the midpoint formula, or the Gauss-Hermite integral for each dimension of $\bm{v}$. 
For example, if $h(\bm{y})$ are the monomials where $\prod_{i=1}^{N}y_{i}^{j_{i}}$, $j_{i}\in \mathbb{N}_{0}$,
the moments of the joint distribution $\tilde{F}_{\bm{Y}}(\bm{y})$ are obtained.
We can use \eqref{formula:intgbysubstitution} in this numerical integration setting.

We consider a method to calculate copula parameters $\hat{m}_{n,i}$
using a joint distribution $F_{\bm{Z}}$ of random variables $\bm{Z}:(Z_{1},\cdots,Z_{N})$.
Let $\alpha_{1}(z),\cdots,\alpha_{N}(z)$ be the monotonically increasing function.
We normalize these random variables as $\bar{\bm{X}}:=(\bar{X_{1}},\cdots,\bar{X}_{N}):=(\alpha_{1}(Z_{1}),\cdots,\alpha_{N}(Z_{N}))$.
Here, all elements of $\bar{\bm{X}}$ have mean 0, variance 1, and take the values of $\mathbb{R}$, respectively.
Let $F_{\bar{\bm{X}}}(x_{1},\cdots,x_{N}):=F_{Z}(\alpha_{1}^{-1}(z_{1}),\cdots,\alpha_{N}^{-1}(z_{N}))$ be the joint distribution of $\bar{\bm{X}}$.
Assuming $F_{\bar{\bm{X}}}(\bm{x})=\tilde{F}_{\mu_{\Phi_{\Sigma}}}(\bm{x})$,
multiplying each density by $\hat{e}_{n,i}(\bm{x})$
and integrating, we obtain
\begin{eqnarray}
  \int_{\mathbb{R}^{N}}\hat{e}_{n,i}(\bm{x})\partial_{\bm{x}}F_{\bar{\bm{X}}}(\bm{x})d\bm{x}&=&\int_{\mathbb{R}^{N}}\hat{e}_{n,i}(\bm{x})\partial_{\bm{x}}\tilde{F}_{\mu_{\Phi_{\Sigma}}}(\bm{x})d\bm{x}\nonumber\\
  &=&\hat{m}_{n,i}.\label{formula:ExtendedMoment}
\end{eqnarray}
From \eqref{formula:ExtendedMoment}, the parameter $\hat{m}_{n,i}$ is obtained by calculating the left-hand side of this equation.
Given that all elements of $\bar{\bm{X}}$ have mean $0$, $\hat{m}_{1,i}=0$.
Furthermore, if $\Sigma$ is considered to coincide with the correlation matrix of $\bar{\bm{Y}}$, we can set $\hat{m}_{2,i}=0$.
In section 5, we show the superiority of the latter setting in two dimensions using numerical comparison
between $\Sigma=\mathbb{I}$ and $\Sigma$, which is matched with the correlation matrix.

When we deal with time series data, they are considered to model a copula which changes with time.
This can be expressed by regarding parameters $\hat{m}_{n,i},\Sigma$ in \eqref{formula:HermiteExpansion} as time-depending variables $\hat{m}_{n,i}(t),\Sigma(t)$. 
When the time $t$ is fixed, the basis transformation from $E_{\mu_{\Phi_{\Sigma}},n}$ to $\hat{E}_{\mu_{\Phi_{\Sigma}},n}$ can be set so that $\frac{\partial}{\partial t}\hat{m}_{n,i}(t)=0$, $(i=2,\cdots,M_{n})$ is satisfied.
In this case, it is possible to consider that the parameters changing in small time are only $\Sigma(t)$ and $\hat{m}_{n,1}(t)$ $(n=1,\cdots)$.
From a different perspective of treating time series data,
estimating higher-order moments from real data generally requires more data than estimating lower-order moments.
To avoid the problem of estimating higher-order moments with less data,
one possible approach is to fix the higher-order moments, for example $\hat{m}_{n,i}(t)$ $(n\geq3)$, and to estimate $\Sigma(t)$ as time dependent parameters.

\subsection{Correction of Probability Density by Projection to a Convex Set}
Although the multivariate Hermite polynomial expansion (\ref{formula:HermiteExpansion}) satisfies non-negativity as an infinite sum,
the expansion truncated by the finite terms does not necessarily satisfy non-negativity.

Takahashi and Tsuzuki \cite{takahashi2016new} have proposed a method to correct a function
which does not satisfy the definition of probability density,
such as non-negativity, using projection onto a convex set in a one-dimensional case.
Here, we extend it to $N$-dimensions.
The proposed method satisfies the conditions of probability density and introduces an additional condition \eqref{formula:ExtendedMoment}, after correction.

Let $\phi$ be an element of the Hilbert space $\mathcal{H}_{\mu}^{N}=L^{2}(\mathbb{R}^{N},\mathcal{M}_{N},\mu)$,
and $\mathcal{K} \subset \mathcal{H}_{\mu}^{N}$ be a nonempty closed convex set.
We define a projection from $\phi$ to $\mathcal{K}$ using
\begin{eqnarray}
  \mbox{Proj}_{\mu,\mathcal{K}}\left(\phi\right):=\left\{\phi^{*}\in\mathcal{K} \middle| \left\|\phi-\phi^{*}\right\|_{\mu}=\inf_{\eta\in\mathcal{K}}\left\|\phi-\eta\right\|_{\mu}\right\}.
\end{eqnarray}
According to the projection theorem in a Hilbert space, there is only one such element.
Hereafter, $\mbox{Proj}_{\mu,\mathcal{K}}\left(\phi\right)$ is also used to denote the element, unless there is a misunderstanding.
If a convex set $\mathcal{K_{=}}$ can be written as $\mathcal{K_{=}}:=\left\{\phi\in\mathcal{H}_{\mu}^{N}\middle| \left<\phi, \phi^{\#}\right>_{\mu}=c \right\}$ with some $\phi^{\#} \in \mathcal{H}_{\mu}^{N}$ and $c\in\mathbb{R}$,
the projection is calculated by
\begin{eqnarray}
  \mbox{Proj}_{\mu,\mathcal{K_{=}}}(\phi):=\phi-\frac{\left<\phi,\phi^{\#}\right>_{\mu}-c}{\left\|\phi^{\#}\right\|_{\mu}^{2}}\phi^{\#}.
\end{eqnarray}
Additionally, if a convex set can be written as $\mathcal{K}_{\leq}:=\left\{\phi\in\mathcal{H}_{\mu}^{N}\middle| \left<\phi, \phi^{\#}\right>_{\mu}\leq c \right\}$,
the projection is calculated by
\begin{eqnarray}
  \mbox{Proj}_{\mu,\mathcal{K}_{\leq}}(\phi):=\phi-\frac{\left\{\left<\phi,\phi^{\#}\right>_{\mu}-c\right\}_{+}}{\left\|\phi^{\#}\right\|_{\mu}^{2}}\phi^{\#}.
\end{eqnarray}
If the inequality is reversed, we consider $\phi^{\#}\to-\phi^{\#}$ and $c\to-c$.

Let $\mathcal{K}_{i} \subset \mathcal{H}_{\mu}^{N},\ i=1,\cdots,k$, $k\in \mathbb{N}_{+}$ be closed convex sets whose projection onto $\mathcal{K}_{i}$ can be easily calculated.
Dykstra's algorithm states a procedure to compute the projection
$\mbox{Proj}_{\mu,\mathcal{K}_{All}}\left(\phi\right)$
onto the non-empty closed convex set, $\mathcal{K}_{All}:=\bigcap_{i=1}^{k}\mathcal{K}_{i}$, which satisfies all conditions.

For $\forall n \in \mathbb{N}_{+}$, we define $[n]_{k}:=((n-1)\, \mbox{mod}\, k) + 1$.
Let $\phi_{0}\in \mathcal{H}_{\mu}^{N}$ be the element whose projection we calculate
and determine the following recurrence formula. 
\begin{eqnarray}
  e_{-(k-1)}&:=&\cdots:=e_{-1}:=e_{0}:=0,\\
  \phi_{n}&:=&\mbox{Proj}_{\mu,\mathcal{K}_{[n]_{k}}}\left(\phi_{n-1}+e_{n-k}\right),\\
  e_{n}&:=&\phi_{n-1}+e_{n-k}-\phi_{n}.
\end{eqnarray}
We obtain
\begin{eqnarray}
  \lim_{n\to\infty}\left\|\phi_{n}-\mbox{Proj}_{\mu,\mathcal{K}_{All}}\left(\phi_{0}\right)\right\|_{\mu}=0,
\end{eqnarray}
from Boyle-Dykstra theorem (See Deutsch \cite{deutsch2001best}).

Let $\tilde{p}$ be an approximation of the probability density which may not satisfy the non-negativity or normalized conditions.
Considering $\tilde{\phi}=\tilde{p}/p_{\mu}$ as the element to be projected and $\mathcal{K}_{i}$ as the constraints of probability density and other conditions,
we obtain a corrected probability density $p_{\mu}^{*}:=\mbox{Proj}_{\mu,\mathcal{K}_{All}}\left(\tilde{\phi}\right)p_{\mu}$, which satisfies all given conditions.

Here, we assume $p_{\mu}(\bm{x})>0$ $(\bm{x}\in\mathbb{R}^{N})$ and the following example is given as a convex set $\mathcal{K}_{i}$. 
\begin{itemize}
  \item Normalized condition of probability density
  \begin{eqnarray}
    \mathcal{K}_{\mu,1,(=1)}:=\left\{\phi\in\mathcal{H}_{\mu}^{N}\middle| \left<\phi, 1\right>_{\mu}=1 \right\}.\label{formula:totalIntg}
  \end{eqnarray}
  \item Non-negativity of probability density
  \begin{eqnarray}
    \mathcal{K}_{\mu,\delta_{\bm{x}},(\geq0)}:=\bigcap_{\bm{x}\in \mathbb{R}^{N}}\left\{\phi\in\mathcal{H}_{\mu}^{N}\middle| \left<\phi, \delta_{\bm{x}}\right>_{\mu}\geq0 \right\}.\label{formula:noneNegative}
  \end{eqnarray}
  where $\delta_{\bm{x}}(y_{1},\cdots,y_{n}):=\delta(y_{1}-x_{1})\cdots\delta(y_{N}-x_{N})$ (delta function) and
  $\mbox{Proj}_{\mathcal{K}_{\mu,\delta_{\bm{x}},(\geq0)}}\left(\phi\right)=\max(\phi,0)$.
  \item Moment matching: Let $r(\bm{x})$ be some polynomial,
  \begin{eqnarray}
    \mathcal{K}_{\mu,r,(=m_{r})}:=\left\{\phi\in\mathcal{H}_{\mu}^{N}\middle| \left<\phi, r\right>_{\mu}=m_{r}\right\}.
  \end{eqnarray}
Particularly, when $r(\bm{x})$ is a monomial, we express the condition that the moment is $m_{r}$.
  In this study, we call this as moment-matching conditions when $r(x)$ is an arbitrary polynomial and $m_{r}$ is the moment in a broader sense.
  \item Matching of marginal distributions : Let $G_{1}(x_{1}),\cdots,G_{N}(x_{N})$ be the marginal distribution, we define
  \begin{eqnarray}
    \mathcal{K}_{\mu,G_{i}}:=\bigcap_{x_{i}\in \mathbb{R}}\left\{\phi\in\mathcal{H}_{\mu}^{N}\middle| \left<\phi, \delta_{x_{i}}\right>_{\mu}=\partial_{x_{i}}G_{i}(x_{i}) \right\},
  \end{eqnarray}
  where $\delta_{x_{i}}(y_{1},\cdots,y_{n}):=\delta(y_{i}-x_{i})$.
\end{itemize}

Given the results of the projection based on the non-negative condition $\mathcal{K}_{\mu,\delta_{\bm{x}},(\geq0)}$
the matching $\mathcal{K}_{\mu,G_{i}}$ of the marginal distributions cannot be simply expressed in an orthonormal basis of Hilbert space. Generally,
numerical calculations are used when applying corrections involving such conditions.
For example, the region is represented as an appropriate grid $\bm{x}_{i}\in \mathbb{R}^{N}\ i=1,\cdots,M$,
and the inner product calculation of $f(\bm{x}),g(\bm{x}) \in \mathcal{H}_{\mu}^{N}$ of real value functions is approximated by the numerical integral
\begin{eqnarray}
  \left<f,g\right>_{\mu}\simeq\sum_{i=1}^{M}w_{i}f(\bm{x}_{i})g(\bm{x}_{i})p_{\mu}(\bm{x}_{i}),
\end{eqnarray}
where $w_{i},\ i=1,\cdots,M$ are the weights of the numerical integration. 
Takahashi and Tsuzuki \cite{takahashi2016new} have shown that approximation using numerical integration corresponds to considering a discrete Hilbert space,
and that calculations using properties of Hilbert space, such as the Dykstra's algorithm, are valid in one dimension.
The discussion of Hilbert spaces is dimension-independent and the same result can be obtained by considering discrete Hilbert spaces in $N$-dimensional case.

As an example of concrete calculation grids and weights,
we propose a method of applying numerical integration (midpoint formula) with equally spaced Cartesian grids in each dimensional direction.
For the $i$-th dimension, we take the equidistant points $U_{i}:=\left\{x_{i,0} + j\Delta_{i}\middle|j=0,1,\cdots,M_{i}\right\},\ x_{i,0}\in \mathbb{R},\ M_{i}\in\mathbb{N}_{+}$.
The grid points are their direct product set $U:=\left\{(x_{1},\cdots,x_{N})\middle| x_{1}\in U_{1},\cdots,x_{N}\in U_{N}\right\}$.
The weights of the numerical integrals at each point are all equal, $w:=\prod_{i=1}^{N}\Delta_{i}$.
The inner product is approximated by
\begin{eqnarray}
  \left<f,g\right>_{\mu}\simeq\sum_{\bm{x_{i}}\in U}w f(\bm{x}_{i})g(\bm{x}_{i})p_{\mu}(\bm{x}_{i}).\label{formula:numericalInnerProduct}
\end{eqnarray}

\subsection{Application of Corrections to Hermite Polynomial Expansions}
We consider applying the correction for probability density proposed in the previous section
to a multivariate Hermite polynomial expansion truncated at $n_{M}$ of \eqref{formula:HermiteExpansion}
\begin{eqnarray}
  \tilde{\phi}_{n_{M}}(\bm{x})&:=&1+\sum_{n=1}^{n_{M}}\sum_{i=1}^{M_{n}}\hat{m}_{n,i}\hat{e}_{n,i}(\bm{x}),\\
  \tilde{p}_{\mu_{\Phi_{\Sigma}},n_{M}}(\bm{x})&:=&\tilde{\phi}_{n_{M}}(\bm{x})\partial_{\bm{x}}\Phi_{\Sigma}(\bm{x}).\label{formula:HermiteNoCorrection}
\end{eqnarray}
Let $\mathcal{S}:=\left\{(n,i)\middle|n=1,\cdots,n_{M}, i=1,\cdots,M_{n}\right\}$ be the set of subscripts $(n,i)$ contained in a finite sum and
define its subset $\mathcal{S}'\subseteq \mathcal{S}$.
We define the following convex sets of properties that the corrected function must satisfy, 
\begin{eqnarray}
  \mathcal{K}_{\Sigma,1,(=1)}&:=&\mathcal{K}_{\mu_{\Phi_{\Sigma}},1,(=1)},\\
  \mathcal{K}_{\Sigma,\delta_{\bm{x}},(\geq0)}&:=&\mathcal{K}_{\mu_{\Phi_{\Sigma}},\delta_{\bm{x}},(\geq0)},\\
  \mathcal{K}_{\Sigma,\hat{e}_{n,i},(=\hat{m}_{n,i})}&:=&\left\{\phi\in\mathcal{H}_{\mu_{\Phi_{\Sigma}}}^{N}\middle| \left<\phi, \hat{e}_{n,i}\right>_{\mu_{\Phi_{\Sigma}}}=\hat{m}_{n,i} \right\}\ \ \ (n,i)\in\mathcal{S}',
\end{eqnarray}
which contain (\ref{formula:totalIntg}), (\ref{formula:noneNegative}), and
we use $\Sigma$ instead of the measure $\mu_{\Phi_{\Sigma}}$ as the first subscript for notational simplicity.
Then, let us define $\mathcal{K}_{\Sigma,All}:=\mathcal{K}_{\Sigma,1,(=1)}\cap\mathcal{K}_{\Sigma,\delta_{\bm{x}},(\geq1)}\cap\left(\bigcap_{(n,i)\in\mathcal{S}'}\mathcal{K}_{\Sigma,\hat{e}_{n,i},(=\hat{m}_{n,i})}\right)$,
and we calculate the correction.
\begin{eqnarray}
  p_{\mu_{\Phi_{\Sigma}},n_{M}}^{*,1}(\bm{x}):=\mbox{Proj}_{\mu_{\Phi_{\Sigma}},\mathcal{K}_{\Sigma,All}}\left(\tilde{\phi}_{n_{M}}\right)(\bm{x})\partial_{\bm{x}}\Phi_{\Sigma}(\bm{x}),\label{formula:HermiteCorrectionNoRot}
\end{eqnarray}
with the Dykstra algorithm.
$p_{\mu_{\Phi_{\Sigma}},n_{M}}^{*,1}(\bm{x})$ satisfies the probability density
and parameter-matching conditions
\begin{eqnarray}
  \int_{\mathbb{R}^{N}}\hat{e}_{n,i}(\bm{x})p_{\mu_{\Phi_{\Sigma}},n_{M}}^{*,1}(\bm{x})d\bm{x}&=&\hat{m}_{n,i},\ \ \ (n,i)\in\mathcal{S}',
\end{eqnarray}
of the expression \eqref{formula:ExtendedMoment} are also satisfied.
If we want to give the marginal distributions condition as $\tilde{\bm{G}}(\bm{x})=\left(\tilde{G}_{1}(x_{1}),\cdots,\tilde{G}_{N}(x_{N})\right)'$,
we can add the convex set
\begin{eqnarray}
  \mathcal{K}_{\Sigma,\tilde{G}_{i}}&:=&\bigcap_{x_{i}\in\mathbb{R}}\left\{\phi\in\mathcal{H}_{\mu_{\Phi_{\Sigma}}}^{N}\middle| \left<\phi, \delta_{x_{i}}\right>_{\mu_{\Phi_{\Sigma}}}=\partial_{x_{i}}\tilde{G}_{i}(x_{i})\right\}\ \ \ i=1,\cdots,N.
\end{eqnarray}
into $\mathcal{K}_{\Sigma,All}$.
Given that the correction generally changes the marginal distributions,
the copula can be constructed using the given marginal distributions without recalculation.

In the calculation of $\mbox{Proj}_{\mu_{\Phi_{\Sigma}},\mathcal{K}_{\Sigma,All}}\left(\tilde{\phi}_{n_{M}}\right)$, we can consider an equivalent calculation on $\mathcal{H}_{\mu_{\Phi_{\mathbb{I}}}}^{N}$ from the formula (\ref{formula:iptransform}). 
Thus, the convex set may be defined as
\begin{eqnarray}
  \mathcal{K}_{\mathbb{I},\hat{e}_{n,i}\circ\Gamma,(=\hat{m}_{n,i})}&:=&\left\{\phi\in\mathcal{H}_{\mu_{\Phi_{\mathbb{I}}}}^{N}\middle| \left<\phi, \hat{e}_{n,i} \circ \Gamma\right>_{\mu_{\Phi_{\mathbb{I}}}}=\hat{m}_{n,i} \right\}\ \ \ (n,i)\in\mathcal{S},\\
  \mathcal{K}_{\mathbb{I},All}&:=&\mathcal{K}_{\mathbb{I},1,(=1)}\cap\mathcal{K}_{\mathbb{I},\delta_{\bm{x}},(\geq0)}\cap\left(\bigcap_{(n,i)\in\mathcal{S}'}\mathcal{K}_{\mathbb{I},\hat{e}_{n,i}\circ\Gamma,(=\hat{m}_{n,i})}\right),
\end{eqnarray}
and correction is calculated as
\begin{eqnarray}
  p_{\mu_{\Phi_{\Sigma}},n_{M}}^{*,2}(\bm{x}):=\left(\mbox{Proj}_{\mu_{\Phi_{\mathbb{I}}}\mathcal{K}_{\mathbb{I},All}}\left(\tilde{\phi}_{n_{M}}\circ \Gamma\right)\circ \Gamma^{-1}\right)(\bm{x})\partial_{\bm{x}}\Phi_{\Sigma}(\bm{x}).\label{formula:HermiteCorrectionPrep}
\end{eqnarray}
Note that $\tilde{\phi}_{n_{M}}\circ \Gamma$ in (\ref{formula:HermiteCorrectionPrep}) is a function which is independent of the correlation matrix $\Sigma$.
If we consider a time-dependent copula such that $\hat{m}_{n,i}$ is a fixed parameter and $\Sigma$ is a time-dependent parameter,
the computation can be reduced using the result of $\mbox{Proj}_{\mu_{\Phi_{\mathbb{I}}}\mathcal{K}_{\mathbb{I},All}}\left(\tilde{\phi}_{n_{M}}\circ \Gamma\right)$ for the change of $\Sigma$.

\subsection{Correction to Product of One-Dimensional Densities}
\label{subsection:1DCorrection}
As a special case, we show the construction of a joint distribution with corrected univariate Hermite expansions.
Let $m_{i,j}, i=1,\cdots,N,j=1,\cdots,n_{i},n_{i}\in \mathbb{N}_{+}$ be the coefficients of univariate Hermite expansions and
we define
\begin{eqnarray}
  \tilde{\phi}_{i}(v_{i})&:=&1+\sum_{j=1}^{n_{i}}m_{i,j}\bar{\mbox{He}}(v_{i}),   (i=1,\cdots,N).
\end{eqnarray}
We consider a situation where $\tilde{\phi}_{\perp}(\Gamma\bm{v})$ can be written as
\begin{eqnarray}
  \tilde{\phi}_{\perp}(\Gamma\bm{v})&:=&\prod_{i=1}^{N}\tilde{\phi}_{i}(v_{i}).
\end{eqnarray}
Let
\begin{eqnarray}
  \mathcal{K}_{1,\bar{\mbox{He}},(=m_{i,j})}&:=&\left\{\phi\in\mathcal{H}_{\mu_{\Phi_{1}}}^{1}\middle| \left<\phi, \bar{\mbox{He}}_{j}\right>_{\mu_{\Phi_{1}}}=m_{i,j} \right\}\ \ \ i=1,\cdots,N,\ j=1,\cdots,n_{i},\\
  \mathcal{K}_{1,All,i}&:=&\mathcal{K}_{1,1,(=1)}\cap\mathcal{K}_{1,\delta_{x},(\geq0)}\cap\left(\bigcap_{j=1}^{n_{i}}\mathcal{K}_{1,\bar{\mbox{He}},(=m_{i,j})}\right),
\end{eqnarray}
be convex sets of conditions to be satisfied.
One-dimensional corrections are applied for each $\tilde{\phi}_{i}$, and the $N$-dimensional probability density is composed by product of them as
\begin{eqnarray}
  \phi_{\perp}^{*}(\bm{v}):=\prod_{i=1}^{N}\mbox{Proj}_{\mathcal{K}_{1,All,i}}(\tilde{\phi}_{i}(v_{i})),\label{formula:HermiteCorrection1DN}\\
  p^{*}_{\perp}:=\left(\phi_{\perp}^{*}\circ \Gamma^{-1}\right)\partial_{\bm{x}}\Phi_{\Sigma}.
\end{eqnarray}

Generally, the computational complexity of numerically calculating the inner product using the formula \eqref{formula:numericalInnerProduct} is $O\left(\prod_{i=1}^{N}M_{i}\right)$,
whereas that of calculating the one-dimensional inner product $N$ times is $O\left(\sum_{i=1}^{N}M_{i}\right)$. The
latter has an advantage in terms of computational complexity over the $N$-dimensional correction.
This method is used to estimate cross currency volatility in numerical examples.

\section{Estimating Volatility Smile of Cross Currency Pair}

\subsection{Relationship between Volatilities of Cross and Straight Currency Pairs}
According to Breeden and Litzenberger \cite{breeden1978prices},
a risk-neutral probability distribution can be extracted from an implied volatility smile.
Considering risk-neutral probability distributions of straight currency pairs as marginal distributions,
a joint distribution is constructed by applying a copula,
and option prices for the cross currency pairs are obtained using joint distribution.
Taylor and Wang \cite{taylor2010option} has followed this method. 
In this subsection, we outline the method described in existing studies. 

Assume a frictionless market with no arbitrage opportunity. 
Let us consider a currency $\mathcal{X},\mathcal{Y}$ and a key currency $\mathcal{Z}$.
We consider $\mathcal{X}$ side the base currency and $\mathcal{Y}$ side as the quote currency in the currency pair $\mathcal{X}-\mathcal{Y}$.
Let $D_{\mathcal{Z}}(T)$ be the discount factor of the key currency $\mathcal{Z}$,
and $S_{\mathcal{X}\mathcal{Y}}(t)$ be the foreign exchange rate for the currency pair $\mathcal{X}-\mathcal{Y}$.
$\sigma_{\mathcal{X}\mathcal{Z}}(K)$ is the implied volatility at maturity $T$ and a strike price $K$ denominated in the currency pair $\mathcal{X}-\mathcal{Z}$.
By changing the currency of these subscripts, the rates of other currencies are represented and we omit subscripts where the currencies do not need to be determined. 

For these rates, the relationship with the inverse currency pair,
$S_{\mathcal{Z}\mathcal{Y}}(t)=\frac{1}{S_{\mathcal{Y}\mathcal{Z}}(t)}$ and
$\sigma_{\mathcal{Z}\mathcal{Y}}(K)=\sigma_{\mathcal{Y}\mathcal{Z}}(1/K)$, holds. 
Generally, the key currency is $\mathcal{Z}=\mbox{USD}$ and the USD-JPY rate is traded in the market.
When the rate for JPY-USD is required, we convert these rates using this relation. 

Let $D(T)$ be the discount factor of the quote currency,
and $V_{\mbox{put}}(K)$ be the put option price of the quote currency at the strike price $K$,
$x:=\log S(T)$ be the log exchange rate at maturity $T$,
and $p(x)$ be the quote currency risk-neutral probability density.
For these variables, the relation
\begin{eqnarray}
  V_{\mbox{put}}(K)&=&D(T)\int_{0}^{\log K}(K-\exp(x))p(x)dx,\\
  \frac{\partial V_{put}}{\partial K}&=&D(T)\int_{-\infty}^{\log K}p(x)dx,
\end{eqnarray}
holds.
Given that $V_{\mbox{put}}(K)$ is a function of an implied volatility, the risk-neutral probability density can be derived from a sufficiently smooth implied volatility smile. 

Let $X_{1}:=\log S_{\mathcal{X}\mathcal{Z}}(T),\ X_{2}:=\log S_{\mathcal{Y}\mathcal{Z}}(T)$
be the log rate at a maturity $T$ under the risk-neutral measure of the currency $\mathcal{Z}$.
We also define a copula $\mathcal{C}$ and the marginal distributions $G_{\mathcal{X}|\mathcal{Z}}(x_{1}),G_{\mathcal{Y}|\mathcal{Z}}(x_{2})$ for $X_{1}$ and $X_{2}$, respectively.
We write the joint distribution of $X_{1}$ and $X_{2}$ under the same measure as
\begin{eqnarray}
  P_{\mathcal{X},\mathcal{Y}|\mathcal{Z}}(x_{1},x_{2}):=\mathcal{C}(G_{\mathcal{X}|\mathcal{Z}}(x_{1}),G_{\mathcal{Y}|\mathcal{Z}}(x_{2})).\label{formula:cross_distribution}
\end{eqnarray}

Considering a call option on the strike price $K$ of the $\mathcal{X}-\mathcal{Y}$ currency pair,
the $\mathcal{Z}$-denominated price of this option is calculated as
\begin{eqnarray}
  V_{\mathcal{X},\mathcal{Y}|\mathcal{Z}}=D_{\mathcal{Z}}(T)\int_{-\infty}^{\infty}\int_{-\infty}^{\infty}\{\exp(x_{1})-K\exp(x_{2})\}_{+}\partial_{\bm{x}}P_{\mathcal{X},\mathcal{Y}|\mathcal{Z}}(x_{1},x_{2})dx_{1}dx_{2},\label{formula:copula_cross}
\end{eqnarray}
and the implied volatility of the cross-currency pair is obtained from the price.
Additionally, by considering the $\mathcal{Z}$-denominated price of the transaction that receives one amount of currency $\mathcal{Y}$ at time $T$,
the relationship
\begin{eqnarray}
  S_{\mathcal{Y}\mathcal{Z}}(0)D_{\mathcal{Y}}(T)&=&D_{\mathcal{Z}}(T)\int_{-\infty}^{\infty}\exp\left(x_{2}\right)\partial_{x_{2}}G_{\mathcal{Y}|\mathcal{Z}}(x_{2})dx_{2}\nonumber\\
  &=&D_{\mathcal{Z}}(T)\int_{-\infty}^{\infty}\int_{-\infty}^{\infty}\exp\left(x_{2}\right)\partial_{\bm{x}}P_{\mathcal{X},\mathcal{Y}|\mathcal{Z}}(x_{1},x_{2})dx_{1}dx_{2},\label{formula:yforward}
\end{eqnarray}
is obtained.

\subsection{Option Pricing of Cross Currency Pair Using Corrected Hermite Polynomial Expansion}
We explain the application of a copula derived from a corrected Hermite polynomial expansion to calculate the option price of cross currency pairs discussed in the previous subsection. 
Hereafter, we assume $N=2$ because we consider two-dimensional probabilities for cross currency. 
In (\ref{formula:HermiteExpansion}), we set
\begin{eqnarray}
  \Sigma &:=& \begin{pmatrix}1&\rho\\\rho&1\end{pmatrix},\label{formula:RotSigma}\\
  \Gamma &:=& \begin{pmatrix}\alpha_{1}&-\alpha_{2}\\\alpha_{1}&\alpha_{2}\end{pmatrix},\label{formula:RotGamma}\\
  \Gamma^{-1} &:=& \begin{pmatrix}\beta_{1}&\beta_{1}\\-\beta_{2}&\beta_{2}\end{pmatrix},\label{formula:RotGammaInv}\\
  \hat{e}_{n,i}(v_{1},v_{2})&:=&e_{n,i}(v_{1},v_{2}):=\bar{\mbox{He}}_{i}(v_{1})\bar{\mbox{He}}_{n-i}(v_{2}), \ \ \ i=0,\cdots,n,
\end{eqnarray}
where $\alpha_{1}:=\sqrt{\frac{1+\rho}{2}},\alpha_{2}:=\sqrt{\frac{1-\rho}{2}}, \beta_{1}:=\sqrt{\frac{1}{2(1+\rho)}}, \beta_{2}:=\sqrt{\frac{1}{2(1-\rho)}}$.
Let $\bm{X}:=(X_{1},X_{2})'$ be the random variables that are not related to the exchange rates.
The joint density $\tilde{p}_{\bm{X}}(x_{1},x_{2})$, expressed using the bivariate Hermite polynomial expansion truncated at $n_{M} (\geq3)$, is
\begin{eqnarray}
  \tilde{p}_{\bm{X}}(x_{1},x_{2})&:=&\left\{1+\sum_{n=3}^{n_{M}}\sum_{i=0}^{n}\hat{m}_{n,i}e_{n,i}(\beta_{1}(x_{1}+x_{2}),\beta_{2}(-x_{1}+x_{2}))\right\}\partial_{\bm{x}}\Phi_{\Sigma}(x_{1},x_{2}),\label{formula:CopulaJointProbDens}
\end{eqnarray}
where $\bm{v}:=(v_{1},v_{2})':=\Gamma\bm{x}$ and $\hat{m}_{1,i}=\hat{m}_{2,i}=0$ so that random variables $(X_{1}, X_{2})$ have mean 0, variance 1, and correlation $\rho$.

Let $\mathcal{S}:=\left\{(n,i)\middle|n=1,\cdots,n_{M},i=0,\cdots,n\right\}$ be the set of subscripts for moment-matching, and
\begin{eqnarray}
  \mathcal{K}_{\mathbb{I},e_{n,i},(=\hat{m}_{n,i})}&:=&\left\{\phi\in\mathcal{H}_{\mu_{\Phi_{\mathbb{I}}}}^{2}\middle| \left<\phi, e_{n,i}\right>_{\mu_{\Phi_{\mathbb{I}}}}=\hat{m}_{n,i} \right\}\ \ \ (n,i)\in\mathcal{S},\\
  \mathcal{K}_{\mathbb{I},All}&:=&\mathcal{K}_{\mathbb{I},1,(=1)}\cap\mathcal{K}_{\mathbb{I},\delta_{\bm{x}},(\geq0)}\cap\left(\bigcap_{(n,i)\in\mathcal{S}}\mathcal{K}_{\mathbb{I},e_{n,i},(=\hat{m}_{n,i})}\right),
\end{eqnarray}
be the convex sets which represent the properties that must be satisfied by probability density.
The probability density, represented by the formula \eqref{formula:HermiteCorrectionPrep}, when corrected using Dykstra's algorithm, is obtained as
\begin{eqnarray}
  \check{p}_{\bm{X}}(x_{1},x_{2}):=\left(\mbox{Proj}_{\mu_{\Phi_{\mathbb{I}}}\mathcal{K}_{\mathbb{I},All}}\left(1+\sum_{n=3}^{n_{M}}\sum_{i=0}^{n}\hat{m}_{n,i}e_{n,i}\right)\circ \Gamma^{-1}\right)(\bm{x})\partial_{\bm{x}}\Phi_{\Sigma}(\bm{x}).\label{formula:AdjustingDens}
\end{eqnarray}
Let the joint distribution of $\check{p}_{\bm{X}}(x_{1},x_{2})$ be $\check{P}_{\bm{X}}(x_{1},x_{2})$ and the marginal distributions be $\check{G}_{X_{1}}(x_{1})$, $\check{G}_{X_{2}}(x_{2})$,
and $\check{\mathcal{C}}_{X}(u_{1},u_{2}):=\check{P}_{\bm{X}}\left(\check{G}_{X_{1}}^{-1}(u_{1}),\check{G}_{X_{2}}^{-1}(u_{2})\right)$ be the copula derived from this probability density.
Note that even though the marginal distributions of $\tilde{p}_{\bm{X}}(x_{1},x_{2})$ and $\check{p}_{\bm{X}}(x_{1},x_{2})$ are different because of the effect of the correction in \eqref{formula:AdjustingDens},
$\check{\mathcal{C}}_{X}(u_{1},u_{2})$ is a valid copula because it is based on the marginal distributions $\check{G}_{X_{1}}(x_{1})$, $\check{G}_{X_{2}}(x_{2})$ recalculated from $\check{p}_{\bm{X}}(x_{1},x_{2})$.

Applying this copula to \eqref{formula:cross_distribution} and \eqref{formula:copula_cross},
the joint distribution of the logarithmic foreign exchange rates becomes $P_{\mathcal{X},\mathcal{Y}|\mathcal{Z}}(x_{1},x_{2}):=\check{P}_{\bm{X}}\left(\check{G}_{X_{1}}^{-1}(G_{\mathcal{X}|\mathcal{Z}}(x_{1})),\check{G}_{X_{2}}^{-1}(G_{\mathcal{Y}|\mathcal{Z}}(x_{2}))\right)$.
By applying the same variable transformation as \eqref{formula:intgbysubstitution}, the option price equation \eqref{formula:copula_cross} becomes
\begin{eqnarray}
  V_{\mathcal{X},\mathcal{Y}|\mathcal{Z}}&=&D_{\mathcal{Z}}(T)\int_{-\infty}^{\infty}\int_{-\infty}^{\infty}\{\exp(x_{1})-K\exp(x_{2})\}_{+}\partial_{\bm{x}}P_{\mathcal{X},\mathcal{Y}|\mathcal{Z}}(x_{1},x_{2})dx_{1}dx_{2}\nonumber\\
  &=&D_{\mathcal{Z}}(T)\int_{-\infty}^{\infty}\int_{-\infty}^{\infty}\left\{\exp\left(G_{\mathcal{X}|\mathcal{Z}}^{-1}\left(\check{G}_{X_{1}}(\alpha_{1}v_{1}-\alpha_{2}v_{2})\right)\right)-K\exp\left(G_{\mathcal{Y}|\mathcal{Z}}^{-1}\left(\check{G}_{X_{2}}(\alpha_{1}v_{1}+\alpha_{2}v_{2})\right)\right)\right\}_{+}\nonumber\\
  &&\times\mbox{Proj}_{\mu_{\Phi_{\mathbb{I}}}\mathcal{K}_{\mathbb{I},All}}\left(1+\sum_{n=3}^{n_{M}}\sum_{i=0}^{n}\hat{m}_{n,i}e_{n,i}\right)\partial_{v_{1}}\Phi(v_{1})\partial_{v_{2}}\Phi(v_{2})dv_{1}dv_{2}.\label{formula:crossVInt}
\end{eqnarray}
This expression can be regarded as an inner product calculation in $\mathcal{H}_{\mu_{\Phi_{\mathbb{I}}}}^{2}$
and can be calculated using the same discretization as the calculation of the projection $\mbox{Proj}_{\mu_{\Phi_{\mathbb{I}}}\mathcal{K}_{\mathbb{I},All}}\left(1+\sum_{n=3}^{n_{M}}\sum_{i=0}^{n}\hat{m}_{n,i}e_{n,i}\right)$. 

\subsection{Change of $\hat{m}_{n,i}$ and Shape of Density}
In this subsection, we show how the density of cross currency pair rates changes with the parameter $\hat{m}_{n,i}$,
using simple calculation with rough approximation.
Let $\nu_{\mathcal{X}\mathcal{Z}}$ and $\sigma_{\mathcal{X}\mathcal{Z}}$ be the mean and standard deviation of $\log(S_{\mathcal{X}\mathcal{Z}}(T))$, respectively,
and $\nu_{\mathcal{Y}\mathcal{Z}}$ and $\sigma_{\mathcal{Y}\mathcal{Z}}$ be the mean and standard deviation of $\log(S_{\mathcal{Y}\mathcal{Z}}(T))$, respectively.
We assume
$\frac{\log(S_{\mathcal{X}\mathcal{Z}}(T))-\nu_{\mathcal{X}\mathcal{Z}}}{\sigma_{\mathcal{X}\mathcal{Z}}}\sim X_{1}$ and
$\frac{\log(S_{\mathcal{Y}\mathcal{Z}}(T))-\nu_{\mathcal{Y}\mathcal{Z}}}{\sigma_{\mathcal{Y}\mathcal{Z}}}\sim X_{2}$,
and obtain
\begin{eqnarray}
  G_{\mathcal{X}|\mathcal{Z}}^{-1}(G_{X_{1}}(x_{1}))&=&\nu_{\mathcal{X}\mathcal{Z}}+\sigma_{\mathcal{X}\mathcal{Z}}x_{1},\\
  G_{\mathcal{Y}|\mathcal{Z}}^{-1}(G_{X_{2}}(x_{2}))&=&\nu_{\mathcal{Y}\mathcal{Z}}+\sigma_{\mathcal{Y}\mathcal{Z}}x_{2}.
\end{eqnarray}
Furthermore, by assuming $\mbox{Proj}_{\mu_{\Phi_{\mathbb{I}}}\mathcal{K}_{\bar{a}}}\left(\left\{1+\sum_{n=3}^{n_{M}}\sum_{i=0}^{n}\hat{m}_{n,i}e_{n,i}\right\}\right)=1+\sum_{n=3}^{n_{M}}\sum_{i=0}^{n}\hat{m}_{n,i}e_{n,i}$, $\sigma\simeq\sigma_{\mathcal{X}\mathcal{Z}}\simeq\sigma_{\mathcal{Y}\mathcal{Z}}$,substituting them into \eqref{formula:crossVInt}, and rearranging,
we get
\begin{eqnarray}
  V_{\mathcal{X},\mathcal{Y}|\mathcal{Z}}&\simeq&
  D_{\mathcal{Z}}(T)\int_{-\infty}^{\infty}\left\{\exp\left(\nu_{\mathcal{X}\mathcal{Z}}-\nu_{\mathcal{Y}\mathcal{Z}}-2\sigma\alpha_{2}v_{2}\right)-K\right\}_{+}\nonumber\\
  &&\times\int_{-\infty}^{\infty}\exp\left(\nu_{\mathcal{Y}\mathcal{Z}}+\sigma(\alpha_{1}v_{1}+\alpha_{2}v_{2})\right)\left(1+\sum_{n=3}^{n_{M}}\sum_{i=0}^{n}\hat{m}_{n,i}e_{n,i}\right)\partial_{v_{1}}\Phi(v_{1})dv_{1}\partial_{v_{2}}\Phi(v_{2})dv_{2}\nonumber\\
  &=&D_{\mathcal{Z}}(T)\exp\left(\nu_{\mathcal{Y}\mathcal{Z}}+\frac{1}{2}\sigma^{2}\left(\alpha_{1}^{2}+\alpha_{2}^{2}\right)\right)\int_{-\infty}^{\infty}\left\{\exp\left(\nu_{\mathcal{X}\mathcal{Z}}-\nu_{\mathcal{Y}\mathcal{Z}}-2\sigma\alpha_{2}v_{2}\right)-K\right\}_{+}\nonumber\\
  &&\times\left(1+\sum_{n=3}^{n_{M}}\sum_{i=0}^{n}\hat{m}_{n,i}\int_{-\infty}^{\infty}\bar{\mbox{He}}_{i}(v_{1})\partial_{v_{1}}\Phi(v_{1}-\sigma\alpha_{1})dv_{1}\bar{\mbox{He}}_{n-i}(v_{2})\right)\partial_{v_{2}}\Phi(v_{2}-\sigma\alpha_{2})dv_{2}\nonumber\\
  &=&D_{\mathcal{Z}}(T)\int_{-\infty}^{\infty}\left\{\exp\left(\nu_{\mathcal{X}\mathcal{Z}}-\nu_{\mathcal{Y}\mathcal{Z}}-2\sigma\alpha_{2}v_{2}\right)-K\right\}_{+}\Lambda(v_{2}-\sigma\alpha_{2})\partial_{v_{2}}\Phi(v_{2}-\sigma\alpha_{2})dv_{2},\label{formula:CrossApprox}
\end{eqnarray}
where $\Lambda(v)$ is
\begin{eqnarray}
  \Lambda(v)&:=&e^{\nu_{\mathcal{Y}\mathcal{Z}}+\frac{1}{2}\sigma^{2}\left(\alpha_{1}^{2}+\alpha_{2}^{2}\right)}\left(1+\sum_{n=3}^{n_{M}}\sum_{i=0}^{n}\hat{m}_{n,i}\frac{(\sigma\alpha_{1})^{i}}{\sqrt{i!}}\bar{\mbox{He}}_{n-i}(v+\sigma\alpha_{2})\right)\nonumber\\
  &=&e^{\nu_{\mathcal{Y}\mathcal{Z}}+\frac{1}{2}\sigma^{2}\left(\alpha_{1}^{2}+\alpha_{2}^{2}\right)}\left(1+\sum_{n=3}^{n_{M}}\sum_{i=0}^{n}\sum_{k=0}^{n-i}\hat{m}_{n,i}\sqrt{\frac{(n-i)!}{i!k!}}\frac{\sigma^{n-k}\alpha_{1}^{i}\alpha_{2}^{n-i-k}}{(n-i-k)!}\bar{\mbox{He}}_{k}(v)\right)\nonumber\\
  &=&e^{\nu_{\mathcal{Y}\mathcal{Z}}+\frac{1}{2}\sigma^{2}\left(\alpha_{1}^{2}+\alpha_{2}^{2}\right)}\left(1+\sum_{k=0}^{n_{M}}\bar{\mbox{He}}_{k}(v)\sum_{n=\max(3,k)}^{n_{M}}\sum_{i=0}^{n-k}\hat{m}_{n,i}\sqrt{\frac{(n-i)!}{i!k!}}\frac{\sigma^{n-k}\alpha_{1}^{i}\alpha_{2}^{n-i-k}}{(n-i-k)!}\right).\label{formula:CrossHermiteExpApprox}
\end{eqnarray}
In the transformation of \eqref{formula:CrossHermiteExpApprox}, the relations obtained using the Taylor expansion of the Hermite polynomial
\begin{eqnarray}
  &\bar{\mbox{He}}_{i}(x+y)=\sum_{k=0}^{i}\sqrt{\frac{i!}{k!}}\frac{x^{i-k}}{(i-k)!}\bar{\mbox{He}}_{k}(y),\\
  &\int_{-\infty}^{\infty}\bar{\mbox{He}}_{i}(v_{1})\partial_{v_{1}}\Phi(v_{1}-\sigma\alpha_{1})dv_{1}=\frac{(\sigma\alpha_{1})^{i}}{\sqrt{i!}},
\end{eqnarray}
are used.
Under the same assumptions as those described in this subsection, \eqref{formula:yforward} become
\begin{eqnarray}
  S_{\mathcal{Y}\mathcal{Z}}(0)D_{\mathcal{Y}}(T)&\simeq&D_{\mathcal{Z}}(T)\int_{-\infty}^{\infty}\Lambda(v_{2}-\sigma\alpha_{2})\partial_{v_{2}}\Phi(v_{2}-\sigma\alpha_{2})dv_{2}.
\end{eqnarray}
Substituting it in \eqref{formula:CrossApprox}, we obtain
\begin{eqnarray}
  V_{\mathcal{X},\mathcal{Y}|\mathcal{Z}}&\simeq&S_{\mathcal{Y}\mathcal{Z}}(0)D_{\mathcal{Y}}(T)\int_{-\infty}^{\infty}\left\{\exp\left(\nu_{\mathcal{X}\mathcal{Z}}-\nu_{\mathcal{Y}\mathcal{Z}}-2\sigma\alpha_{2}v_{2}\right)-K\right\}_{+}\bar{\Lambda}(v_{2})dv_{2},\\
  \bar{\Lambda}(v_{2})&:=&\frac{\Lambda(v_{2}-\sigma\alpha_{2})\partial_{v_{2}}\Phi(v_{2}-\sigma\alpha_{2})}{\int_{-\infty}^{\infty}\Lambda(v_{2}-\sigma\alpha_{2})\partial_{v_{2}}\Phi(v_{2}-\sigma\alpha_{2})dv_{2}}.
\end{eqnarray}
By considering $\log S_{\mathcal{X}\mathcal{Y}}(T)\simeq\nu_{\mathcal{X}\mathcal{Z}}-\nu_{\mathcal{Y}\mathcal{Z}}-2\sigma\alpha_{2}v_{2}$,
the option price becomes the integral of the product of the payoff $\left\{S_{\mathcal{X}\mathcal{Y}}(T)-K\right\}_{+}$
and function $\bar{\Lambda}(v_{2})$, which satisfies the condition of probability density.

Given that $\bar{\Lambda}(v_{2})$ takes the form of a Hermite polynomial expansion from \eqref{formula:CrossHermiteExpApprox},
the parameters $\hat{m}_{n,i}$ can be adjusted according to the moments in this probability density.
Particularly, the coefficients of any Hermite polynomial expansion can be represented
by changing only the parameters $\hat{m}_{n,0}$, $n=3,\cdots,n_{M}$.
However, if $\hat{m}_{n,0}=0$, $n=3,\cdots,n_{M}$ is satisfied,
the coefficients of the Hermite polynomial expansion converge to zero when $\alpha_{1}$ goes to 0.
Regardless of the other parameters $\hat{m}_{n,i}$ $(i\neq0)$, the distribution asymptotically approaches a normal distribution in the case of $\alpha_{1}\to0$.

\begin{figure}[H]
  \centering
  \includegraphics[keepaspectratio, scale=1.0]{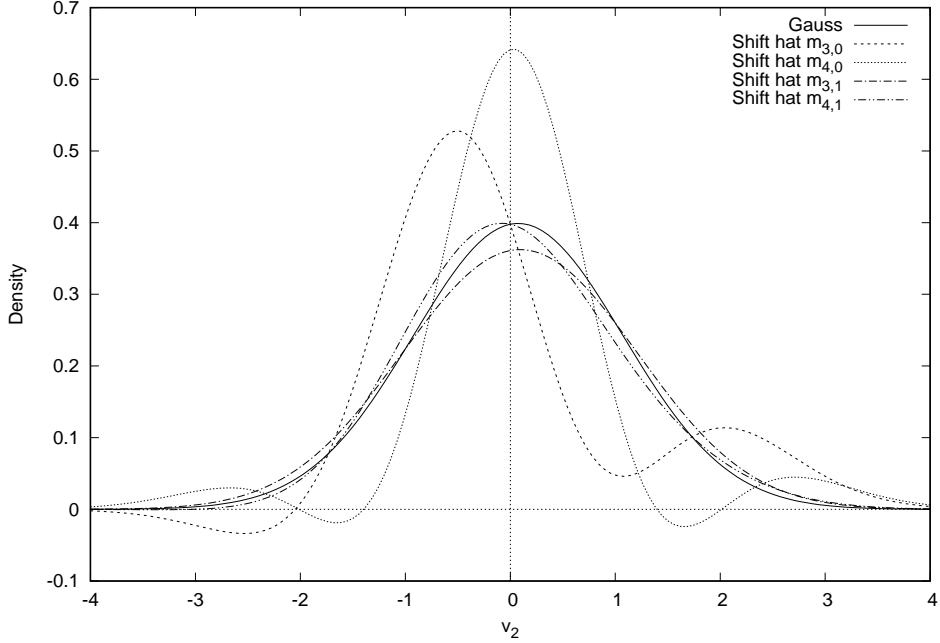}
  \caption{Change of Cross Density by Hermite Expansion Parameter.}
  \label{figure:Chapter4_SimpleDens}
\end{figure}

Under the aforementioned assumptions, the shape of the density $\bar{\Lambda}(v_{2})$ with $\rho=0.5$,
and each changing parameter $\hat{m}_{n,i}$, is shown in Figure \ref{figure:Chapter4_SimpleDens}.
"Gauss" takes all coefficients as 0, "Shift hat $m_{3,0}$" to "Shift hat $m_{4,1}$" are
the conditions for each parameter, $\hat{m}_{3,0}$, $\hat{m}_{4,0}$, $\hat{m}_{3,1}$, and $\hat{m}_{4,1}$, which are all set to 1.
It can be seen that the changes in the shapes of the densities when $\hat{m}_{3,1}$ and $\hat{m}_{4,1}$ are shifted
are smaller than those of $\hat{m}_{3,0}$ and $\hat{m}_{4,0}$.
Other cases of changing parameters are excluded from the graph because they show almost the same lines as the "Gauss" density.

Based on the results discussed in this subsection, 
we adopt a setting where all parameters are set to 0 except for $\hat{m}_{n,0}$,
and confirm the fitting of the proposed method to the volatility smile of the cross-currency pair in the numerical example of the next section.

\section{Numerical Examples}
\subsection{Approximation of Joint Distribution by Corrected Hermite Polynomial Expansion}
We show numerical examples of the approximation of several joint distributions using the multivariate Hermite expansion
and its correction proposed in Section \ref{section:CopulaWithHermiteExpansion}.

The target joint distributions are obtained by composing some copulas $\mathcal{C}(u_{1},u_{2})$ and standard normal distributions as $\mathcal{C}(\Phi(x_{1}),\Phi(x_{2}))$.
The parameters of the proposed copula are calculated using \eqref{formula:ExtendedMoment}.
Next, we calculate joint density using \eqref{formula:HermiteNoCorrection} as "No Correction" and {\eqref{formula:HermiteCorrectionNoRot}} as "Applying Correction" in these parameters,
and plot them as contours.

The setting of the proposed copula is
$\hat{e}_{n,i}(v_{1},v_{2}):=e_{n,i}(v_{1},v_{2}):=\bar{\mbox{He}}_{i}(v_{1})\bar{\mbox{He}}_{n-i}(v_{2}), \ \ \ i=0,\cdots,n$ and $n_{M}=4$.
We prepare case (a) as $\Sigma = \Gamma = \mathbb{I}$,
and (b) as $\rho=\int_{\mathbb{R}^{2}}x_{1}x_{2}\partial_{\bm{x}}\mathcal{C}(\Phi(x_{1}),\Phi(x_{2}))d\bm{x}$ and using \eqref{formula:RotSigma}, \eqref{formula:RotGamma}.
In the correction, the projection satisfies the conditions that the value of the integral in the whole range is 1, the non-negativity condition of the probability density,
and the matching condition of the parameter $\hat{m}_{n,i}$ for the terms up to $n_{M}=4$.
To adjust the density function, we calculate 200 equal sections for each direction in the region $[-6,6]\times[-6,6]$. 
For the copula to be approximated, we use the two-dimensional Clayton, Frank, Gumbel, and Placett copulas with the parameters being set to reflect that Spearman's rank correlation is 0.6.

\begin{figure}[H]
  \begin{tabular}{c}
    \begin{minipage}[t]{1.0\hsize}
      \centering
      \includegraphics[keepaspectratio, scale=0.8]{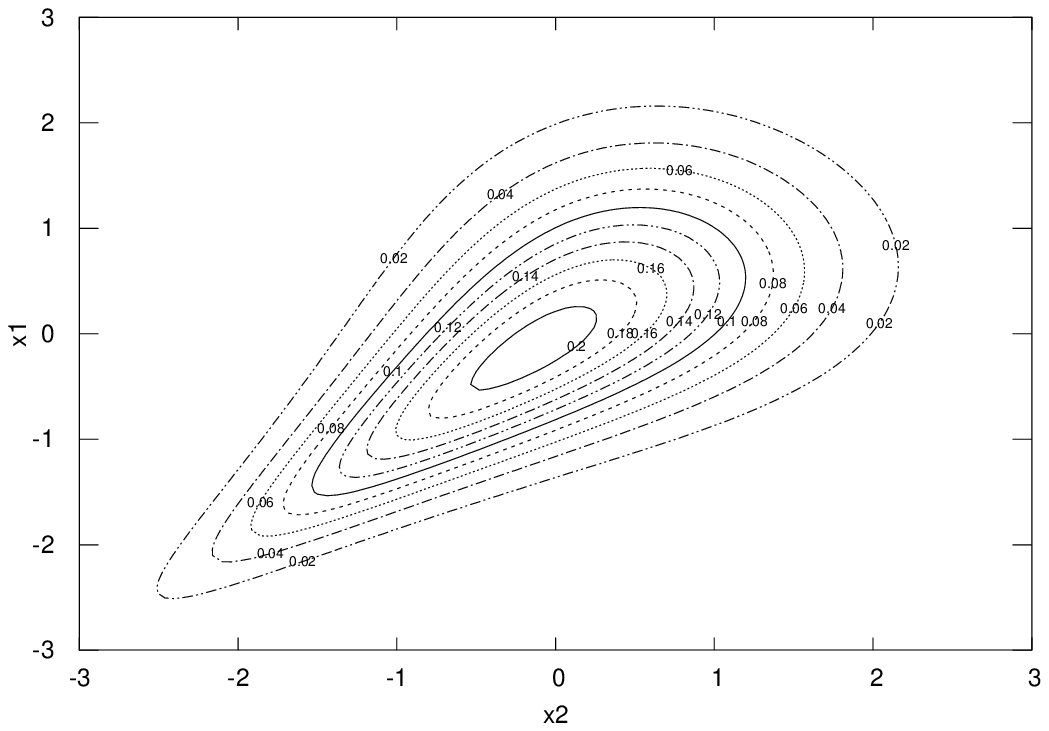}
      \caption{Density by Clayton copula.}
      \label{figure:Exam1_Perp04Org}
    \end{minipage}
  \end{tabular}
  \begin{tabular}{cc}
    \begin{minipage}[t]{0.5\hsize}
      \centering
      \includegraphics[keepaspectratio, scale=0.6]{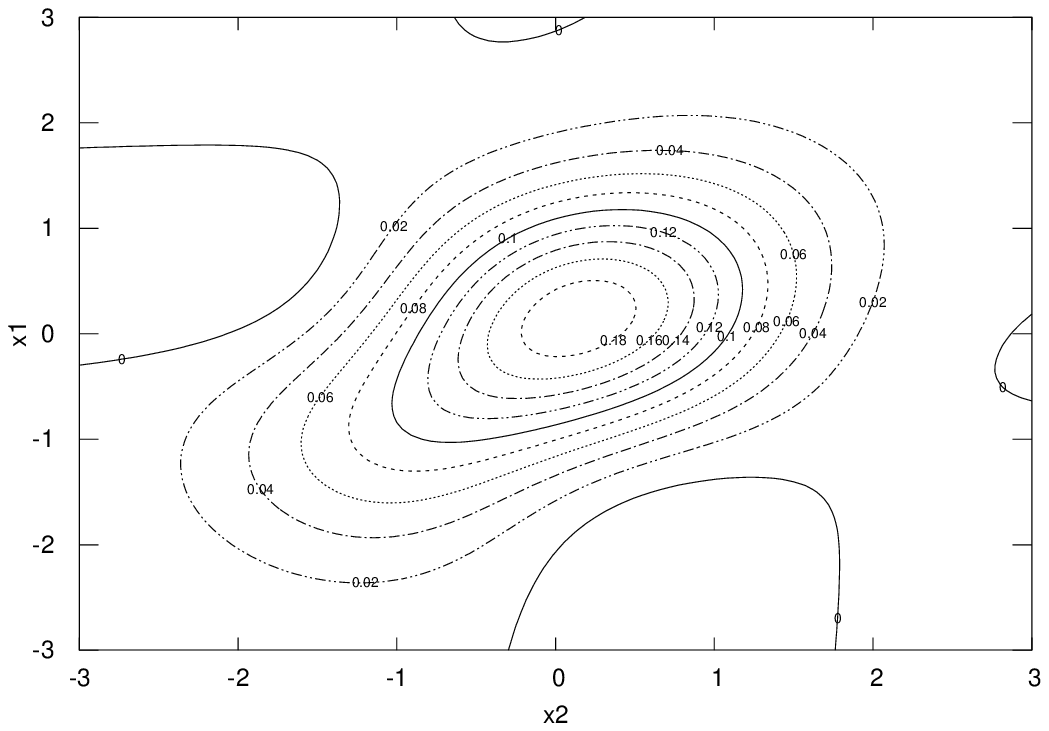}
      \caption{Expansion using (a) and (No correction) fitting to Clayton copula distribution}
      \label{figure:Exam1_Perp04Raw}
    \end{minipage} &
    \begin{minipage}[t]{0.5\hsize}
      \centering
      \includegraphics[keepaspectratio, scale=0.6]{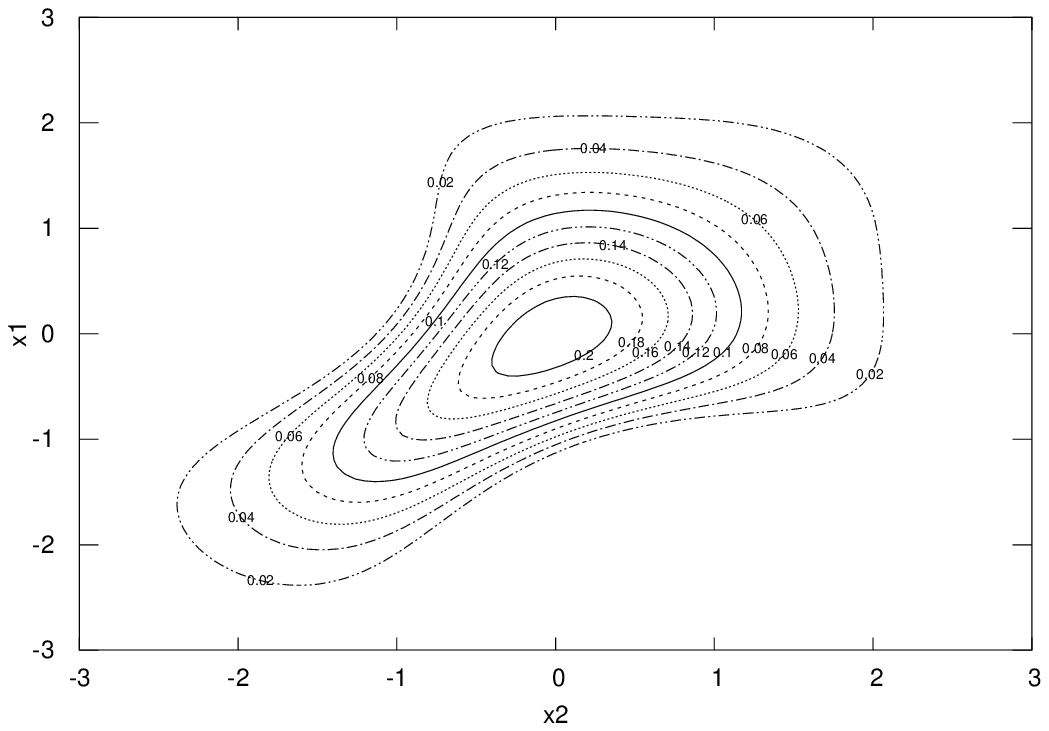}
      \caption{Expansion using (a) and (Applying correction) fitting to Clayton copula distribution}
      \label{figure:Exam1_Perp04Corr}
    \end{minipage}\\
    \begin{minipage}[t]{0.5\hsize}
      \centering
      \includegraphics[keepaspectratio, scale=0.6]{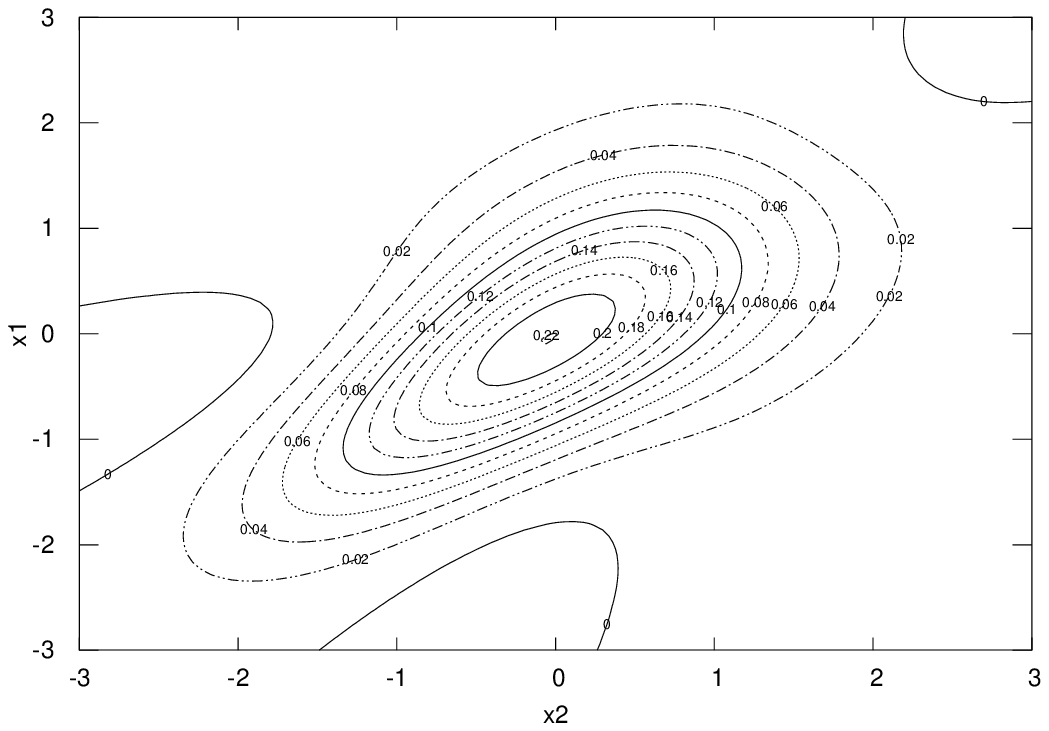}
      \caption{Expansion using (b) and (No correction) fitting to Clayton copula distribution}
      \label{figure:Exam1_Rot04Raw}
    \end{minipage} &
    \begin{minipage}[t]{0.5\hsize}
      \centering
      \includegraphics[keepaspectratio, scale=0.6]{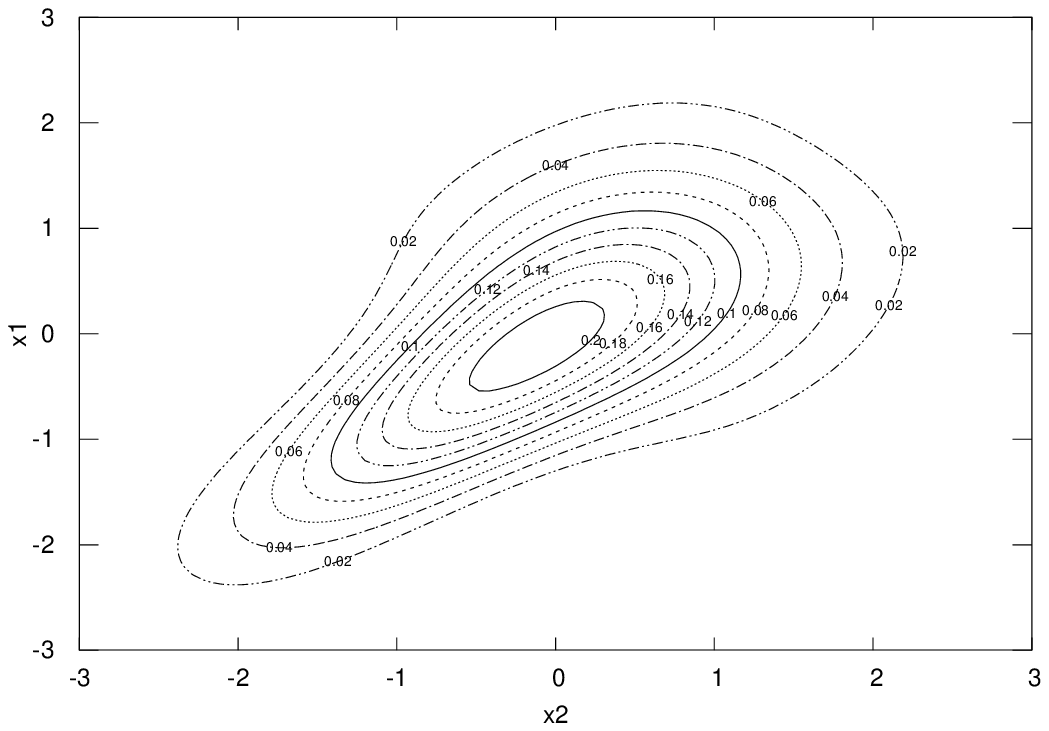}
      \caption{Expansion using (b) and (Applying correction) fitting to Clayton copula distribution}
      \label{figure:Exam1_Rot04Corr}
    \end{minipage}
  \end{tabular}
\end{figure}

Figure \ref{figure:Exam1_Perp04Org} is a contour plot of the probability density when the Clayton copula is used to construct a joint density whose marginal distributions are standard normal distributions.
Figure \ref{figure:Exam1_Perp04Raw} and Figure \ref{figure:Exam1_Perp04Corr} show contour plots of \eqref{formula:HermiteNoCorrection} (No correction) and \eqref{formula:HermiteCorrectionNoRot} (Applying correction) in the case (a).
Figure \ref{figure:Exam1_Rot04Raw} and Figure \ref{figure:Exam1_Rot04Corr} show contour plots of \eqref{formula:HermiteNoCorrection} (No correction) and \eqref{formula:HermiteCorrectionNoRot} (Applying correction) in the case (b).
Figures \ref{figure:Exam1_Perp14Org} - \ref{figure:Exam1_Rot14Corr},
Figures \ref{figure:Exam1_Perp24Org} - \ref{figure:Exam1_Rot24Corr} and
Figures \ref{figure:Exam1_Perp34Org} - \ref{figure:Exam1_Rot34Corr} in appendix
are contour plots for the Frank, Gumbel, and Placett copulas under the same settings as those of the Clayton copula.

For any case with target copulas, the proposed method with $n_{M}=4$ reproduces the shape of the original joint densities and presents good approximations.
Additionally, it can be seen that the non-negative constraint is satisfied when the correction is applied.
In case (a), the contour lines tend to be less smooth, while they are smoother and more similar to the original density in case (b).

Table \ref{table:MomentClaytonOrg} shows the moments up to the eighth order, when the probability density is constructed using the Clayton copula with the standard normal distribution as the marginal distributions.
Table \ref{table:MomentClaytonCorr} shows the moments up to the eighth order, when $n_{M}=4$ and $\Sigma$ are set to case (a).
Tables \ref{table:MomentFrankOrg} - \ref{table:MomentPlacettCorr} in Appendix show the moments for the Frank, Gumbel, and Placett copulas, under the same settings as Tables \ref{table:MomentClaytonOrg} and \ref{table:MomentClaytonCorr}.
It is confirmed that the moments up to the fourth order are consistent with the original distribution. Moreover, the moments related to the correction conditions are preserved even when the corrections are applied.

The moments of marginal distributions (columns of $x_{1}^{0}$ and rows of $x_{2}^{0}$) in Table \ref{table:MomentClaytonCorr} differ from the moments of the standard normal distribution in the higher-order terms, for which the moment condition is not imposed.
This indicates that the marginal distributions may change because of the correction.
Therefore, it is necessary to recalculate the marginal distributions or to correct them by introducing the condition that the marginal distributions should also coincide with those of the original joint distribution
to construct a copula from the corrected joint density.

Accordingly, the proposed method can properly approximate the different simultaneous distributions
and probability density can be corrected without changing the moments related to the specified conditions.

\begin{table}[h]
  \caption{Moments of Clayton Copula with Normal Marginal}
  \label{table:MomentClaytonOrg}
  \centering
  \begin{tabular}{|c|ccccccccc|}
    \hline
               &$x_{1}^{0}$&$x_{1}^{1}$&$x_{1}^{2}$&$x_{1}^{3}$&$x_{1}^{4}$&$x_{1}^{5}$&$x_{1}^{6}$&$x_{1}^{7}$&$x_{1}^{8}$\\\hline
    $x_{2}^{0}$&1.000&-0.000&1.000&-0.000&3.000&-0.000&15.000&-0.000&105.000\\
    $x_{2}^{1}$&-0.000&0.611&-0.324&1.818&-1.766&8.994&-12.172&62.311&\\
    $x_{2}^{2}$&1.000&-0.324&1.811&-1.885&8.145&-13.263&54.973&&\\
    $x_{2}^{3}$&-0.000&1.818&-1.885&7.670&-13.818&50.886&&&\\
    $x_{2}^{4}$&3.000&-1.766&8.145&-13.818&50.239&&&&\\
    $x_{2}^{5}$&-0.000&8.994&-13.263&50.886&&&&&\\
    $x_{2}^{6}$&15.000&-12.172&54.973&&&&&&\\
    $x_{2}^{7}$&-0.000&62.311&&&&&&&\\
    $x_{2}^{8}$&105.000&&&&&&&&\\
    \hline
  \end{tabular}
\end{table}
\begin{table}[h]
  \caption{Moments of Fourth Hermite Expansion with Correction from Clayton Copula}
  \label{table:MomentClaytonCorr}
  \centering
  \begin{tabular}{|c|ccccccccc|}
    \hline
               &$x_{1}^{0}$&$x_{1}^{1}$&$x_{1}^{2}$&$x_{1}^{3}$&$x_{1}^{4}$&$x_{1}^{5}$&$x_{1}^{6}$&$x_{1}^{7}$&$x_{1}^{8}$\\\hline
    $x_{2}^{0}$&1.000&-0.000&1.000&-0.000&3.000&-0.170&14.350&-2.291&92.643\\
    $x_{2}^{1}$&-0.000&0.611&-0.324&1.818&-1.383&8.540&-7.619&54.865&\\
    $x_{2}^{2}$&1.000&-0.324&1.811&-1.200&7.479&-6.120&44.569&&\\
    $x_{2}^{3}$&-0.000&1.818&-1.200&6.811&-6.115&37.183&&&\\
    $x_{2}^{4}$&3.000&-1.383&7.479&-6.115&35.883&&&&\\
    $x_{2}^{5}$&-0.170&8.540&-6.120&37.183&&&&&\\
    $x_{2}^{6}$&14.350&-7.619&44.569&&&&&&\\
    $x_{2}^{7}$&-2.291&54.865&&&&&&&\\
    $x_{2}^{8}$&92.643&&&&&&&&\\
    \hline
  \end{tabular}
\end{table}

\subsection{Numerical Examples of Cross Currency Volatility}

This section compares the method for estimating cross currency volatilities using copulas proposed in this study with
the method that uses well-known classical copulas, as presented in Taylor and Wang \cite{taylor2010option}.

\subsubsection{Calibration to Cross Currency Volatility Smile}
First, we examine the fitting to the volatility smile of cross currency pairs.
Supported by CARF, we obtain the volatility data of ATM, 25 Delta Call, 25 Delta Put, 10 Delta Call, and 10 Delta Put for EUR-USD and USD-JPY, provided by Bloomberg.
Next, we interpolate and extrapolate the data to be continuous to the second derivative without any arbitrage, using the method by Kahale \cite{kahale2005smile}.
Furthermore, we calculate the volatilities of the ATM, 25 Delta Call, 25 Delta Put, 10 Delta Call, and 10 Delta Put of the EUR-JPY using the Gauss, Frank, Placett, Clayton, and Gumbel copulas, and the proposed copula.
We calibrate the parameters of the copula to minimize the mean of squares of the difference between the original and estimated volatilities for ATM, 25 Delta Call, 25 Delta Put, 10 Delta Call, and 10 Delta Put of the EUR-JPY, obtained from Bloomberg. 
In the case with more than two parameters (proposed copula), the quasi-Newton method is used for minimization. For the case with
one parameter (other copulas), the Brent method (See Press et al. \cite{PresTeukVettFlan92}) is used.

We set $N_{M}=6$ in the formula \eqref{formula:crossVInt}.
$\rho,\hat{m}_{3,0}, \hat{m}_{4,0}, \hat{m}_{5,0}, and \hat{m}_{6,0 }$ are chosen as the calibration parameters,
and the other parameters $\hat{m}_{n,i}$ are set to 0.
Non-negativity of the probability density, normalized condition of the probability density, and moments match of $\hat{m}_{i,0}, (1\leq i \leq 6)$ including $\hat{m}_{1,0}=0$ and $\hat{m}_{2,0}=0$.
They are chosen as the properties for the correction
$\mbox{Proj}_{\mu_{\Phi_{\mathbb{I}}}\mathcal{K}_{\mathbb{I},All}}\left(1+\sum_{n=3}^{n_{M}}\sum_{i=0}^{n}\hat{m}_{n,i}e_{n,i}\right)$ $\hat{m}_{i,0}, (1\leq i \leq 6)$.
In this situation, we use the correction of a one-dimensional product discussed in subsection \ref{subsection:1DCorrection}.

Hereafter, the volatilities set for ATM, 25 Delta Call, 25 Delta Put, 10 Delta Call, and 10 Delta Put are called ``smiles,''
and estimating the parameters of each copula using the method described here is called calibrating to the smile.
We also scale the coefficients according to the parameters of the proposed copula, using $\check{m}_{i}=i!\times\hat{m}_{i,0}$, $i=1,\cdots,6$.

\begin{figure}[H]
  \begin{tabular}{cc}
    \begin{minipage}[t]{0.5\hsize}
      \centering
      \includegraphics[keepaspectratio, scale=0.6]{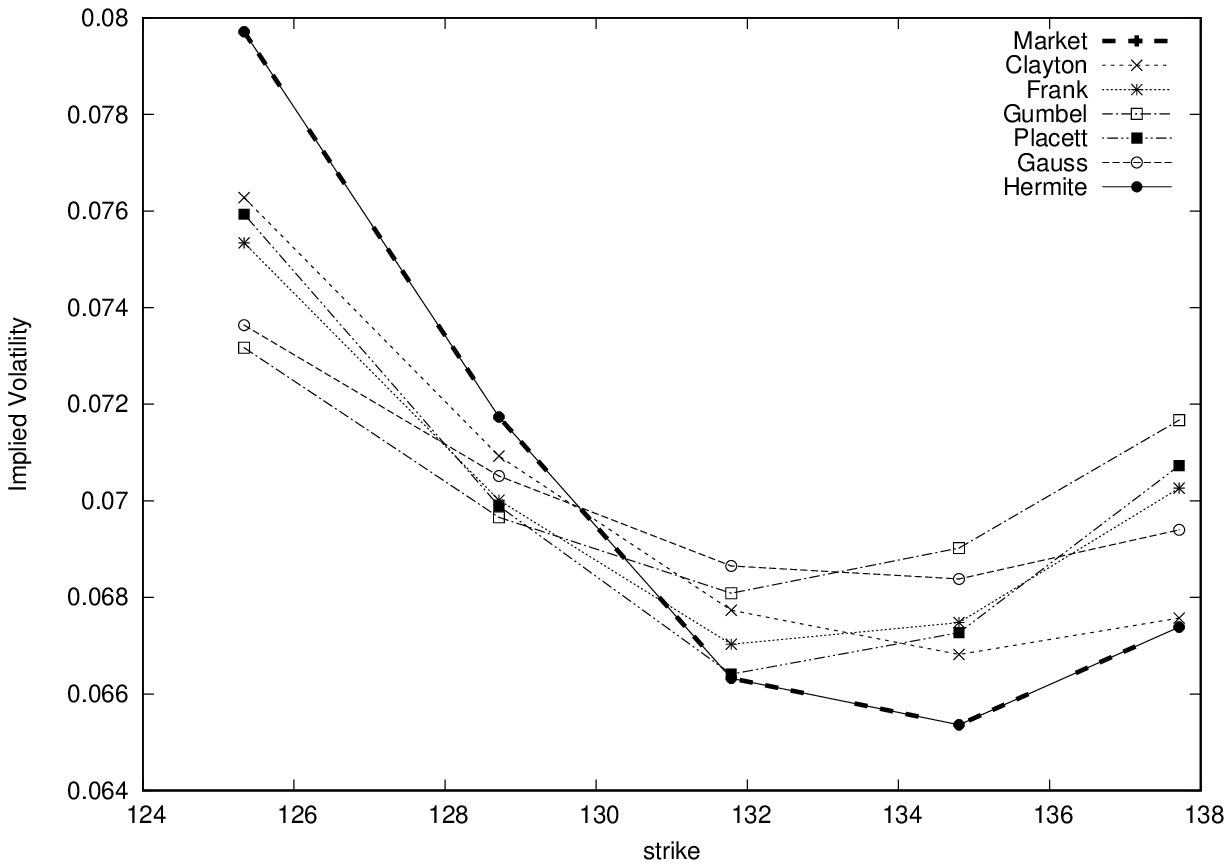}
      \caption{EUR-JPY 3M Implied volatility (Oct 29 2021)}
      \label{figure:Exam2_20211029Smile3M}
    \end{minipage} &
    \begin{minipage}[t]{0.5\hsize}
      \centering
      \includegraphics[keepaspectratio, scale=0.6]{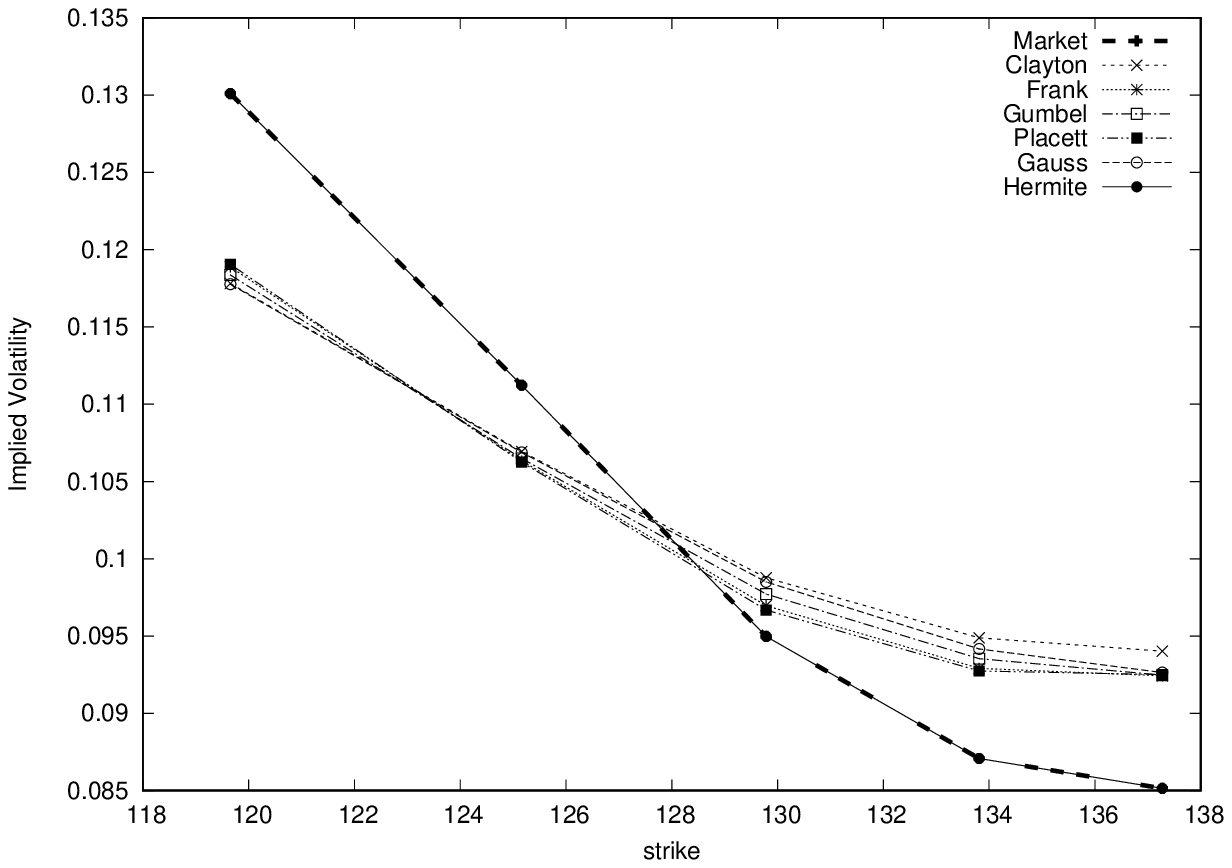}
      \caption{EUR-JPY 3M Implied volatility (Feb 28 2022)}
      \label{figure:Exam2_20220228Smile3M}
    \end{minipage}
  \end{tabular}
\end{figure}
\begin{figure}[H]
  \begin{tabular}{cc}
    \begin{minipage}[t]{0.5\hsize}
      \centering
      \includegraphics[keepaspectratio, scale=0.6]{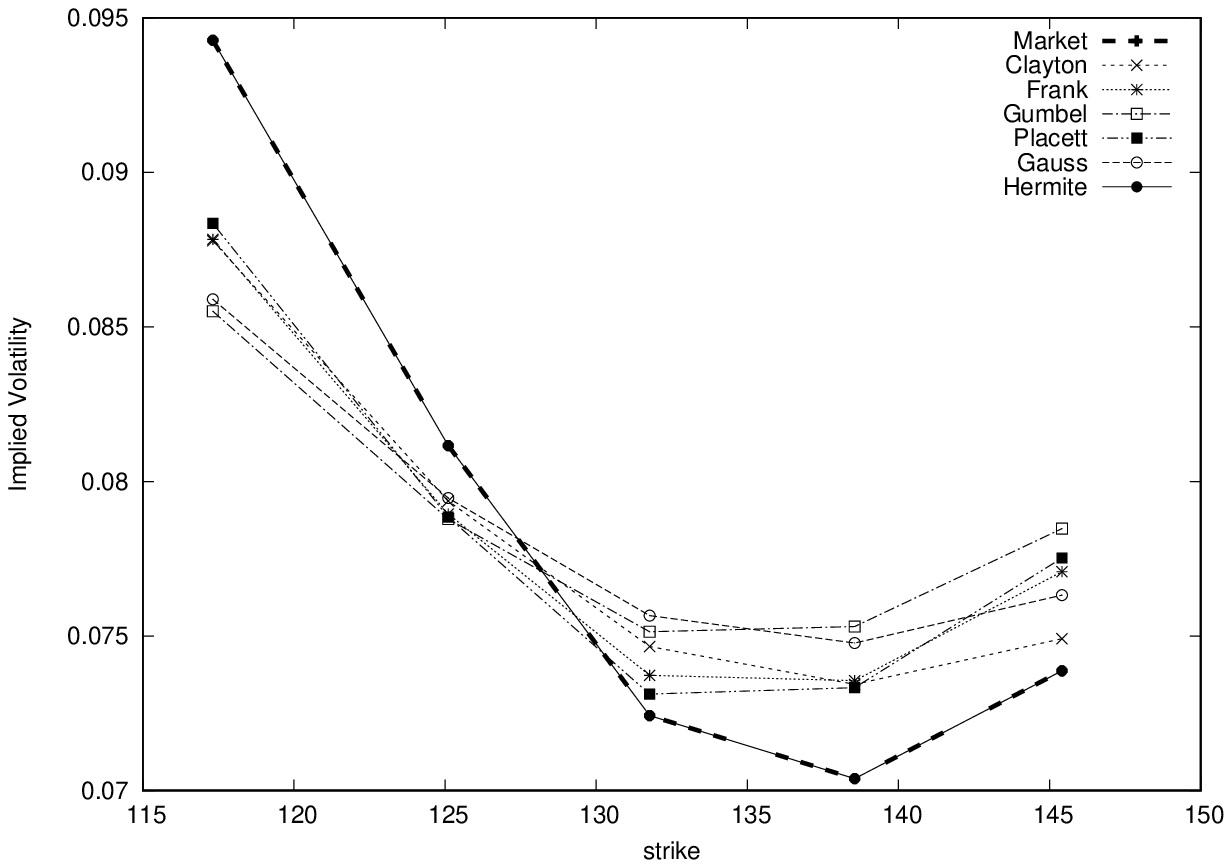}
      \caption{EUR-JPY 1Y Implied volatility (Oct 29 2021)}
      \label{figure:Exam2_20211029Smile1Y}
    \end{minipage} &
    \begin{minipage}[t]{0.5\hsize}
      \centering
      \includegraphics[keepaspectratio, scale=0.6]{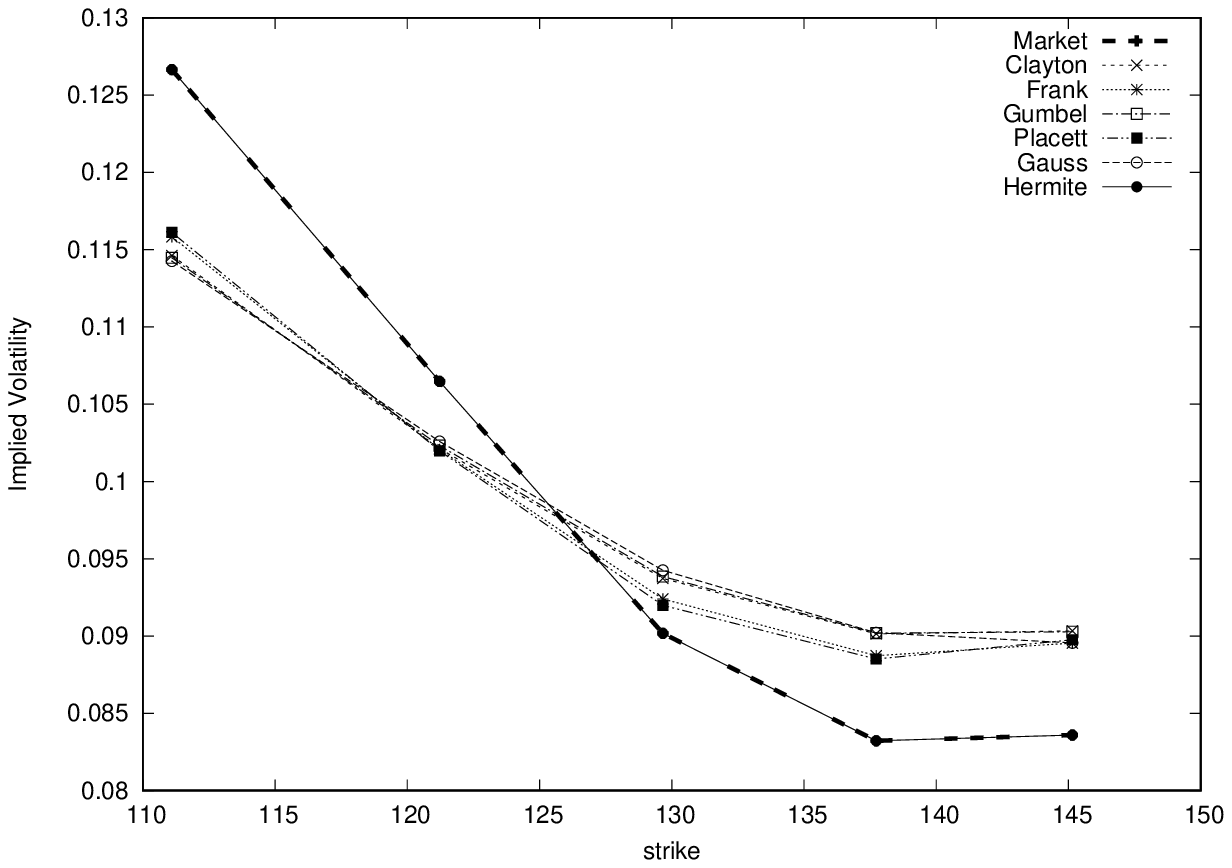}
      \caption{EUR-JPY 1Y Implied volatility (Feb 28 2022)}
      \label{figure:Exam2_20220228Smile1Y}
    \end{minipage}
  \end{tabular}
\end{figure}
\begin{figure}[H]
  \begin{tabular}{cc}
    \begin{minipage}[t]{0.5\hsize}
      \centering
      \includegraphics[keepaspectratio, scale=0.6]{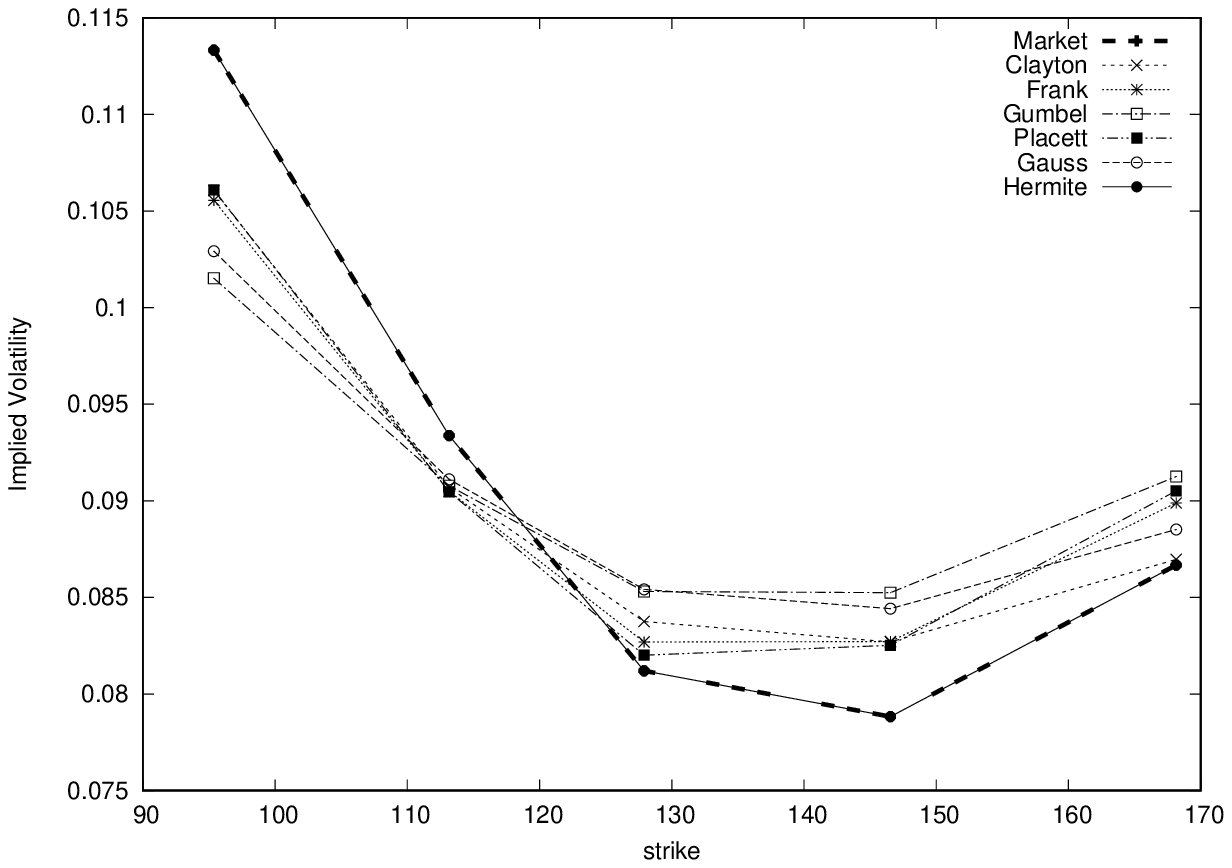}
      \caption{EUR-JPY 5Y Implied volatility (Oct 29 2021)}
      \label{figure:Exam2_20211029Smile5Y}
    \end{minipage} &
    \begin{minipage}[t]{0.5\hsize}
      \centering
      \includegraphics[keepaspectratio, scale=0.6]{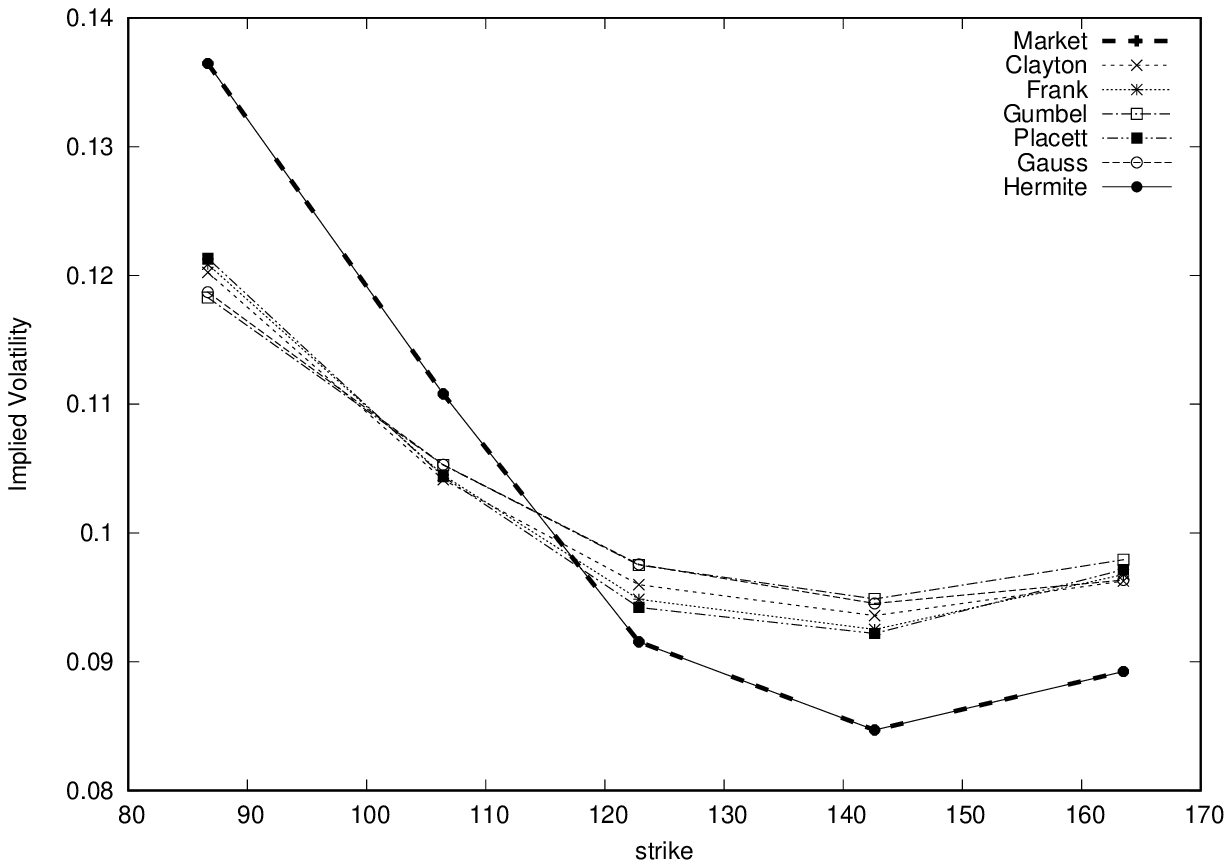}
      \caption{EUR-JPY 5Y Implied volatility (Feb 28 2022)}
      \label{figure:Exam2_20220228Smile5Y}
    \end{minipage}
  \end{tabular}
\end{figure}

\begin{table}[H]
  \caption{Fitting parameters for EUR-JPY Smile}
  \label{table:OtherCopulaParams}
  \centering
  \begin{tabular}{|c|c|c|c|c|c|c|}
    \hline
    Date & Tenor & Clayton & Frank & Gumbel & Placett & Gauss\\\hline
    Oct 29 2021 & 3M & 0.8160 & 3.0926 & 1.3357 & 4.2196 & 0.4203\\
    Feb 28 2022 & 3M & 0.2601 & 1.3456 & 1.1283 & 1.9490 & 0.1928\\\hline
    Oct 29 2021 & 1Y & 0.6031 & 2.6733 & 1.2842 & 3.5460 & 0.3657\\
    Feb 28 2022 & 1Y & 0.3859 & 1.8095 & 1.1679 & 2.4207 & 0.2541\\\hline
    Oct 29 2021 & 5Y & 0.7708 & 3.1600 & 1.2982 & 4.2894 & 0.4068\\
    Feb 28 2022 & 5Y & 0.5922 & 2.5364 & 1.2211 & 3.3346 & 0.3342\\
    \hline
  \end{tabular}
\end{table}

\begin{table}[H]
  \caption{Fitting Parameters of Hermite Expansion Copula for EUR-JPY Smile}
  \label{table:HermiteExpansionParams}
  \centering
  \begin{tabular}{|c|c|c|c|c|c|c|}
    \hline
    Date & Tenor & $\rho$ & $\check{m}_{3}$ & $\check{m}_{4}$ & $\check{m}_{5}$ & $\check{m}_{6}$\\\hline
    Oct 29 2021 & 3M & 0.3884 & -0.2960 & 0.5528 & 0.0924 & -2.4115\\
    Feb 28 2022 & 3M & 0.2080 & -0.5156 & 0.6395 & 0.5010 & -2.4400\\\hline
    Oct 29 2021 & 1Y & 0.3661 & -0.3535 & 0.9641 & 0.0827 & -2.0537\\
    Feb 28 2022 & 1Y & 0.2679 & -0.4929 & 0.7944	& 0.3854	& -3.2577\\\hline
    Oct 29 2021 & 5Y & 0.4068 & -0.4090 & 1.2771 & -0.1257 & -2.6293\\
    Feb 28 2022 & 5Y & 0.3477 & -0.7098 & 1.3640 & 0.2541 & -5.1991\\
    \hline
  \end{tabular}
\end{table}

The results of calibrating to EUR-JPY smiles of 3M, 1Y, and 5Y on October 29, 2021 are shown in Figures \ref{figure:Exam2_20211029Smile3M}, \ref{figure:Exam2_20211029Smile1Y}, and \ref{figure:Exam2_20211029Smile5Y}.
The results of calibrating to smiles of 3M, 1Y and 5Y on February 28, 2022 are shown in Figures \ref{figure:Exam2_20220228Smile3M}, \ref{figure:Exam2_20220228Smile1Y}, and \ref{figure:Exam2_20220228Smile5Y}.
The date October 29, 2021 is selected because it was marked with a period of relatively calm market movement throughout the following month,
while February 28 2022 was selected as a time of market turbulence caused by the combination of the Ukrainian war and an increased interest rate in the U.S. over the following month.

"Market" in Figures \ref{figure:Exam2_20211029Smile3M} -\ref{figure:Exam2_20220228Smile5Y} is the original smile of the EUR-JPY.
"Gauss", "Frank", "Placett", "Clayton", and "Gumbel" are the implied volatilities of the respective copulas.
"Hermite" is the implied volatility of the proposed method.
Tables \ref{table:OtherCopulaParams} and \ref{table:HermiteExpansionParams} are the copula parameters obtained through the calibration.

The smiles estimated using the "Hermite" are much closer to the original smiles than those estimated using other copulas.
This indicates that the selected parameters can sufficiently replicate the original volatility smile.
However, when other copulas are used, large differences from the original volatility smile are observed.
It is difficult to express a strongly skewed shape using these classical copulas.

The parameter $\rho$ of the proposed method is close to that of the Gauss copula,
which is consistent with the fact that it is an extension of the Gauss copula.

Additionally, negative values are obtained as the coefficient $\check{m}_{6}$ for the highest order.
This indicates that there exists a negative density region in the non-corrected Hermite polynomial expansion;
nonetheless, the correction method satisfies non-negativity and the copula is obtained. 

\subsubsection{Estimation of Monthly Cross Currency Volatility Smile}
In the market, the liquidity of the volatility of cross currency pairs is generally less than that of the straight currency pairs.
For example, there could be a situation where the smiles for the straight currency pairs can be obtained daily
but a smile for a cross currency pair can only be obtained monthly.
We also consider the case that the smile of a cross currency pair can be obtained monthly
but the ATM volatility can be obtained daily.

In this section, we examine the same copulas as in the previous section under two different settings.
One setting (c) is to calibrate the copula parameters according to the smile of the cross-currency pair at the end of the previous month.
Next, one of the parameters is calibrated to match the ATM volatility of the cross-currency pair for each business day of the current month. Accordingly, we calculate the smile.
In the other setting (d), the copula parameters are calibrated according to the smile of the cross-currency pair at the end of the previous month,
and the parameters are used to calculate the smile of the crosscurrency pair for each business day in the current month.

First, we verify which parameters of the proposed copula should be used for daily calibrations in setting (c).

\begin{figure}[H]
  \centering
  \includegraphics[keepaspectratio, scale=1.0]{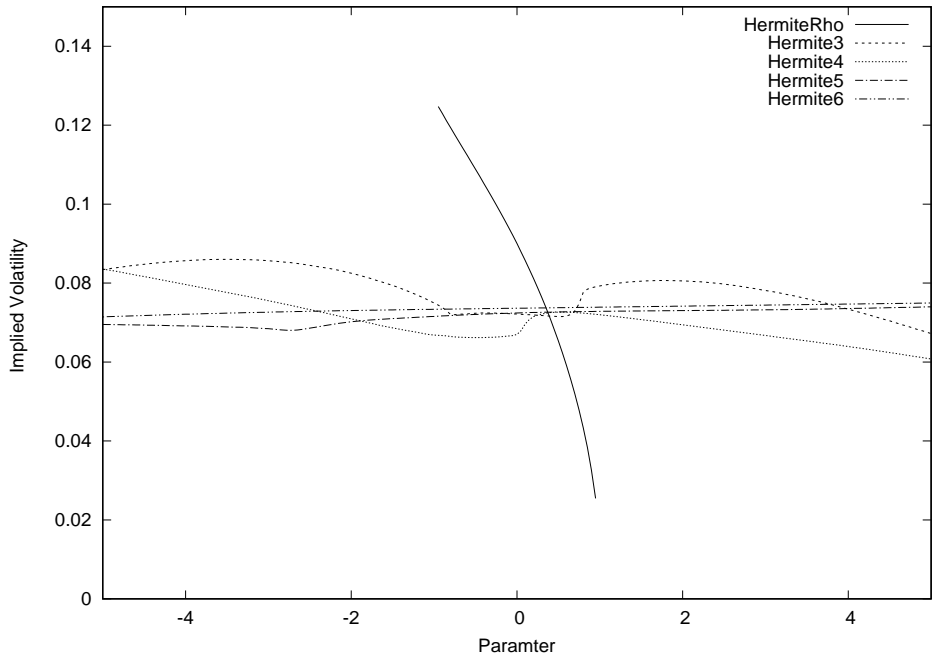}
  \caption{Volatility of Parameter shift (Oct 29 2021)}
  \label{figure:Exam2_20221029OneParamShift1Y}
\end{figure}

Figure \ref{figure:Exam2_20221029OneParamShift1Y} plots the change in EUR-JPY ATM volatility when only one parameter is shifted
from the parameters set on October 29, 2021 with 1Y tenor in Table \ref{table:HermiteExpansionParams}.

"HermiteRho", "Hermite3", "Hermite4", "Hermite5", and "Hermite6", correspond to the case where each parameter $\rho$, $\check{m}_{3}$, $\check{m}_{4}$, $\check{m}_{5}$, and $\check{m}_{6}$ has moved.
The horizontal axis denotes the value of each parameter and the vertical axis denotes the value of ATM volatility.

$\check{m}_{3}$, $\check{m}_{4}$, and $\check{m}_{5}$ are not monotonically changing graphs.
There may be more than one solution for the root of the given ATM volatility,
which leads to parameter instability when fitting daily ATM volatility.

\begin{figure}[H]
  \centering
  \includegraphics[keepaspectratio, scale=1.0]{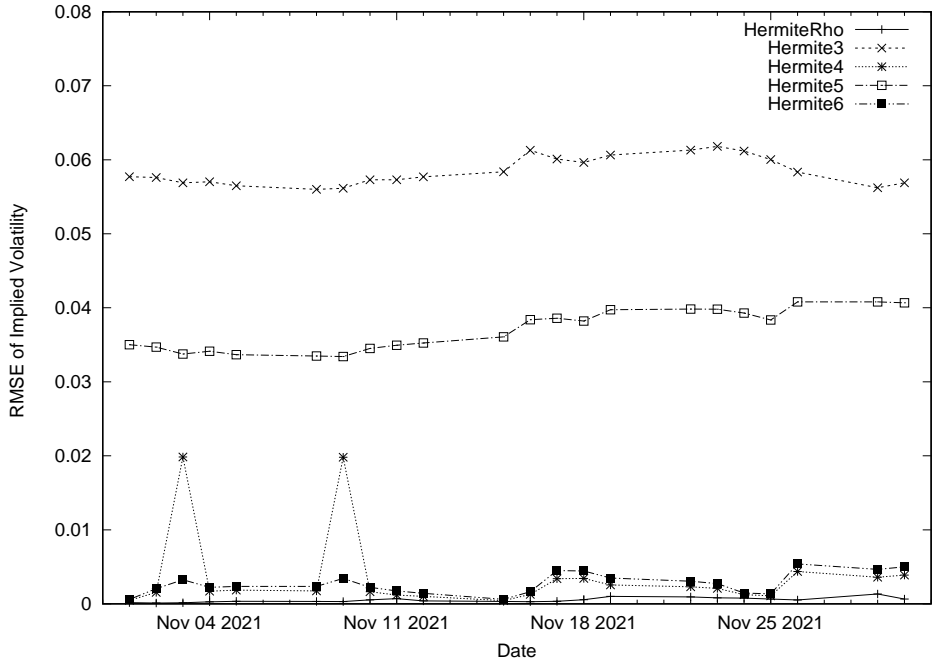}
  \caption{EUR-JPY 1Y Daily RMSE with calibration of Hermite parameters (Nov 2021)}
  \label{figure:Exam2_202211OneParam1Y}
\end{figure}

Figure \ref{figure:Exam2_202211OneParam1Y} plots the root mean squared error (RMSE) between the real smile of EUR-JPY with 1Y tenor and estimated smile
with recalibrated parameters from each business day in Nov 2021.
Recalibration is carried out by adjusting the ATM volatility of EUR-JPY for each business day and
by changing one of the parameters that were calibrated according to the smile on October 29, 2021.
The Brent method is used for recalibration.

The relations between the legend and parameters in Figure \ref{figure:Exam2_202211OneParam1Y} are the same as those in Figure \ref{figure:Exam2_20221029OneParamShift1Y}.
The RMSE of daily calibrations with $\rho$ is the smallest and stable throughout the whole term.
These results suggest that $\rho$ should be used for calibrating to ATM volatility.

\begin{figure}[H]
  \begin{tabular}{cc}
    \begin{minipage}[t]{0.5\hsize}
      \centering
      \includegraphics[keepaspectratio, scale=0.6]{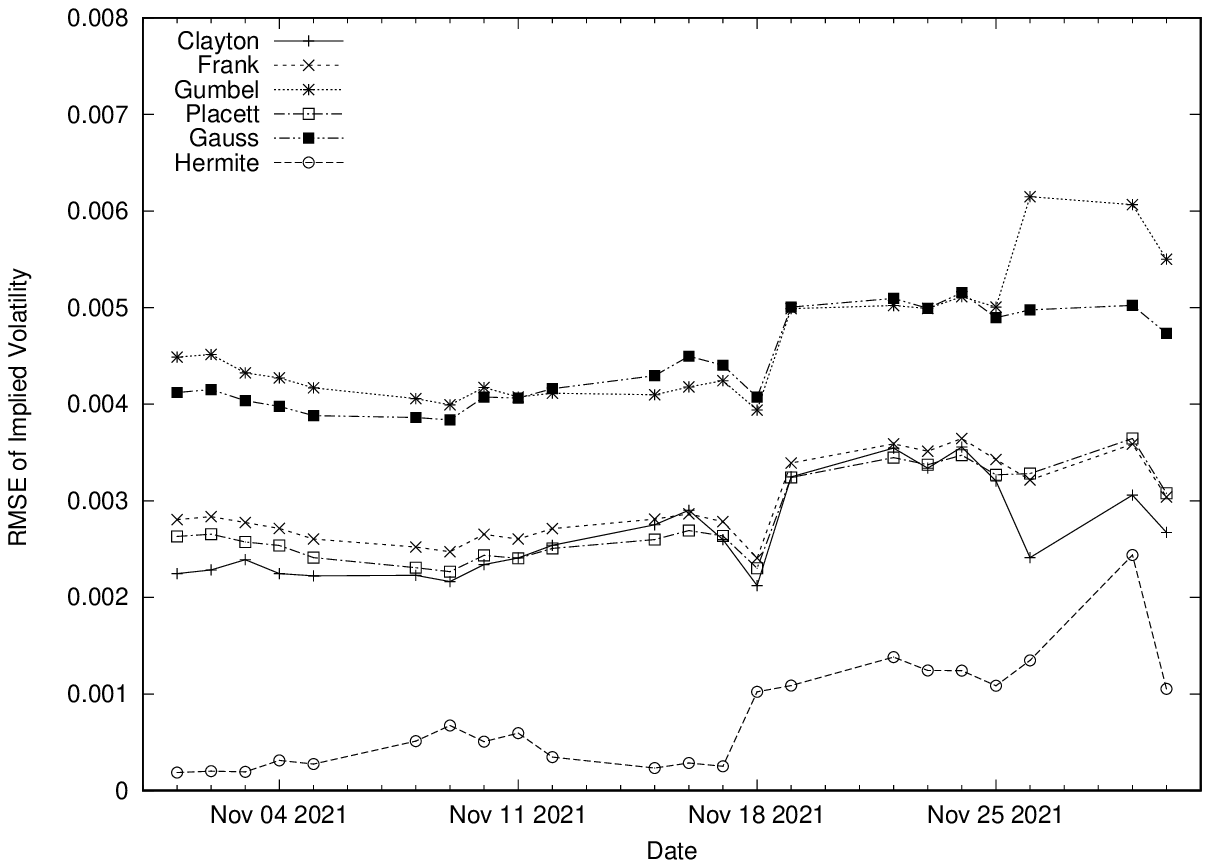}
      \caption{Daily RMSE of tenor 3M and (c) in Nov 2021}
      \label{figure:Exam2_202111ForecastC3M}
    \end{minipage} &
    \begin{minipage}[t]{0.5\hsize}
      \centering
      \includegraphics[keepaspectratio, scale=0.6]{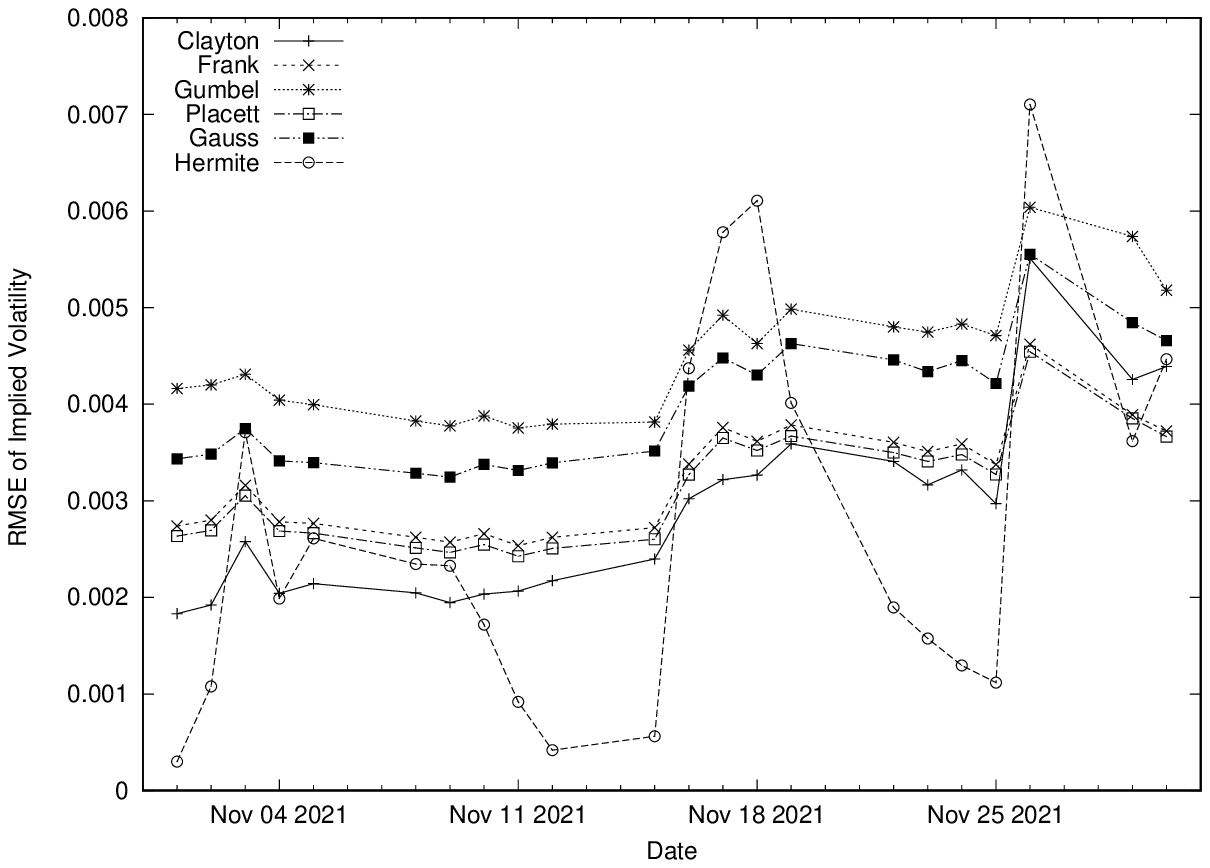}
      \caption{Daily RMSE of tenor 3M and (d) in Nov 2021}
      \label{figure:Exam2_202111ForecastD3M}
    \end{minipage}
  \end{tabular}
\end{figure}
\begin{figure}[H]
  \begin{tabular}{cc}
    \begin{minipage}[t]{0.5\hsize}
      \centering
      \includegraphics[keepaspectratio, scale=0.6]{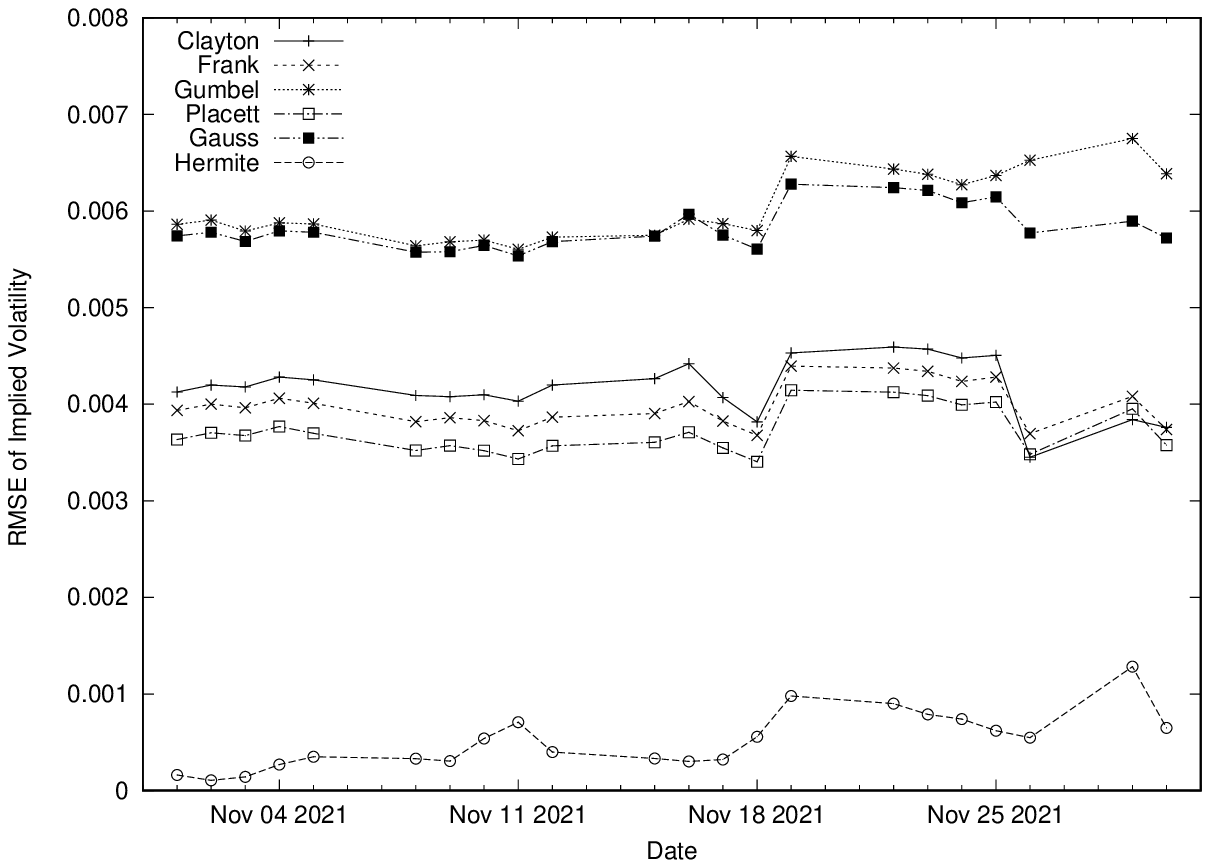}
      \caption{Daily RMSE of tenor 1Y and (c) in Nov 2021}
      \label{figure:Exam2_202111ForecastC1Y}
    \end{minipage} &
    \begin{minipage}[t]{0.5\hsize}
      \centering
      \includegraphics[keepaspectratio, scale=0.6]{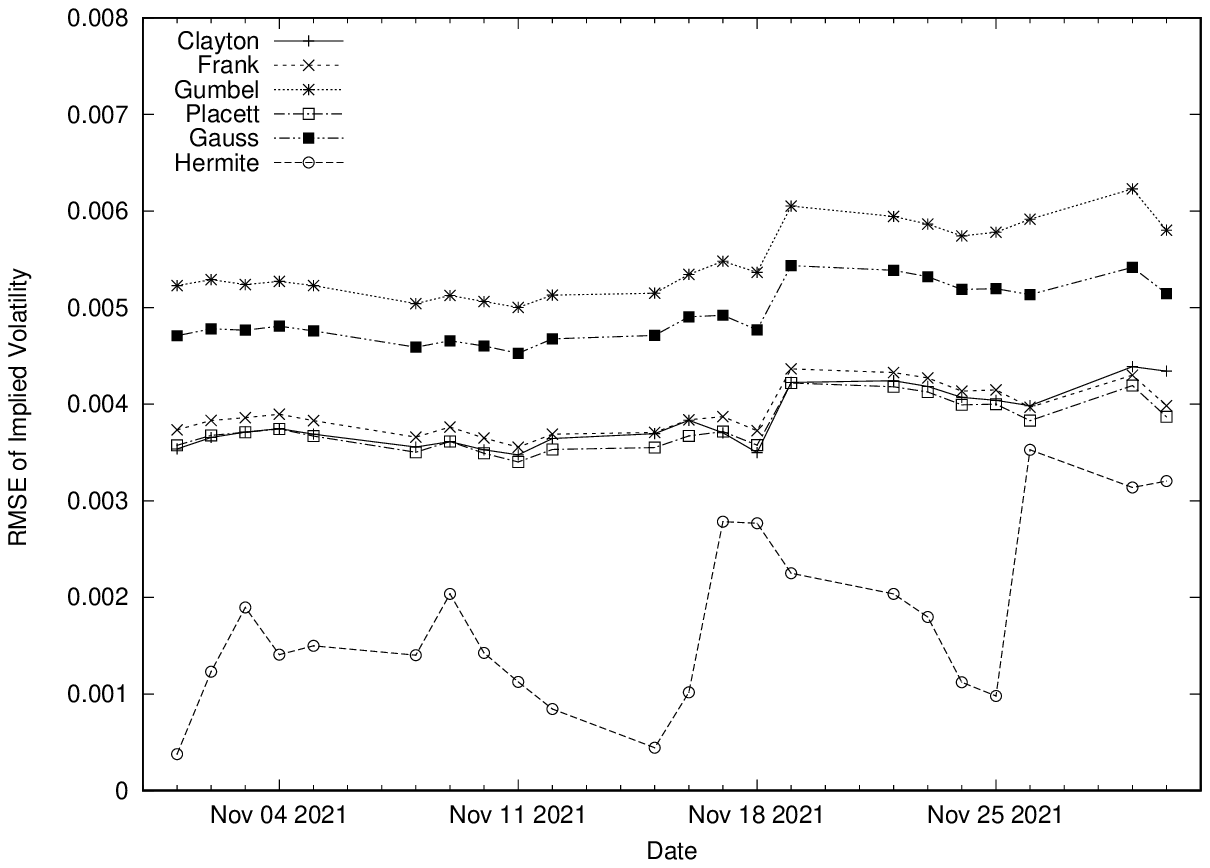}
      \caption{Daily RMSE of tenor 1Y and (d) in Nov 2021}
      \label{figure:Exam2_202111ForecastD1Y}
    \end{minipage}
  \end{tabular}
\end{figure}
\begin{figure}[H]
  \begin{tabular}{cc}
    \begin{minipage}[t]{0.5\hsize}
      \centering
      \includegraphics[keepaspectratio, scale=0.6]{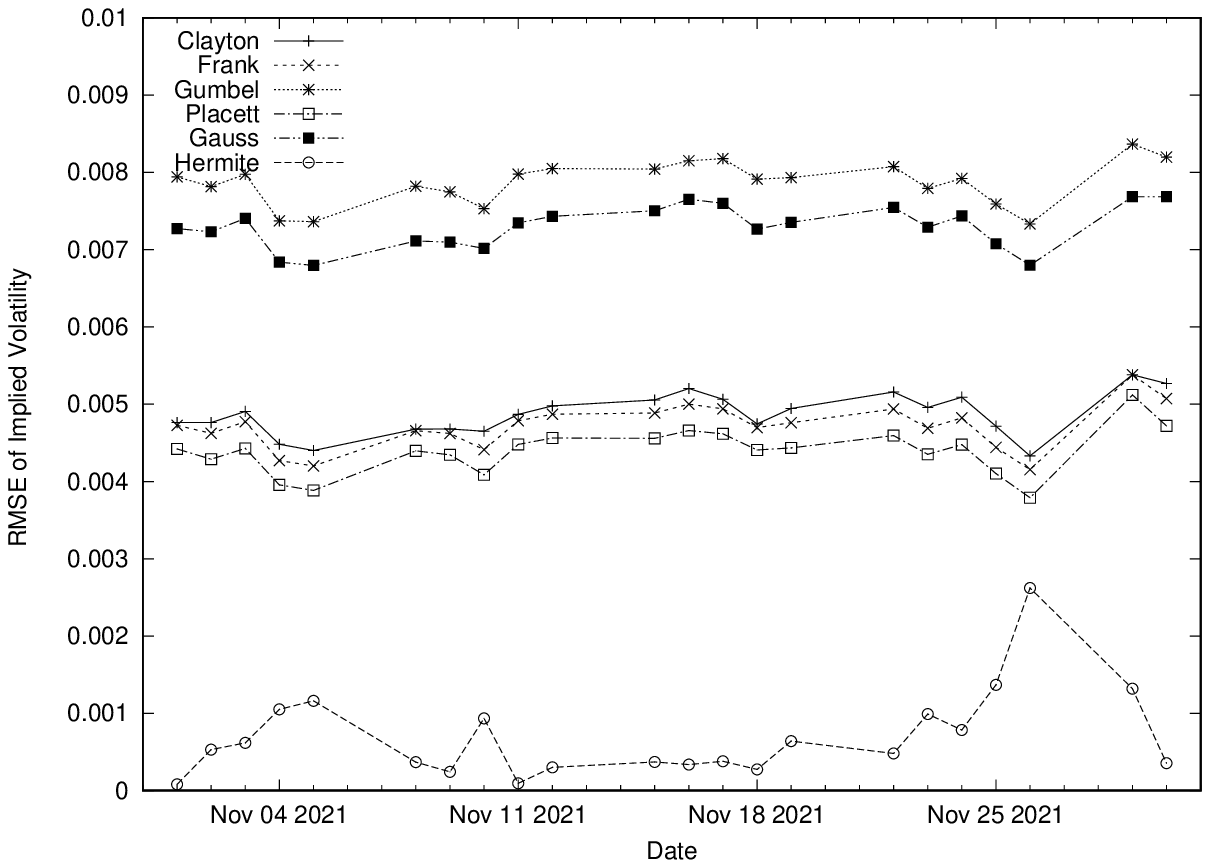}
      \caption{Daily RMSE of tenor 5Y and (c) in Nov 2021}
      \label{figure:Exam2_202111ForecastC5Y}
    \end{minipage} &
    \begin{minipage}[t]{0.5\hsize}
      \centering
      \includegraphics[keepaspectratio, scale=0.6]{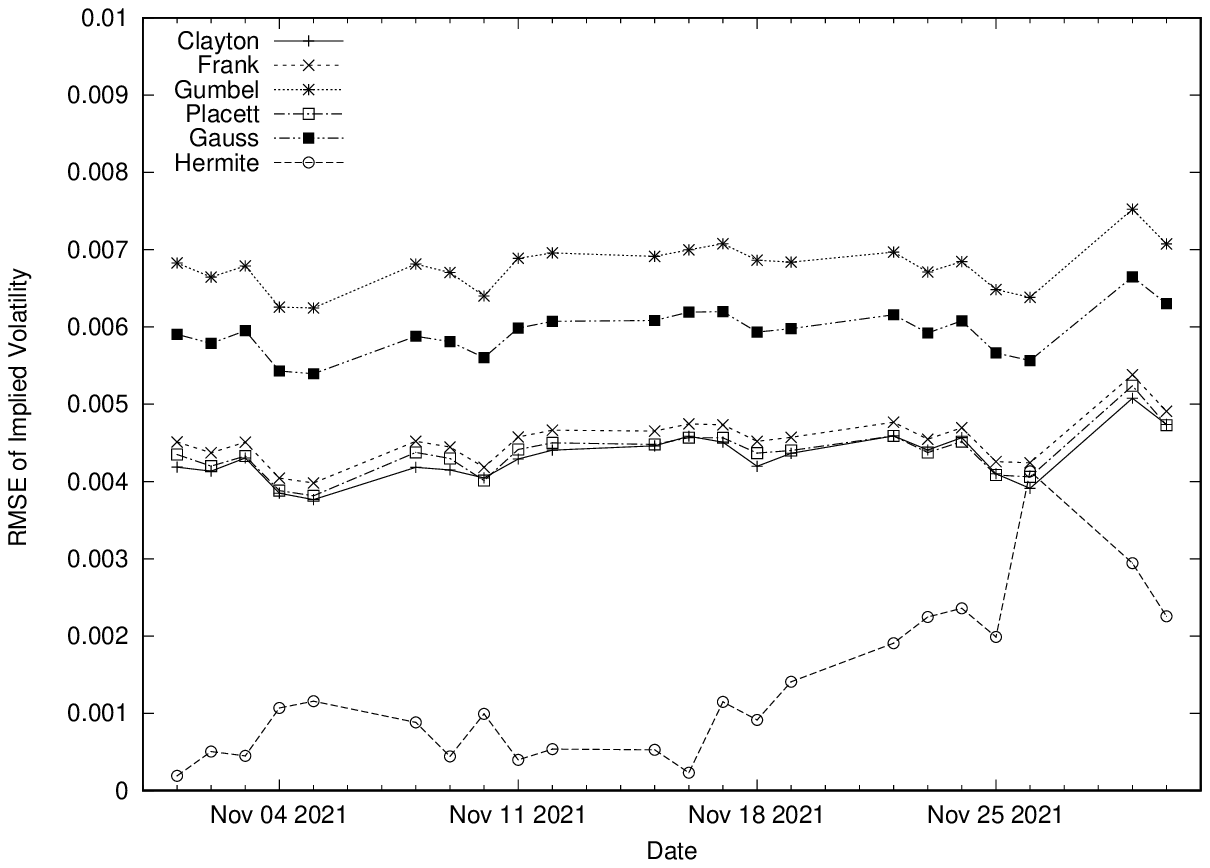}
      \caption{Daily RMSE of tenor 5Y and (d) in Nov 2021}
      \label{figure:Exam2_202111ForecastD5Y}
    \end{minipage}
  \end{tabular}
\end{figure}

Figures \ref{figure:Exam2_202111ForecastC3M} - \ref{figure:Exam2_202111ForecastD5Y} show the daily RMSE in November 2021
whose market data are 3M, 1Y, and 5Y tenor and settings are (c) and (d).
Figures \ref{figure:Exam2_202203ForecastC3M} - \ref{figure:Exam2_202203ForecastD5Y} in Appendix
are the daily RMSE in March 2022 whose tenor and settings are the same as those in Figures \ref{figure:Exam2_202111ForecastC3M} - \ref{figure:Exam2_202111ForecastD5Y}.

The RMSE using the proposed copula show smaller RMSE for many days
compared to those that have been computed using other copulas.
Additionally, the RMSE in setting (c) is much lower than that in setting (d) across many days.
This fact implies that the copula parameter $\rho$ can accurately explain the volatility smile fluctuations in the cross-currency pair,
which are not explained by the volatility smile fluctuations in straight currency pairs.

\begin{figure}[H]
  \begin{tabular}{cc}
    \begin{minipage}[t]{0.5\hsize}
      \centering
      \includegraphics[keepaspectratio, scale=0.6]{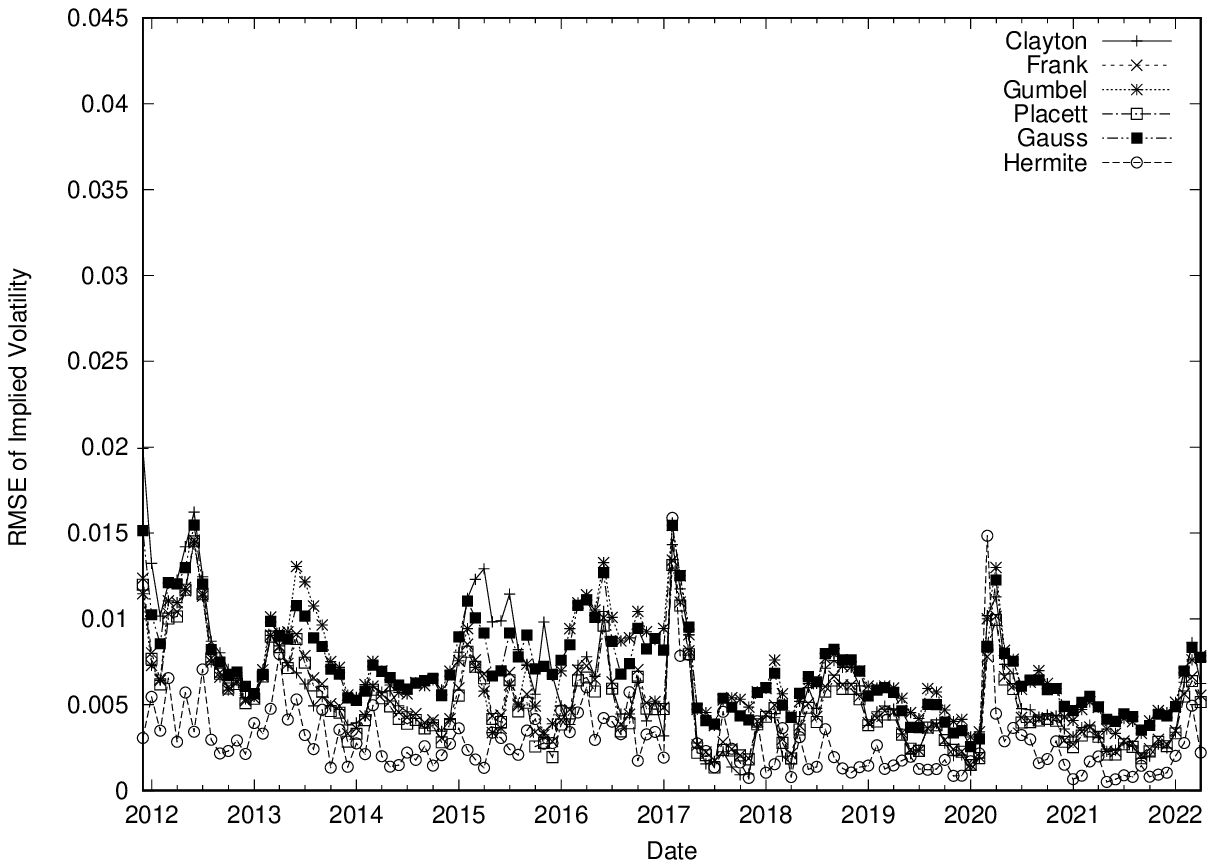}
      \caption{Monthly RMSE tenor 3M and (c)}
      \label{figure:Exam2_MonthlyC3M}
    \end{minipage} &
    \begin{minipage}[t]{0.5\hsize}
      \centering
      \includegraphics[keepaspectratio, scale=0.6]{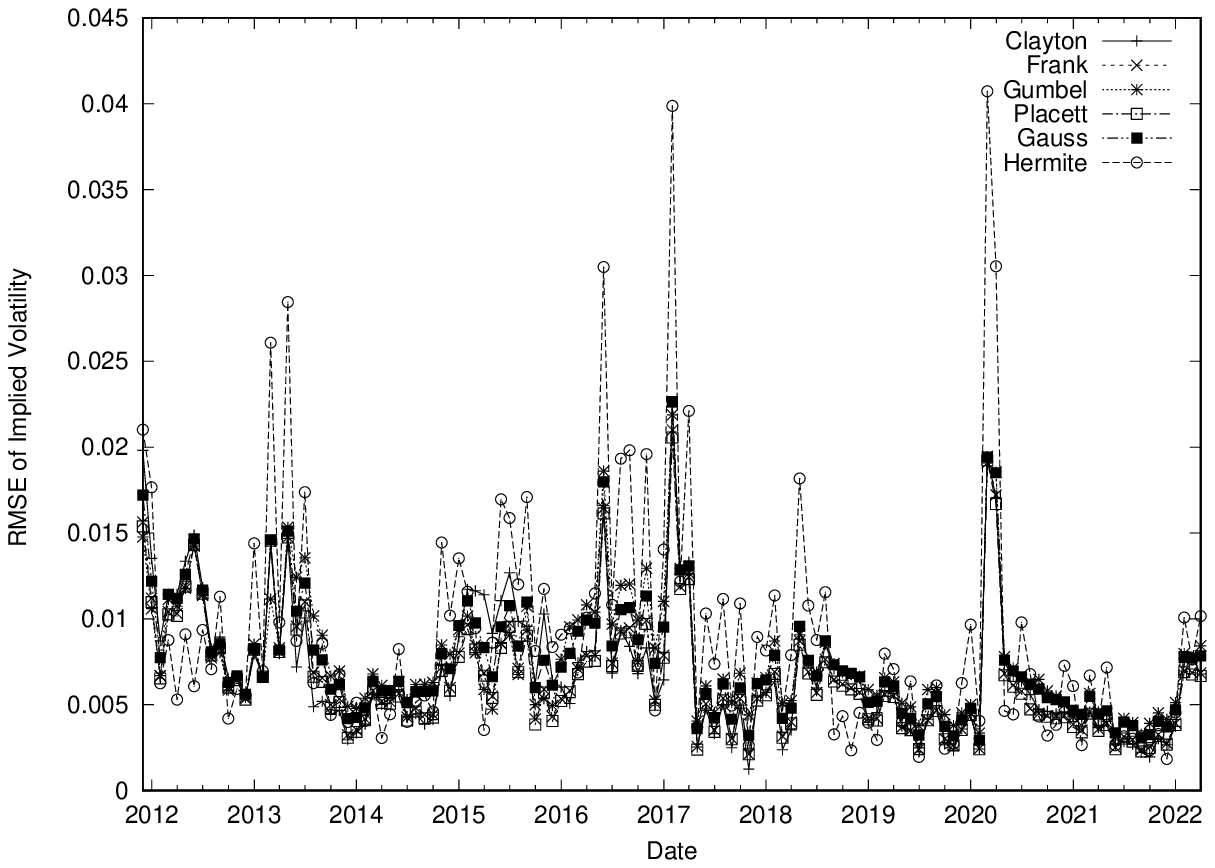}
      \caption{Monthly RMSE tenor 3M and (d)}
      \label{figure:Exam2_MonthlyD3M}
    \end{minipage}
  \end{tabular}
\end{figure}
\begin{figure}[H]
  \begin{tabular}{cc}
    \begin{minipage}[t]{0.5\hsize}
      \centering
      \includegraphics[keepaspectratio, scale=0.6]{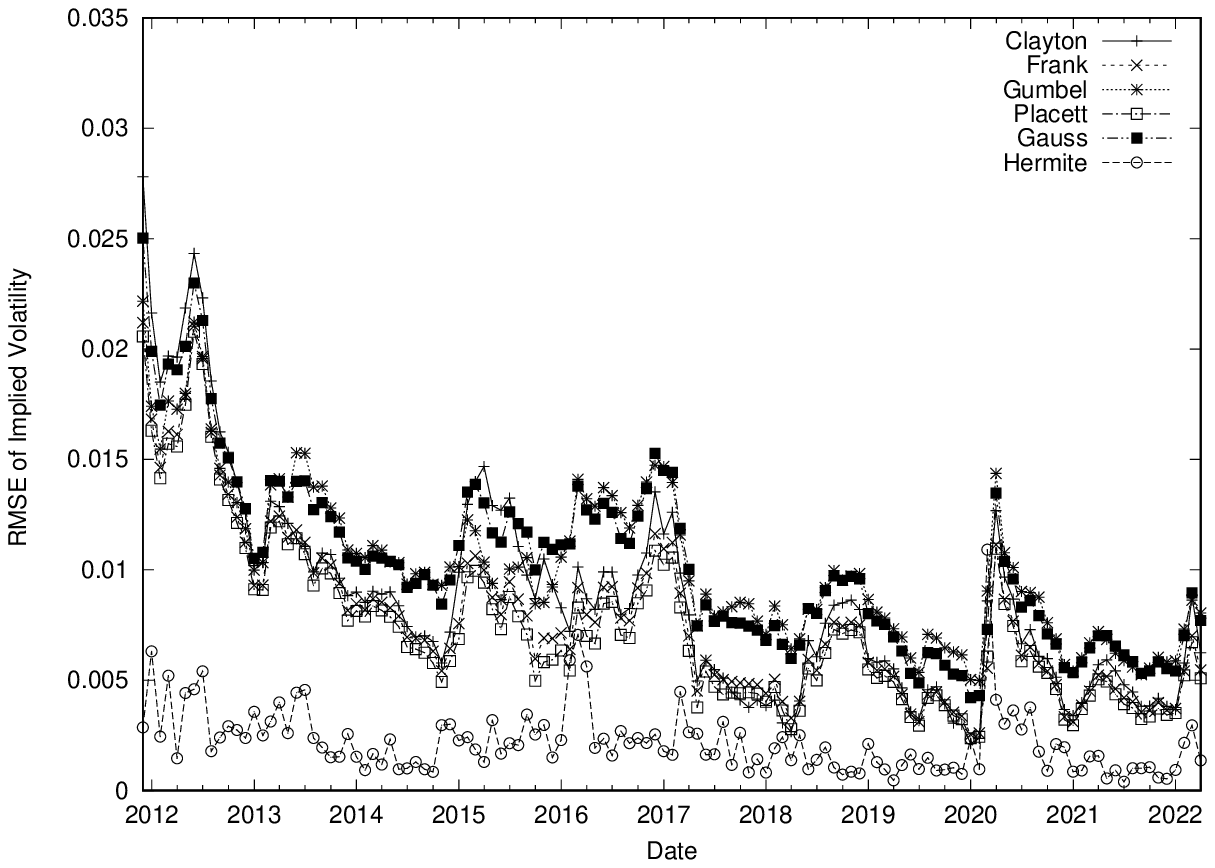}
      \caption{Monthly RMSE tenor 1Y and (c)}
      \label{figure:Exam2_MonthlyC1Y}
    \end{minipage} &
    \begin{minipage}[t]{0.5\hsize}
      \centering
      \includegraphics[keepaspectratio, scale=0.6]{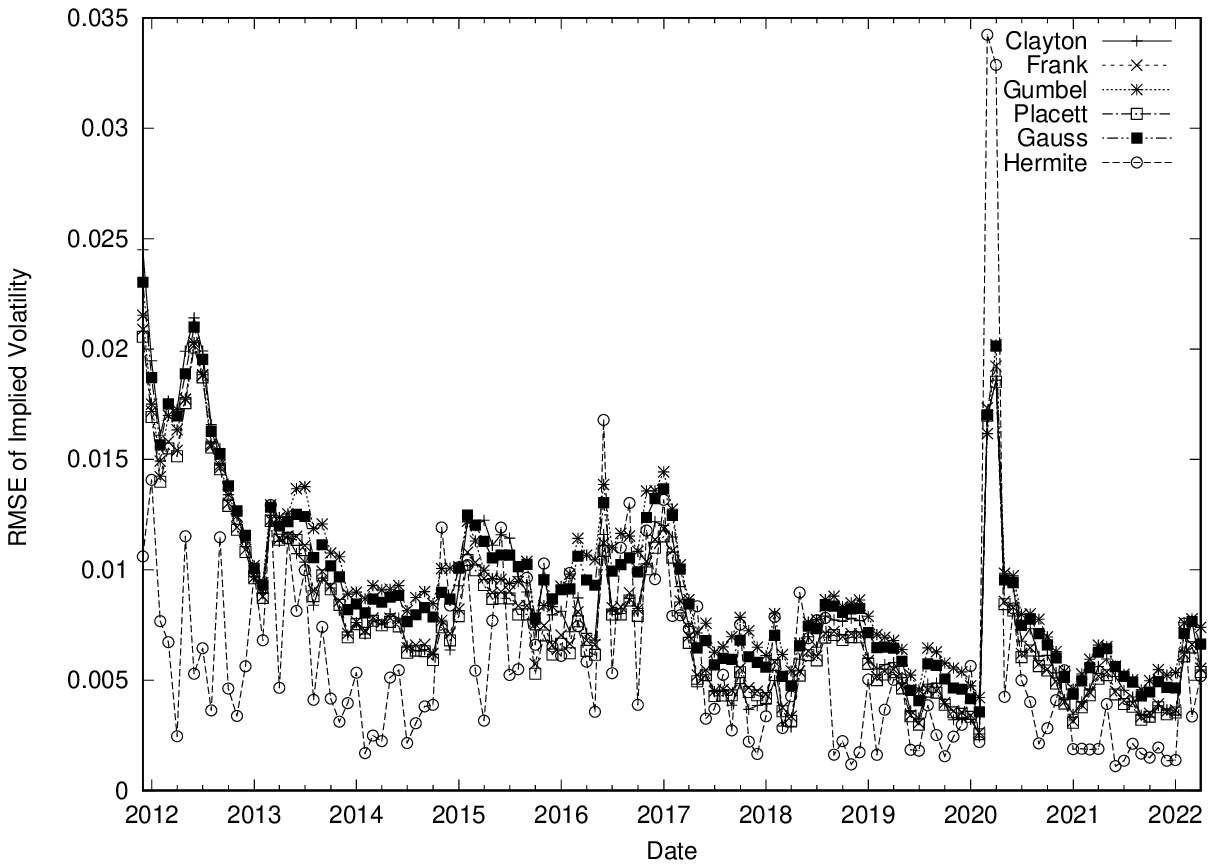}
      \caption{Monthly RMSE tenor 1Y and (d)}
      \label{figure:Exam2_MonthlyD1Y}
    \end{minipage}
  \end{tabular}
\end{figure}
\begin{figure}[H]
  \begin{tabular}{cc}
    \begin{minipage}[t]{0.5\hsize}
      \centering
      \includegraphics[keepaspectratio, scale=0.6]{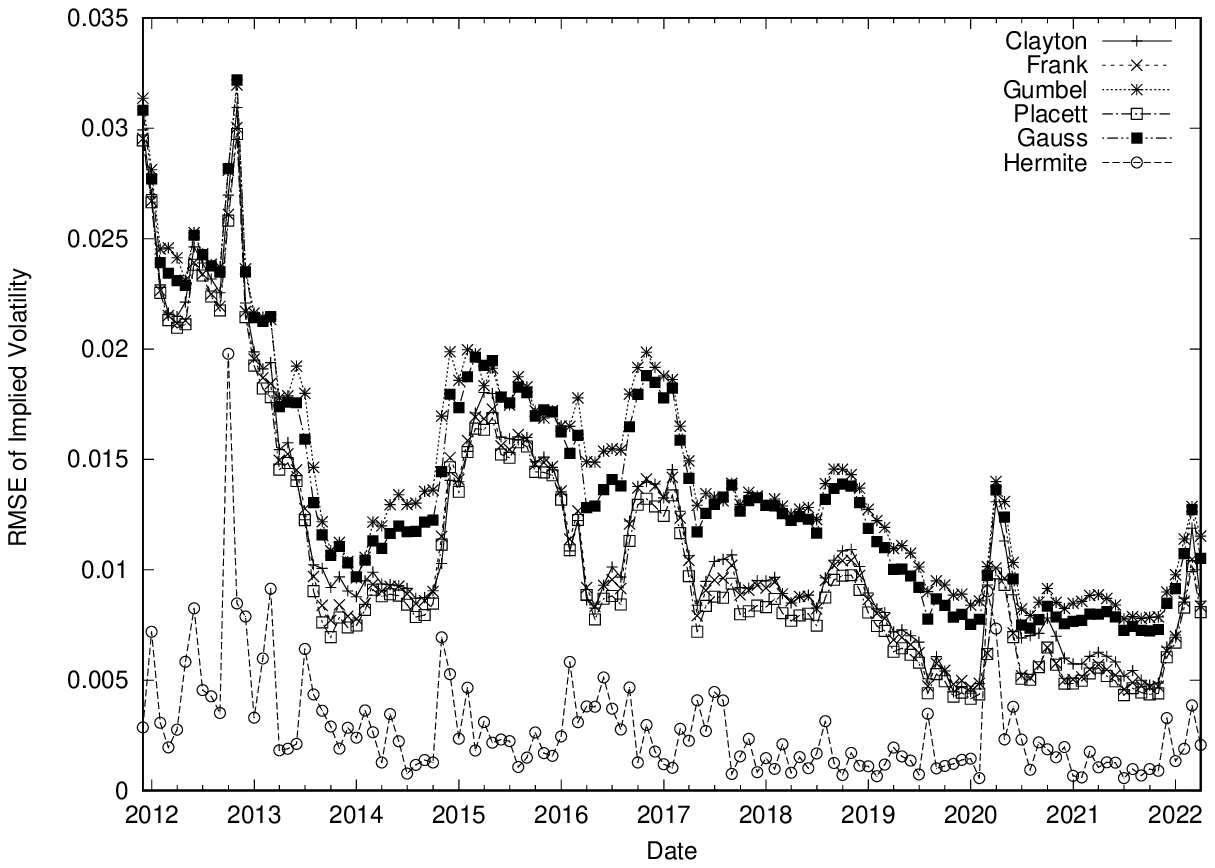}
      \caption{Monthly RMSE tenor 5Y and (c)}
      \label{figure:Exam2_MonthlyC5Y}
    \end{minipage} &
    \begin{minipage}[t]{0.5\hsize}
      \centering
      \includegraphics[keepaspectratio, scale=0.6]{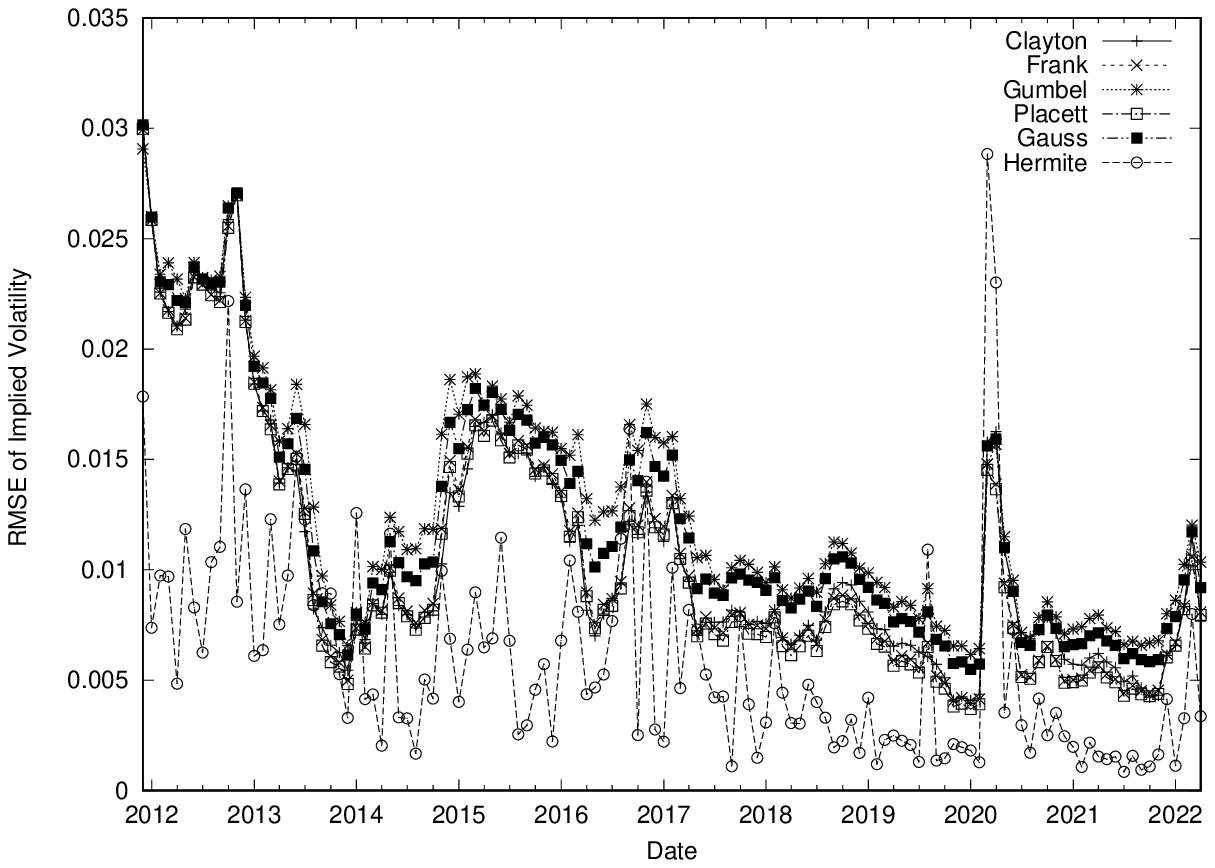}
      \caption{Monthly RMSE tenor 5Y and (d)}
      \label{figure:Exam2_MonthlyD5Y}
    \end{minipage}
  \end{tabular}
\end{figure}

Figures \ref{figure:Exam2_MonthlyC3M} - \ref{figure:Exam2_MonthlyD5Y} show the monthly RMSE
calculated using 3M, 1Y, and 5Y tenor market data from January, 2007 to April, 2022, for each business day in the month. 
In setting (c), after calibrating the smile of the cross-currency pair at the end of the previous month,
one parameter ($\rho$ for proposed methods)
is recalibrated to match the ATM volatility of the cross-currency pair for each business day,
and the smile is calculated using the recalibrated parameter.
In setting (d), the copula parameters are calibrated according to the smile of the cross-currency pair at the end of the previous month,
and the same parameters are used to calculate the smile of the cross-currency pair for each business day.

For the maturities 1Y and 5Y, the errors of the proposed method are smaller than those of the other copulas for most months.
In the case of 3M maturity, it is found that the proposed method has smaller errors for most months in setting (c), and the errors of the proposed method are similar to those of the other copulas in setting (d).
Therefore, it can be inferred that the parameter $\rho$ in our method plays an important role in explaining the daily fluctuations in volatility smiles of cross currency pairs,
especially in short-term maturities.

Therefore, the proposed method is more effective than the method that uses well-known copulas.

\section{Conclusion}
In this study, we proposed a method for constructing copulas using multivariate Hermite polynomial expansion,
a method for estimating parameters. We also proposed a correction for satisfying properties of density function using projection onto convex set in Hilbert spaces.
Furthermore, we proposed its application to a method for estimating the volatility smiles of cross currency pairs and compared the results with those of 
the volatility smiles of straight currency pairs.
We derive that the proposed copula parameters can adjust the higher-order moments
under risk-neutral measures for cross currency pairs.
Numerical experiments showed that the proposed copula can approximate many types of joint distributions.
Additionally, the application to the estimation of the volatility of the cross currency pairs shows that the proposed copula provides better estimates than the other copulas.

\bibliography{HermiteCopula}
\bibliographystyle{jplain}

\appendix
\def\thesection{Appendix \Alph{section}}
\section{Tables and Figures}
In Appendix, tables and graphs of Sections 5.1 and 5.2 are presented.

\begin{table}[H]
  \caption{Moments of Frank Copula with Normal Marginal}
  \label{table:MomentFrankOrg}
  \centering
  \begin{tabular}{|c|ccccccccc|}
    \hline
               &$x_{1}^{0}$&$x_{1}^{1}$&$x_{1}^{2}$&$x_{1}^{3}$&$x_{1}^{4}$&$x_{1}^{5}$&$x_{1}^{6}$&$x_{1}^{7}$&$x_{1}^{8}$\\\hline
    $x_{2}^{0}$&1.000&0.000&1.000&0.000&3.000&0.000&15.000&0.000&105.000\\
    $x_{2}^{1}$&0.000&0.570&0.000&1.423&0.000&6.114&-0.000&37.736&\\
    $x_{2}^{2}$&1.000&-0.000&1.361&-0.000&4.559&-0.000&23.865&&\\
    $x_{2}^{3}$&0.000&1.423&-0.000&3.781&-0.000&16.729&&&\\
    $x_{2}^{4}$&3.000&0.000&4.559&-0.000&15.886&&&&\\
    $x_{2}^{5}$&0.000&6.114&-0.000&16.729&&&&&\\
    $x_{2}^{6}$&15.000&-0.000&23.865&&&&&&\\
    $x_{2}^{7}$&0.000&37.736&&&&&&&\\
    $x_{2}^{8}$&105.000&&&&&&&&\\
    \hline
  \end{tabular}
\end{table}
\begin{table}[H]
  \caption{Moments of Fourth Hermite Expansion with Correction from Frank Copula}
  \label{table:MomentFrankCorr}
  \centering
  \begin{tabular}{|c|ccccccccc|}
    \hline
               &$x_{1}^{0}$&$x_{1}^{1}$&$x_{1}^{2}$&$x_{1}^{3}$&$x_{1}^{4}$&$x_{1}^{5}$&$x_{1}^{6}$&$x_{1}^{7}$&$x_{1}^{8}$\\\hline
    $x_{2}^{0}$&1.000&-0.000&1.000&0.000&3.000&-0.000&15.072&-0.000&107.551\\
    $x_{2}^{1}$&-0.000&0.570&-0.000&1.423&-0.000&5.875&-0.000&32.682&\\
    $x_{2}^{2}$&1.000&-0.000&1.361&-0.000&4.970&-0.000&29.094&&\\
    $x_{2}^{3}$&-0.000&1.423&-0.000&3.541&-0.000&14.229&&&\\
    $x_{2}^{4}$&3.000&-0.000&4.970&-0.000&20.087&&&&\\
    $x_{2}^{5}$&-0.000&5.875&-0.000&14.229&&&&&\\
    $x_{2}^{6}$&15.072&-0.000&29.094&&&&&&\\
    $x_{2}^{7}$&0.000&32.682&&&&&&&\\
    $x_{2}^{8}$&107.551&&&&&&&&\\
    \hline
  \end{tabular}
\end{table}

\begin{table}[H]
  \centering
  \caption{Moments of Gumbel Copula with Normal Marginal}
  \label{table:MomentGumbelOrg}
  \centering
  \begin{tabular}{|c|ccccccccc|}
    \hline
               &$x_{1}^{0}$&$x_{1}^{1}$&$x_{1}^{2}$&$x_{1}^{3}$&$x_{1}^{4}$&$x_{1}^{5}$&$x_{1}^{6}$&$x_{1}^{7}$&$x_{1}^{8}$\\\hline
    $x_{2}^{0}$&1.000&0.000&1.000&0.000&3.000&0.000&15.000&0.000&105.000\\
    $x_{2}^{1}$&0.000&0.622&0.175&1.911&1.014&9.716&7.333&68.788&\\
    $x_{2}^{2}$&1.000&0.175&1.900&1.138&8.647&8.580&58.767&&\\
    $x_{2}^{3}$&0.000&1.911&1.138&8.082&9.113&53.277&&&\\
    $x_{2}^{4}$&3.000&1.014&8.647&9.113&52.236&&&&\\
    $x_{2}^{5}$&0.000&9.716&8.580&53.277&&&&&\\
    $x_{2}^{6}$&15.000&7.333&58.767&&&&&&\\
    $x_{2}^{7}$&0.000&68.788&&&&&&&\\
    $x_{2}^{8}$&105.000&&&&&&&&\\
    \hline
  \end{tabular}
\end{table}
\begin{table}[H]
  \caption{Moments of Fourth Hermite Expansion with Correction from Gumbel Copula}
  \label{table:MomentGumbelCorr}
  \centering
  \begin{tabular}{|c|ccccccccc|}
    \hline
               &$x_{1}^{0}$&$x_{1}^{1}$&$x_{1}^{2}$&$x_{1}^{3}$&$x_{1}^{4}$&$x_{1}^{5}$&$x_{1}^{6}$&$x_{1}^{7}$&$x_{1}^{8}$\\\hline
    $x_{2}^{0}$&1.000&0.000&1.000&0.000&3.000&0.103&14.934&0.936&103.464\\
    $x_{2}^{1}$&0.000&0.622&0.175&1.911&0.745&9.242&4.234&60.994&\\
    $x_{2}^{2}$&1.000&0.175&1.900&0.625&8.029&2.984&49.564&&\\
    $x_{2}^{3}$&0.000&1.911&0.625&6.961&3.298&37.786&&&\\
    $x_{2}^{4}$&3.000&0.745&8.029&3.298&38.846&&&&\\
    $x_{2}^{5}$&0.103&9.242&2.984&37.786&&&&&\\
    $x_{2}^{6}$&14.934&4.234&49.564&&&&&&\\
    $x_{2}^{7}$&0.936&60.994&&&&&&&\\
    $x_{2}^{8}$&103.464&&&&&&&&\\
    \hline
  \end{tabular}
\end{table}

\begin{table}[H]
  \caption{Moments of Placett Copula with Normal Marginal}
  \label{table:MomentPlacettOrg}
  \centering
  \begin{tabular}{|c|ccccccccc|}
    \hline
               &$x_{1}^{0}$&$x_{1}^{1}$&$x_{1}^{2}$&$x_{1}^{3}$&$x_{1}^{4}$&$x_{1}^{5}$&$x_{1}^{6}$&$x_{1}^{7}$&$x_{1}^{8}$\\\hline
    $x_{2}^{0}$&1.000&0.000&1.000&0.000&3.000&-0.000&15.000&0.000&105.000\\
    $x_{2}^{1}$&0.000&0.579&0.000&1.507&0.000&6.624&0.000&41.358&\\
    $x_{2}^{2}$&1.000&0.000&1.533&0.000&5.442&0.000&29.348&&\\
    $x_{2}^{3}$&0.000&1.507&0.000&4.486&-0.000&21.085&&&\\
    $x_{2}^{4}$&3.000&0.000&5.442&0.000&20.823&&&&\\
    $x_{2}^{5}$&0.000&6.624&0.000&21.085&&&&&\\
    $x_{2}^{6}$&15.000&0.000&29.348&&&&&&\\
    $x_{2}^{7}$&0.000&41.358&&&&&&&\\
    $x_{2}^{8}$&105.000&&&&&&&&\\
    \hline
  \end{tabular}
\end{table}
\begin{table}[H]
  \caption{Moments of Fourth Hermite Expansion with Correction from Placett Copula}
  \label{table:MomentPlacettCorr}
  \centering
  \begin{tabular}{|c|ccccccccc|}
    \hline
               &$x_{1}^{0}$&$x_{1}^{1}$&$x_{1}^{2}$&$x_{1}^{3}$&$x_{1}^{4}$&$x_{1}^{5}$&$x_{1}^{6}$&$x_{1}^{7}$&$x_{1}^{8}$\\\hline
    $x_{2}^{0}$&1.000&0.000&1.000&0.000&3.000&0.000&15.070&0.000&106.761\\
    $x_{2}^{1}$&-0.000&0.579&0.000&1.507&-0.000&6.427&-0.000&37.382&\\
    $x_{2}^{2}$&1.000&-0.000&1.533&-0.000&6.031&0.000&36.969&&\\
    $x_{2}^{3}$&-0.000&1.507&-0.000&4.051&-0.000&17.464&&&\\
    $x_{2}^{4}$&3.000&-0.000&6.031&0.000&26.791&&&&\\
    $x_{2}^{5}$&0.000&6.427&0.000&17.464&&&&&\\
    $x_{2}^{6}$&15.070&-0.000&36.969&&&&&&\\
    $x_{2}^{7}$&0.000&37.382&&&&&&&\\
    $x_{2}^{8}$&106.761&&&&&&&&\\
    \hline
  \end{tabular}
\end{table}

\begin{figure}[H]
  \begin{tabular}{c}
    \begin{minipage}[t]{1.0\hsize}
      \centering
      \includegraphics[keepaspectratio, scale=0.8]{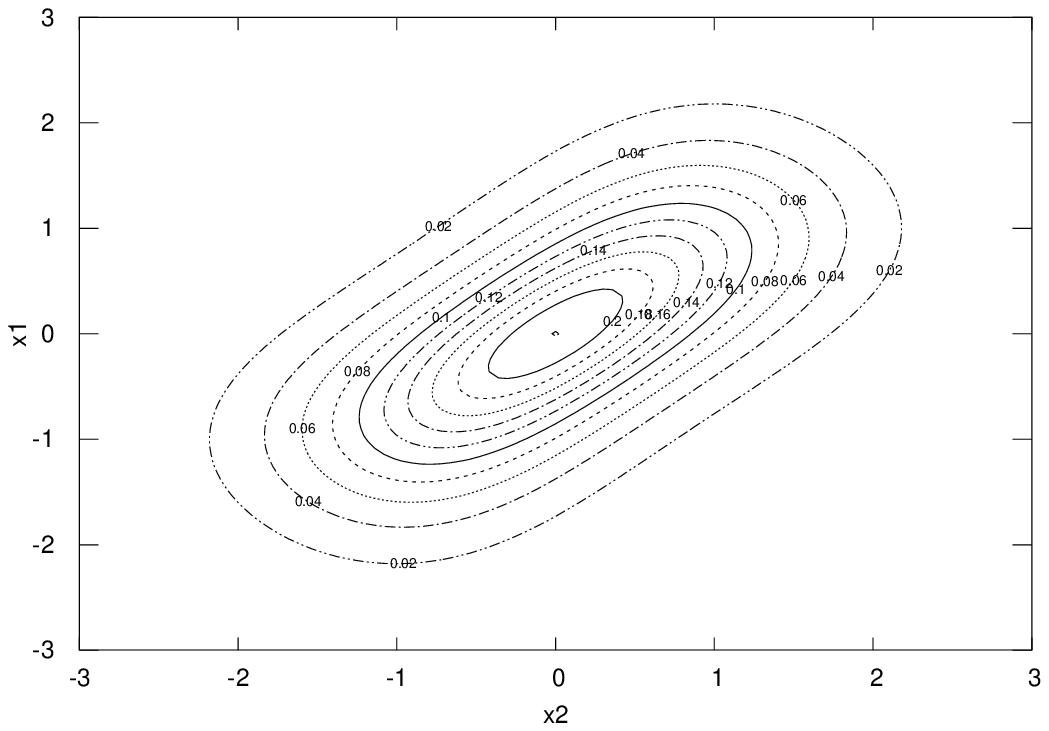}
      \caption{Density by Frank copula.}
      \label{figure:Exam1_Perp14Org}
    \end{minipage}
  \end{tabular}
  \begin{tabular}{cc}
    \begin{minipage}[t]{0.5\hsize}
      \centering
      \includegraphics[keepaspectratio, scale=0.6]{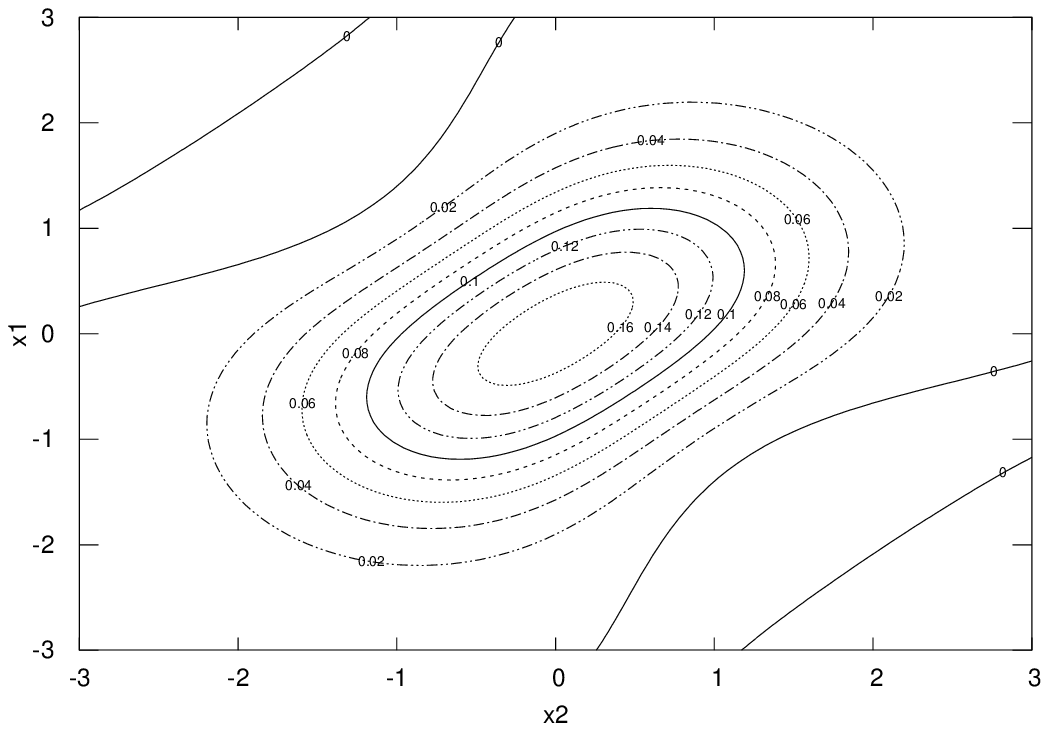}
      \caption{Expansion using (a) and (No correction) fitting to Frank copula distribution}
      \label{figure:Exam1_Perp14Raw}
    \end{minipage} &
    \begin{minipage}[t]{0.5\hsize}
      \centering
      \includegraphics[keepaspectratio, scale=0.6]{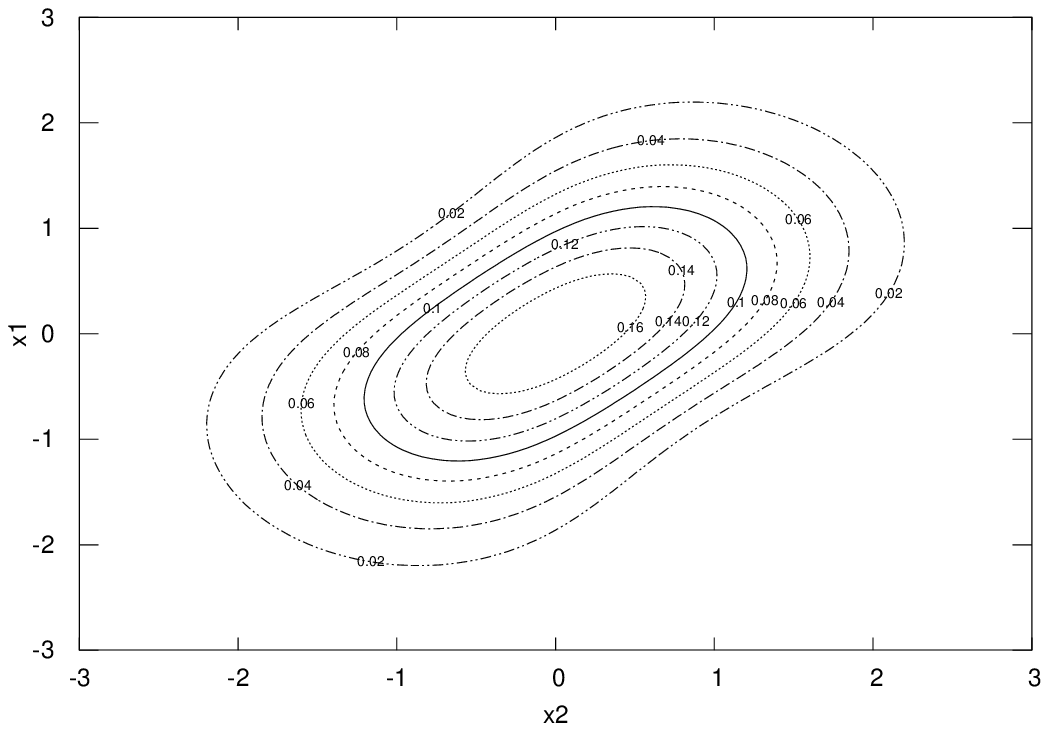}
      \caption{Expansion using (a) and (Applying correction) fitting to Frank copula distribution}
      \label{figure:Exam1_Perp14Corr}
    \end{minipage}\\
    \begin{minipage}[t]{0.5\hsize}
      \centering
      \includegraphics[keepaspectratio, scale=0.6]{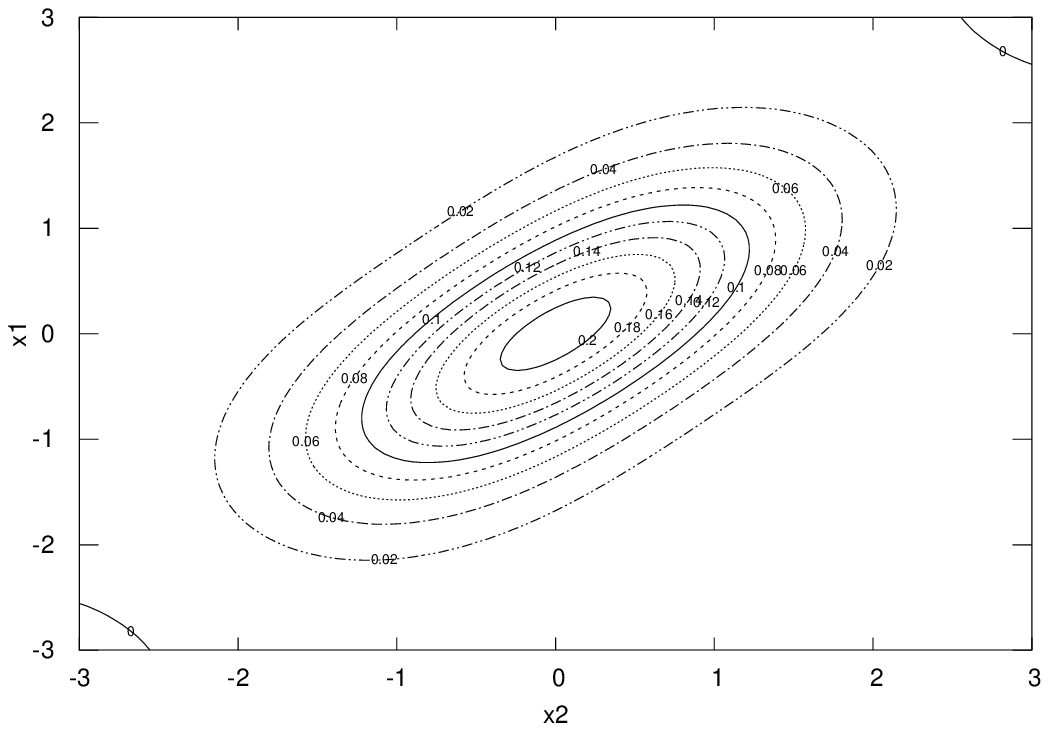}
      \caption{Expansion using (b) and (No correction) fitting to Frank copula distribution}
      \label{figure:Exam1_Rot14Raw}
    \end{minipage} &
    \begin{minipage}[t]{0.5\hsize}
      \centering
      \includegraphics[keepaspectratio, scale=0.6]{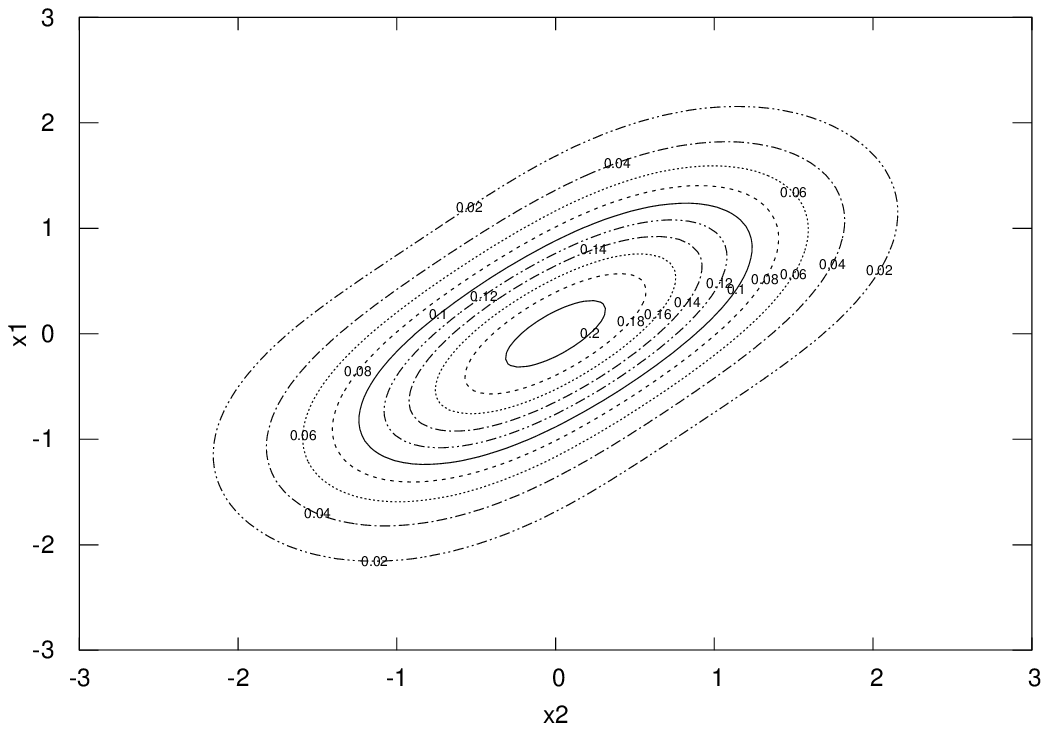}
      \caption{Expansion using (b) and (Applying correction) fitting to Frank copula distribution}
      \label{figure:Exam1_Rot14Corr}
    \end{minipage}
  \end{tabular}
\end{figure}

\begin{figure}[H]
  \begin{tabular}{c}
    \begin{minipage}[t]{1.0\hsize}
      \centering
      \includegraphics[keepaspectratio, scale=0.8]{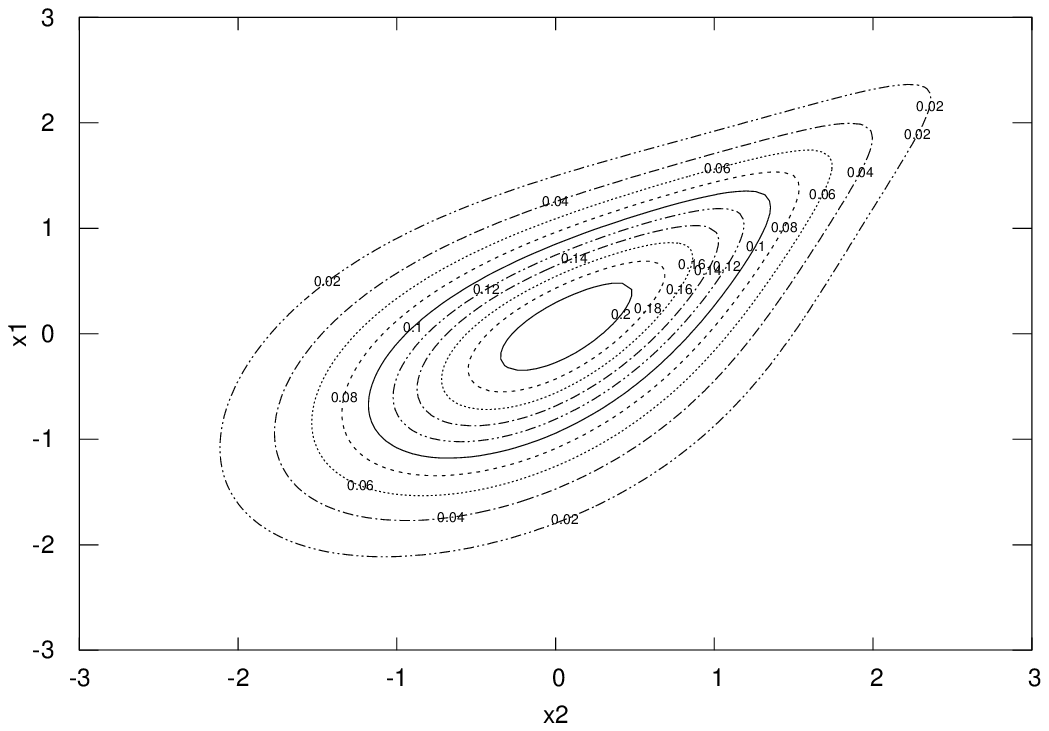}
      \caption{Density by Gumbel copula.}
      \label{figure:Exam1_Perp24Org}
    \end{minipage}
  \end{tabular}
  \begin{tabular}{cc}
    \begin{minipage}[t]{0.5\hsize}
      \centering
      \includegraphics[keepaspectratio, scale=0.6]{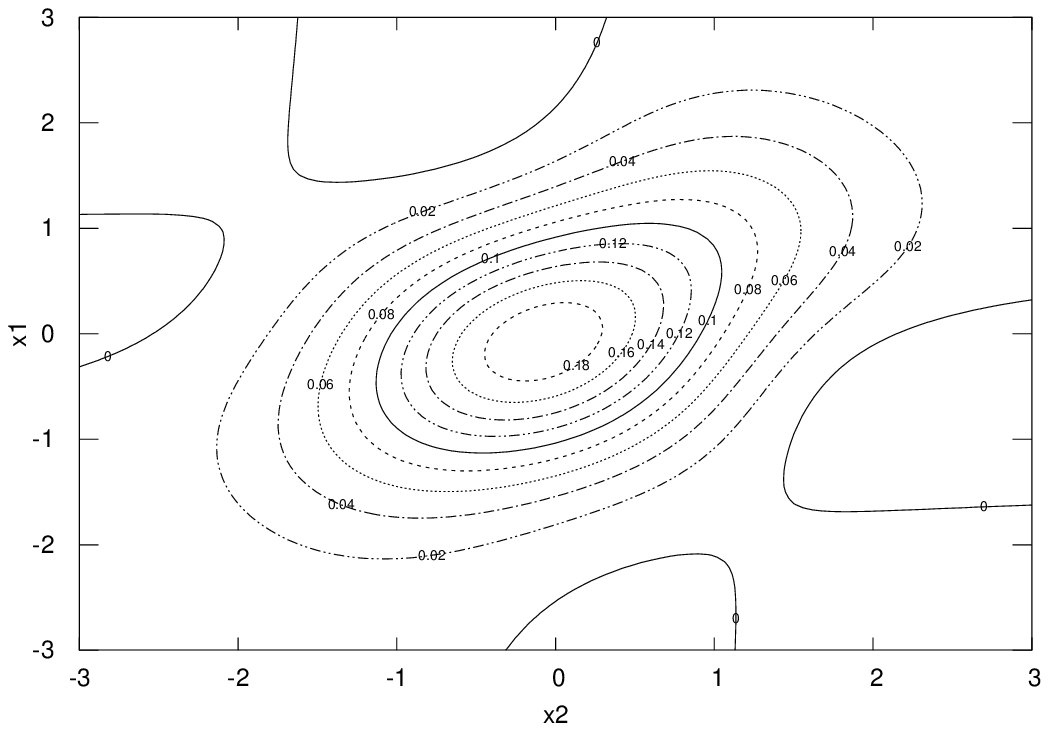}
      \caption{Expansion using (a) and (No correction) fitting to Gumbel copula distribution}
      \label{figure:Exam1_Perp24Raw}
    \end{minipage} &
    \begin{minipage}[t]{0.5\hsize}
      \centering
      \includegraphics[keepaspectratio, scale=0.6]{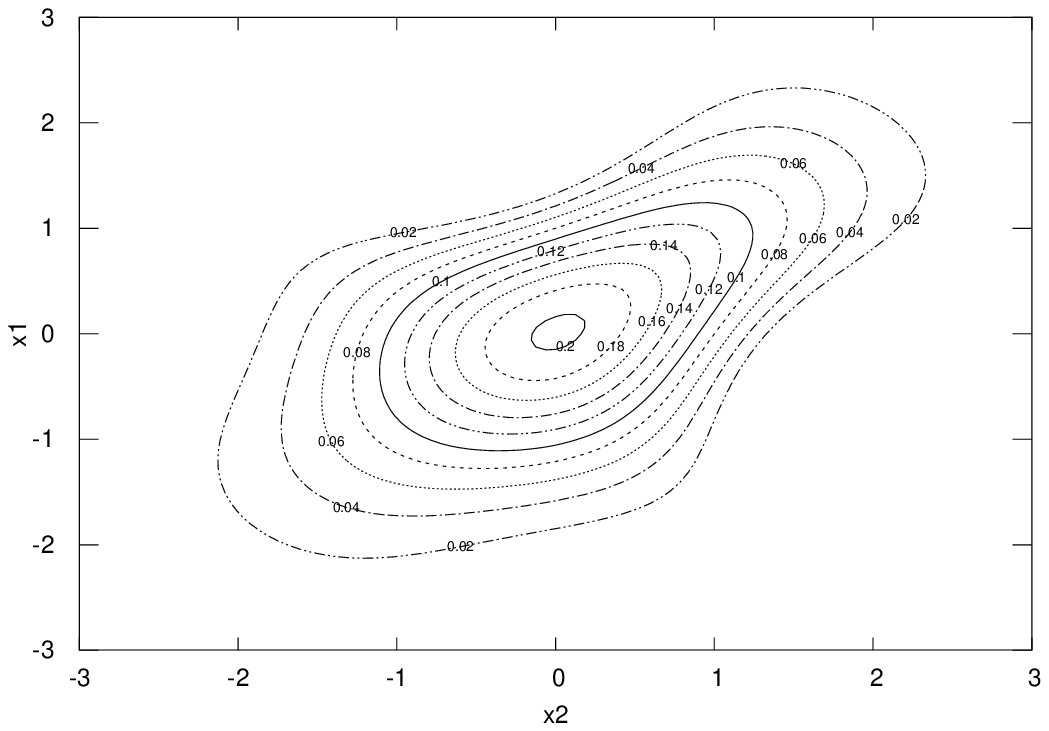}
      \caption{Expansion using (a) and (Applying correction) fitting to Gumbel copula distribution}
      \label{figure:Exam1_Perp24Corr}
    \end{minipage}\\
    \begin{minipage}[t]{0.5\hsize}
      \centering
      \includegraphics[keepaspectratio, scale=0.6]{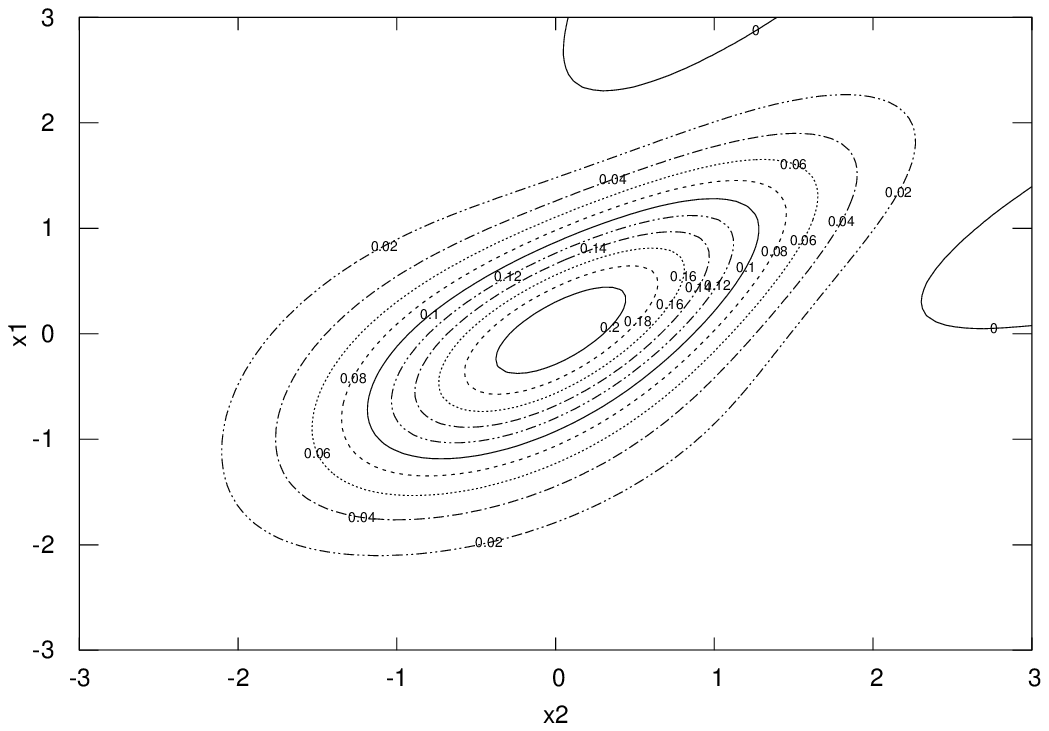}
      \caption{Expansion using (b) and (No correction) fitting to Gumbel copula distribution}
      \label{figure:Exam1_Rot24Raw}
    \end{minipage} &
    \begin{minipage}[t]{0.5\hsize}
      \centering
      \includegraphics[keepaspectratio, scale=0.6]{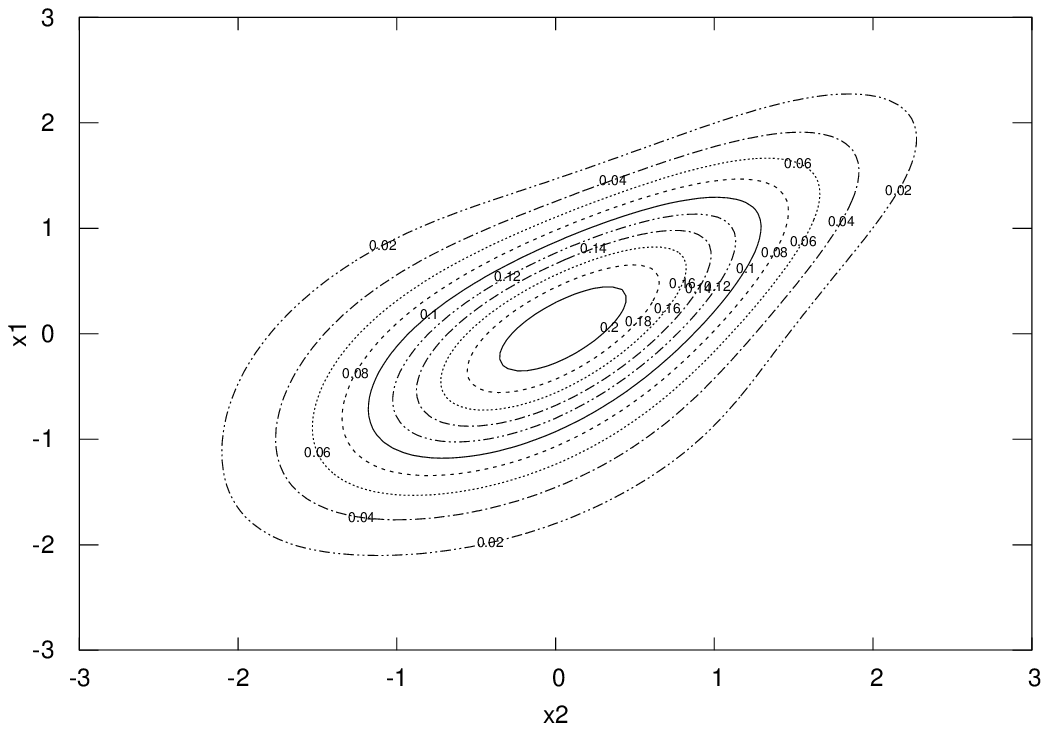}
      \caption{Expansion using (b) and (Applying correction) fitting to Gumbel copula distribution}
      \label{figure:Exam1_Rot24Corr}
    \end{minipage}
  \end{tabular}
\end{figure}

\begin{figure}[H]
  \begin{tabular}{c}
    \begin{minipage}[t]{1.0\hsize}
      \centering
      \includegraphics[keepaspectratio, scale=0.8]{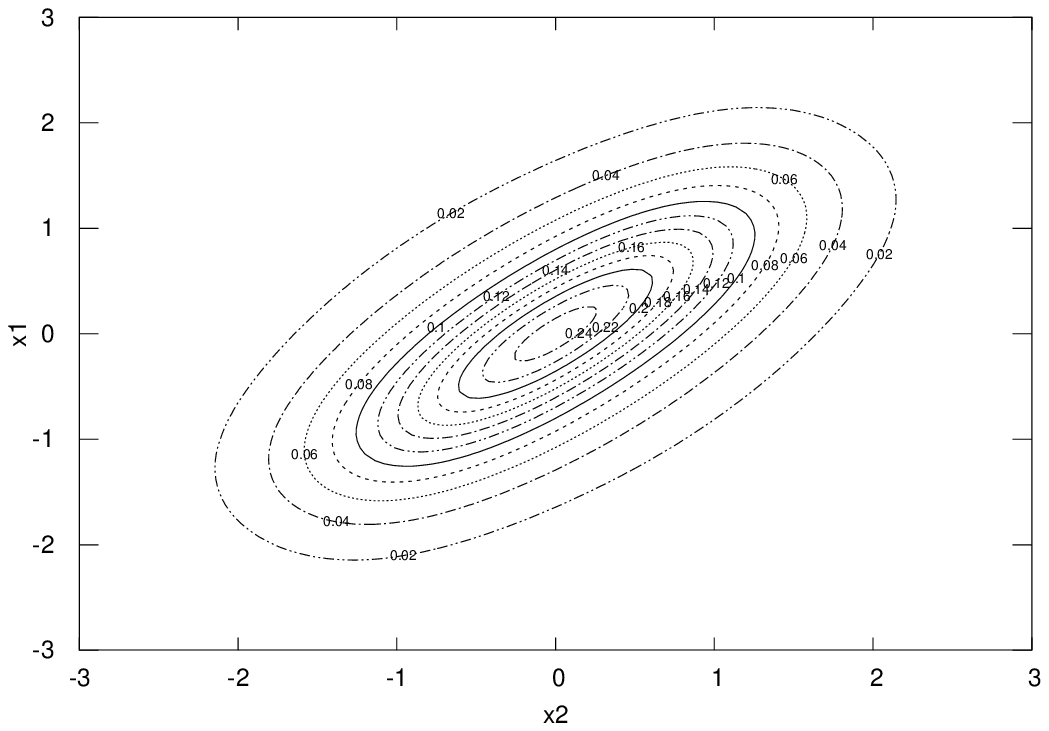}
      \caption{Density by Plackett copula.}
      \label{figure:Exam1_Perp34Org}
    \end{minipage}
  \end{tabular}
  \begin{tabular}{cc}
    \begin{minipage}[t]{0.5\hsize}
      \centering
      \includegraphics[keepaspectratio, scale=0.6]{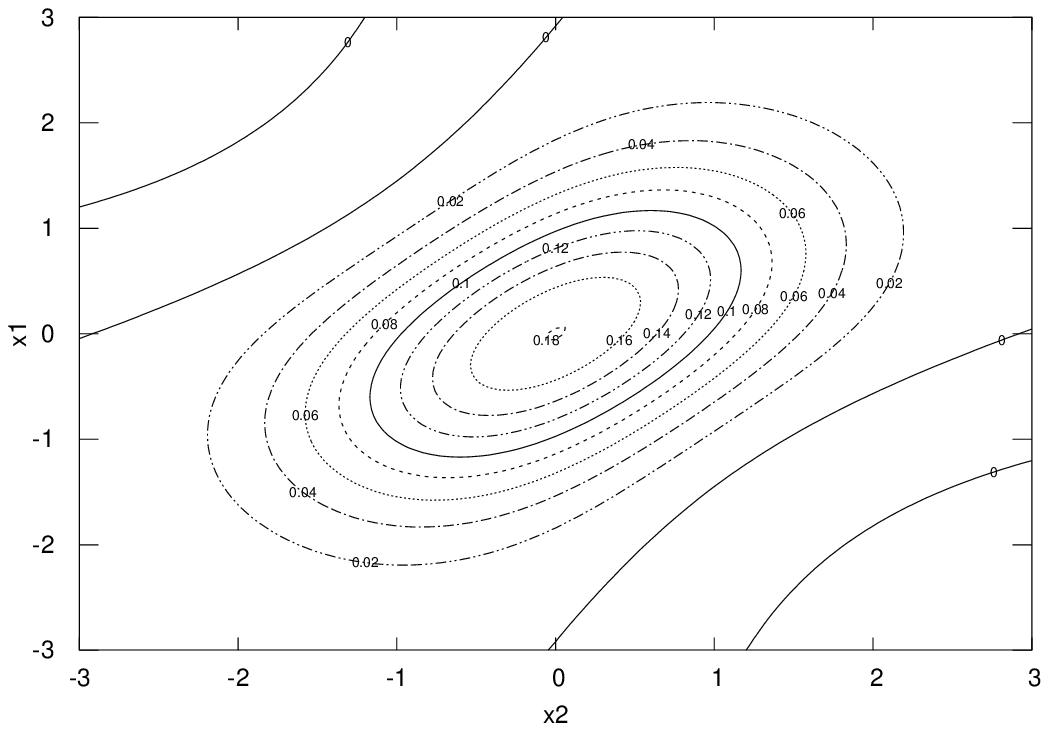}
      \caption{Expansion using (a) and (No correction) fitting to Plackett copula distribution}
      \label{figure:Exam1_Perp34Raw}
    \end{minipage} &
    \begin{minipage}[t]{0.5\hsize}
      \centering
      \includegraphics[keepaspectratio, scale=0.6]{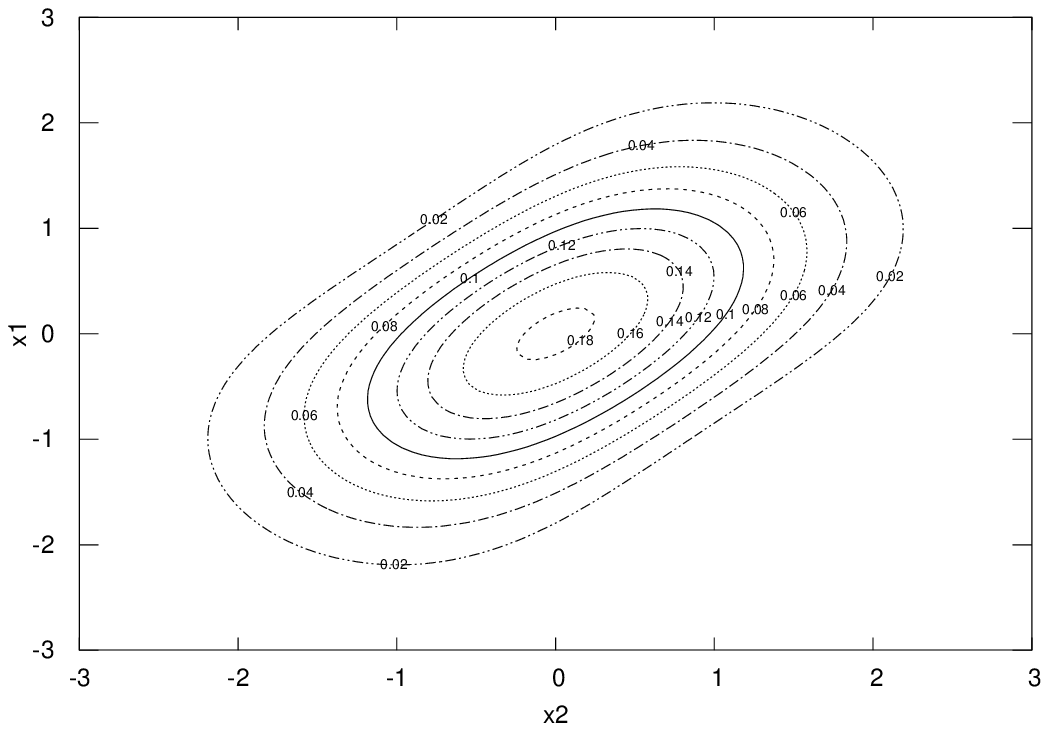}
      \caption{Expansion using (a) and (Applying correction) fitting to Plackett copula distribution}
      \label{figure:Exam1_Perp34Corr}
    \end{minipage}\\
    \begin{minipage}[t]{0.5\hsize}
      \centering
      \includegraphics[keepaspectratio, scale=0.6]{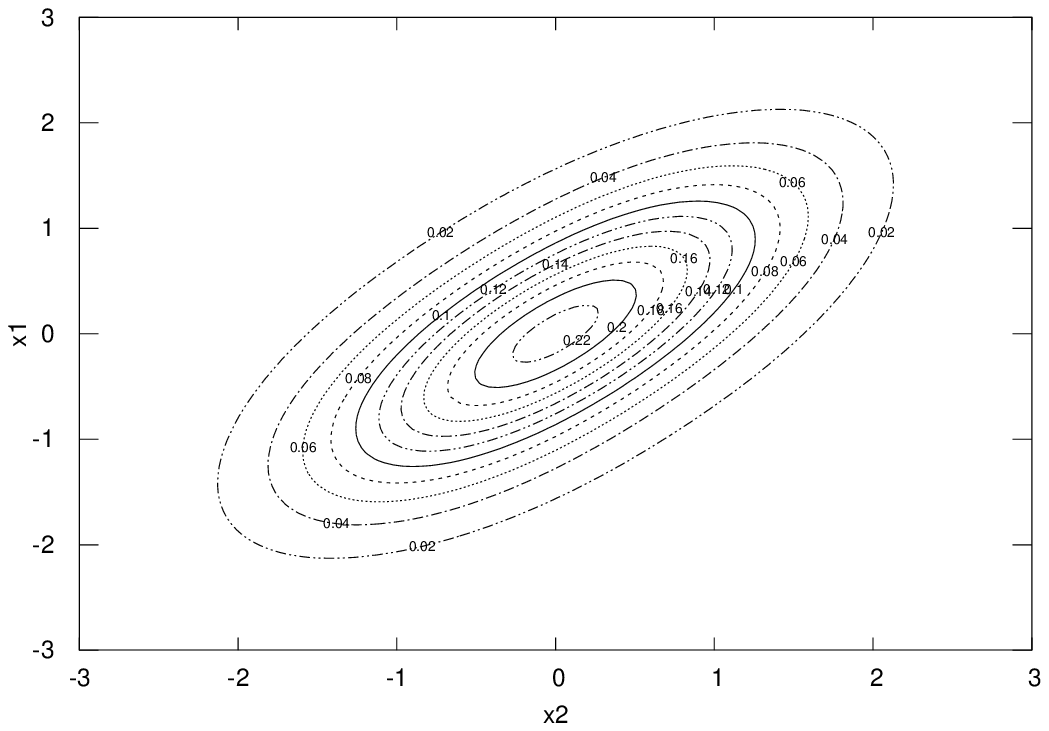}
      \caption{Expansion using (b) and (No correction) fitting to Plackett copula distribution}
      \label{figure:Exam1_Rot34Raw}
    \end{minipage} &
    \begin{minipage}[t]{0.5\hsize}
      \centering
      \includegraphics[keepaspectratio, scale=0.6]{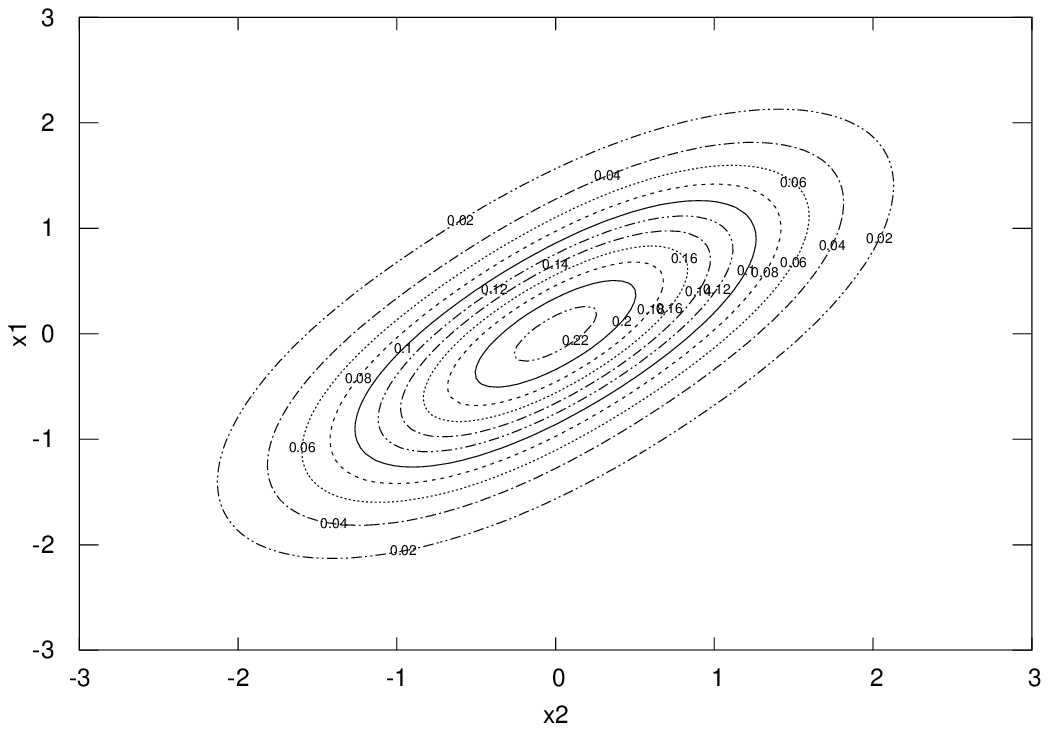}
      \caption{Expansion using (b) and (Applying correction) fitting to Plackett copula distribution}
      \label{figure:Exam1_Rot34Corr}
    \end{minipage}
  \end{tabular}
\end{figure}

\begin{figure}[H]
  \begin{tabular}{cc}
    \begin{minipage}[t]{0.5\hsize}
      \centering
      \includegraphics[keepaspectratio, scale=0.6]{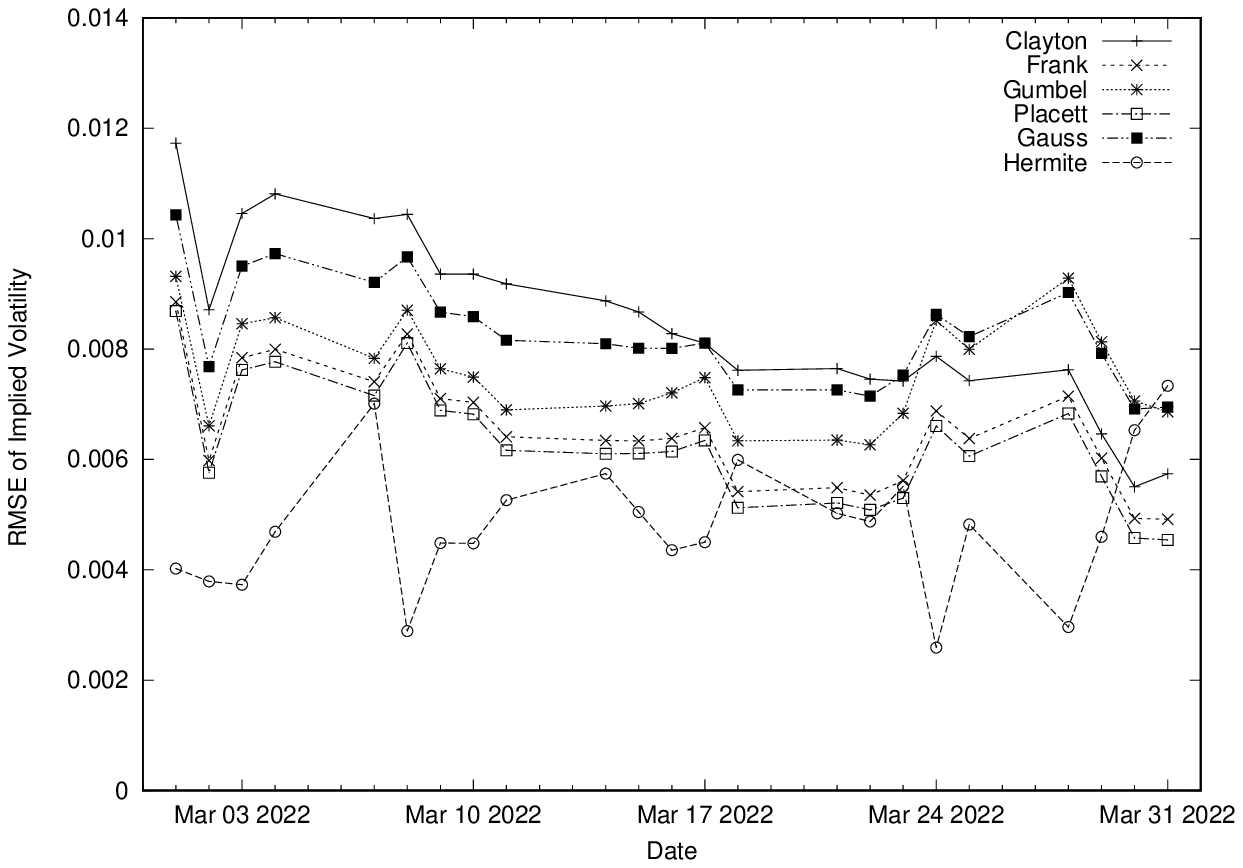}
      \caption{Daily RMSE of tenor 3M and (c) in March, 2022}
      \label{figure:Exam2_202203ForecastC3M}
    \end{minipage} &
    \begin{minipage}[t]{0.5\hsize}
      \centering
      \includegraphics[keepaspectratio, scale=0.6]{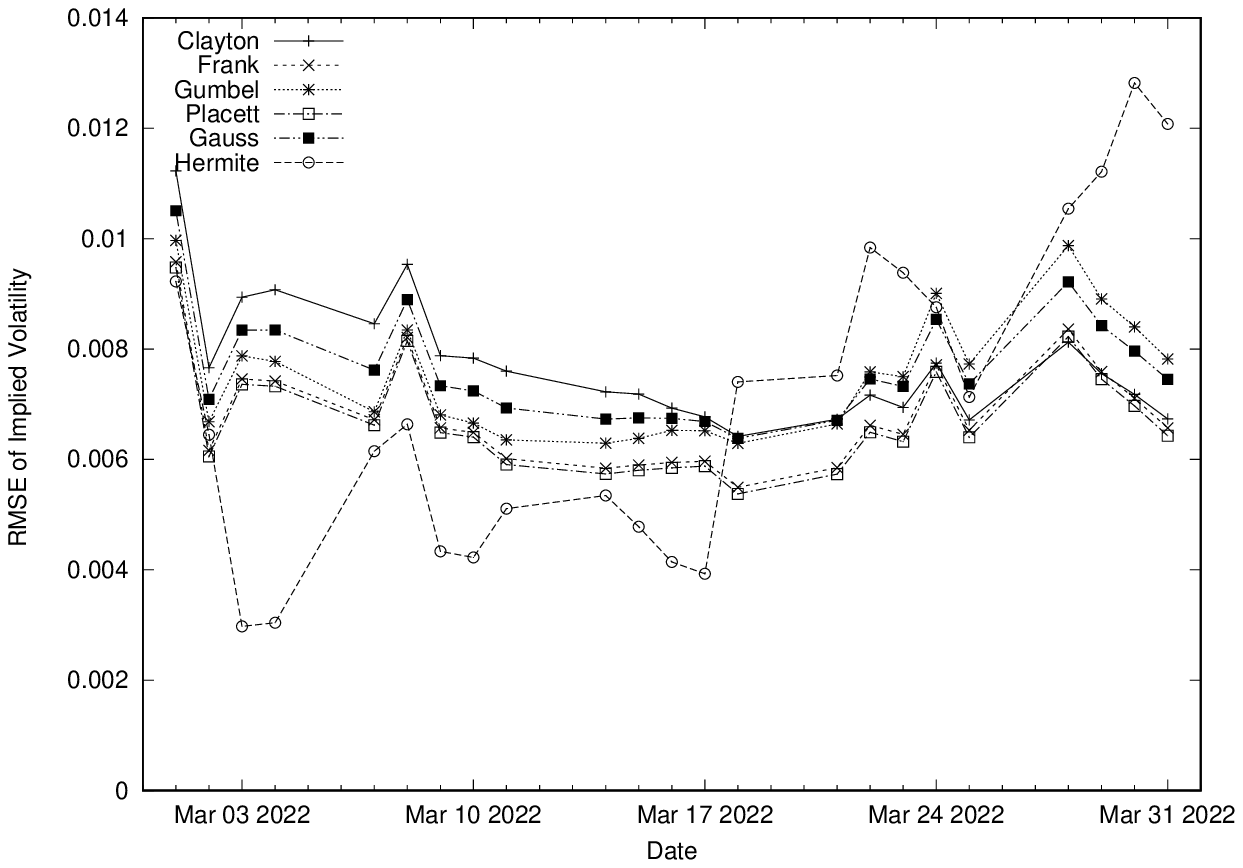}
      \caption{Daily RMSE of tenor 3M and (d) in March, 2022}
      \label{figure:Exam2_202203ForecastD3M}
    \end{minipage}
  \end{tabular}
\end{figure}
\begin{figure}[H]
  \begin{tabular}{cc}
    \begin{minipage}[t]{0.5\hsize}
      \centering
      \includegraphics[keepaspectratio, scale=0.6]{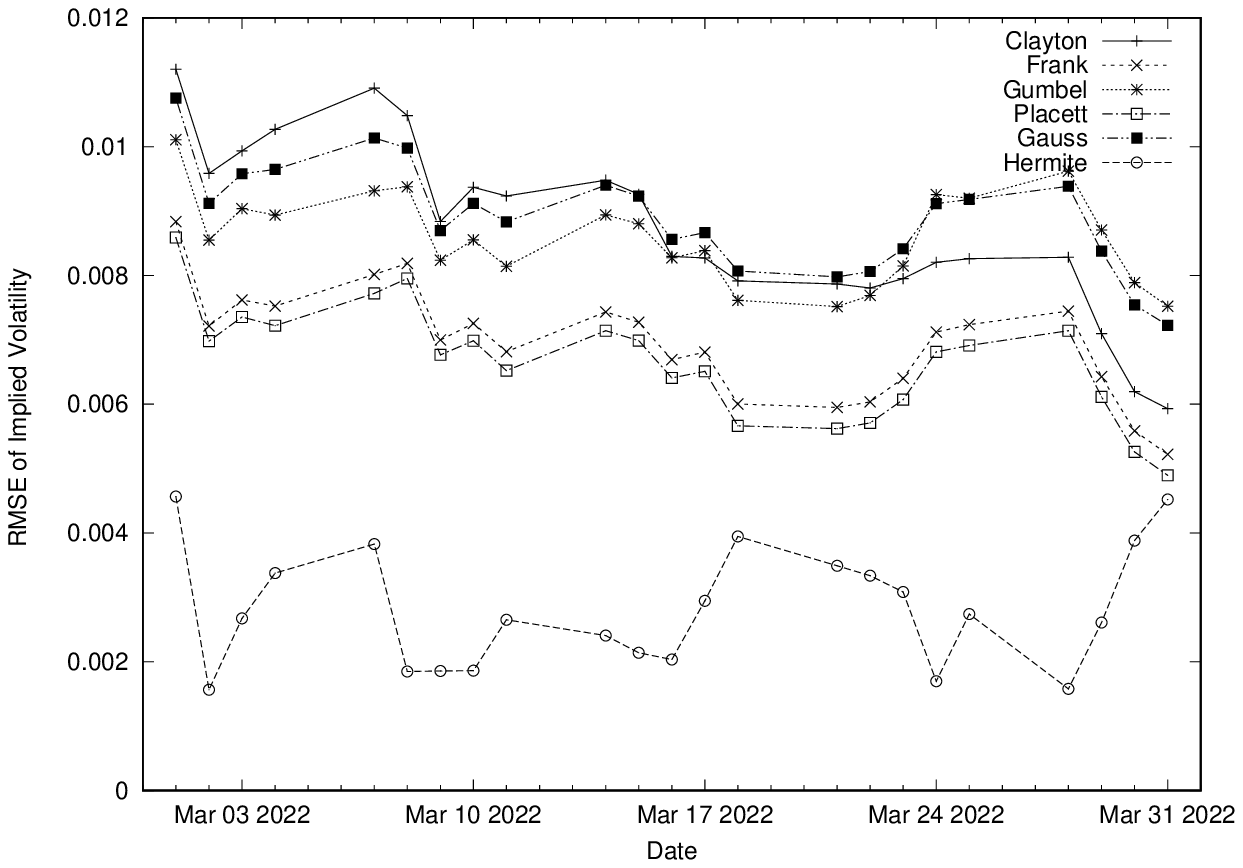}
      \caption{Daily RMSE of tenor 1Y and (c) in March, 2022}
      \label{figure:Exam2_202203ForecastC1Y}
    \end{minipage} &
    \begin{minipage}[t]{0.5\hsize}
      \centering
      \includegraphics[keepaspectratio, scale=0.6]{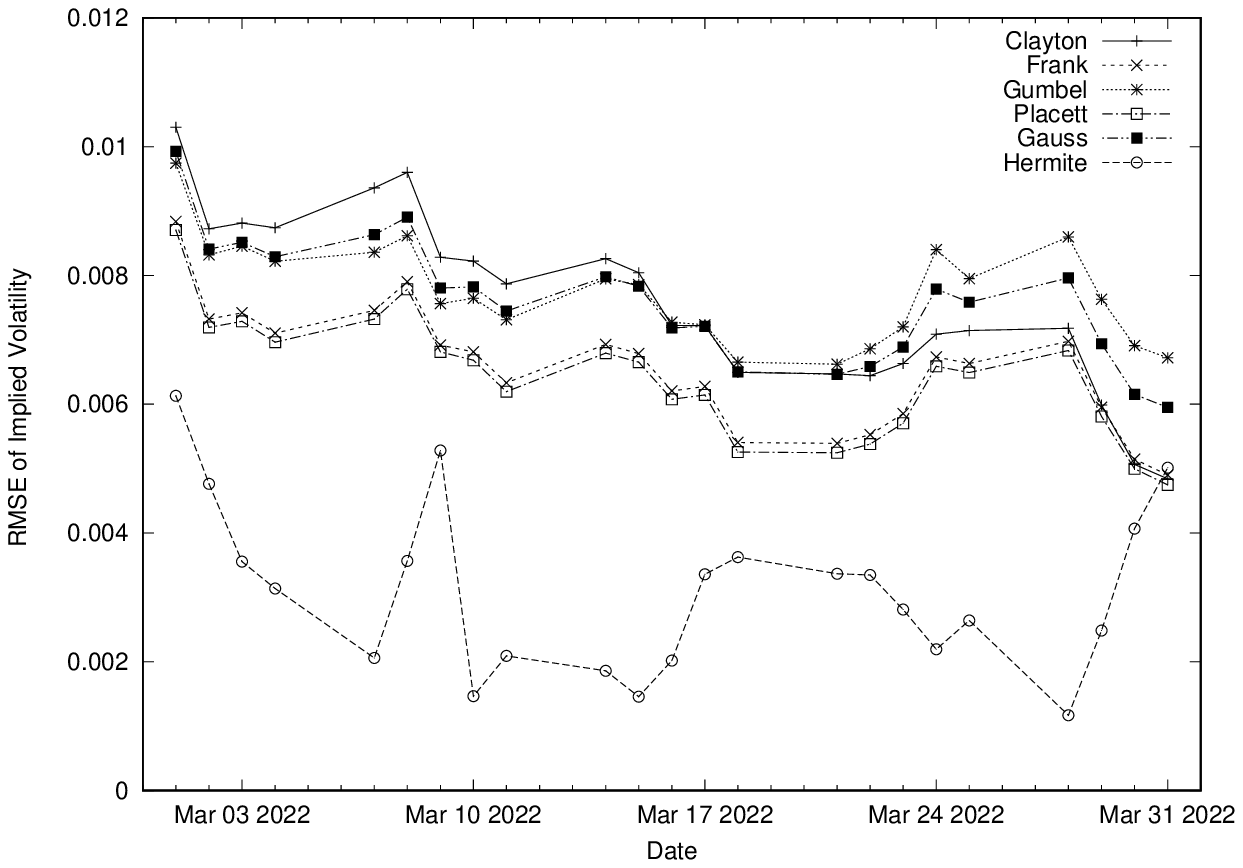}
      \caption{Daily RMSE of tenor 1Y and (d) in March, 2022}
      \label{figure:Exam2_202203ForecastD1Y}
    \end{minipage}
  \end{tabular}
\end{figure}
\begin{figure}[H]
  \begin{tabular}{cc}
    \begin{minipage}[t]{0.5\hsize}
      \centering
      \includegraphics[keepaspectratio, scale=0.6]{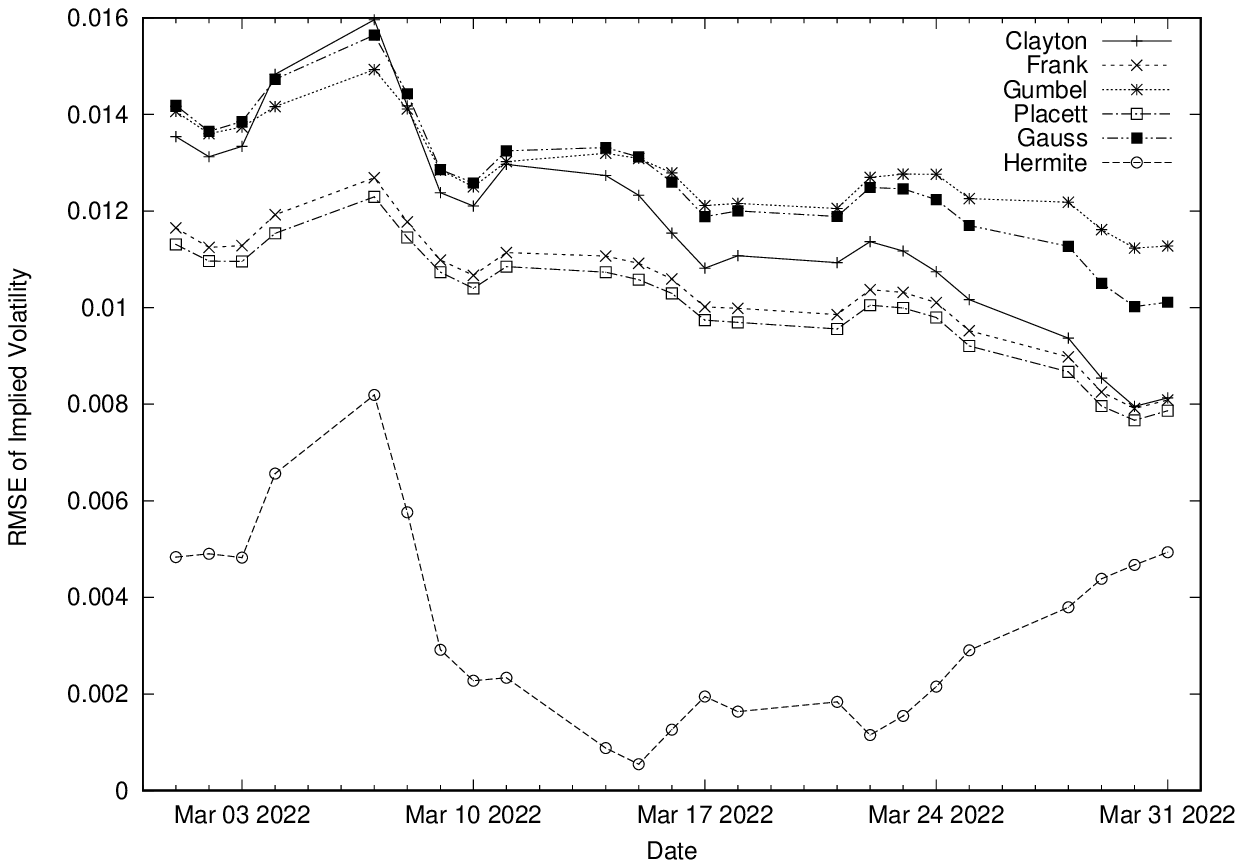}
      \caption{Daily RMSE of tenor 5Y and (c) in March, 2022}
      \label{figure:Exam2_202203ForecastC5Y}
    \end{minipage} &
    \begin{minipage}[t]{0.5\hsize}
      \centering
      \includegraphics[keepaspectratio, scale=0.6]{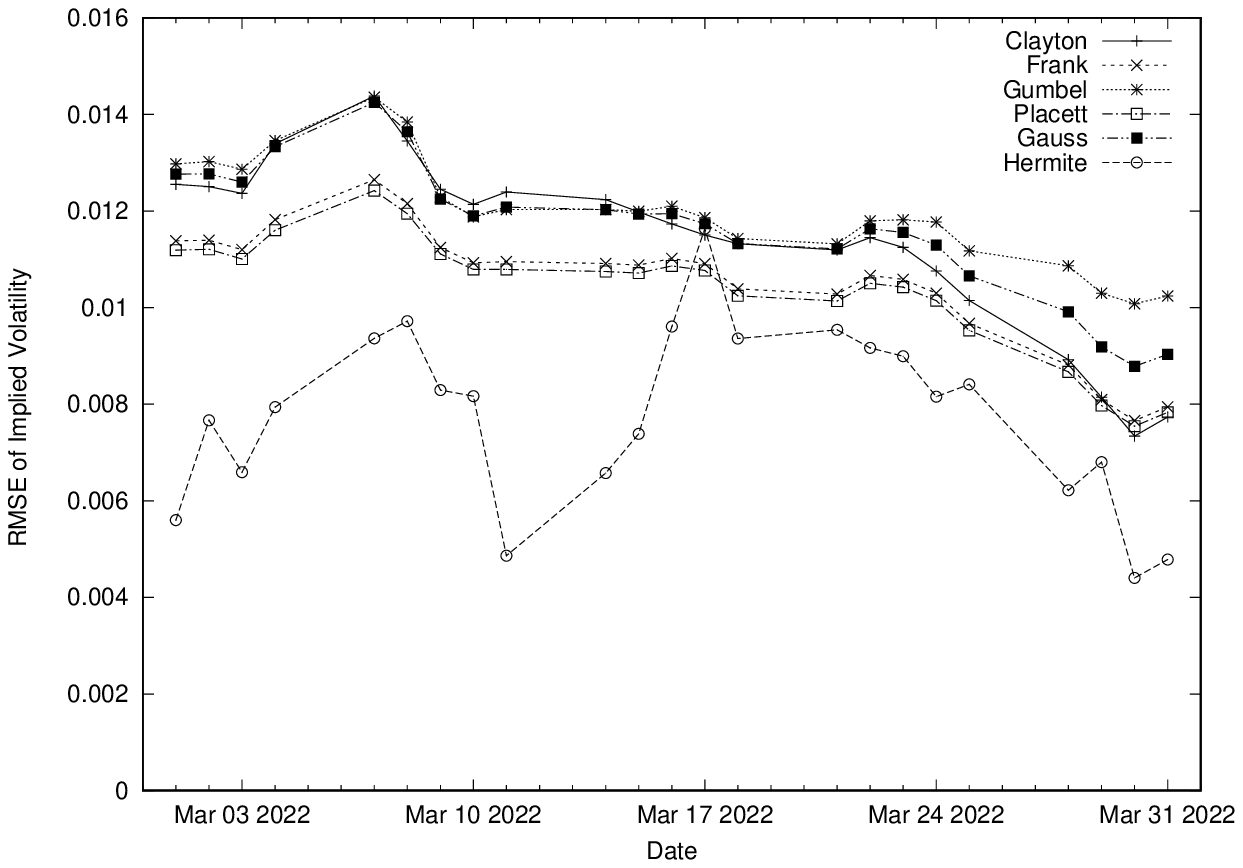}
      \caption{Daily RMSE of tenor 5Y and (d) in March, 2022}
      \label{figure:Exam2_202203ForecastD5Y}
    \end{minipage}
  \end{tabular}
\end{figure}

\end{document}